\documentclass[aps,prx, amsfonts, amssymb, amsmath, reprint, showkeys, nofootinbib, superscriptaddress]{revtex4-1}

\usepackage[english]{babel}
\usepackage[utf8]{inputenc}
\usepackage{soul}
\usepackage[dvipsnames]{xcolor}
\usepackage{graphicx}

\usepackage{amsfonts}
\usepackage{physics}
\usepackage{comment}
\usepackage{MnSymbol}

\newcommand{\hhh}{\textbf{H}}

\newcommand{\aaa}{\textbf{a}}

\newcommand{\nnn}{\textbf{n}}
\newcommand{\bbb}{\textbf{b}}

\newcommand{\ppphi}{\boldsymbol{\phi}}

\newcommand{\rrr}{\boldsymbol{\rho}}

\usepackage[pdftex, pdftitle={Article}, pdfauthor={Author}]{hyperref} \hypersetup{
    colorlinks = true,
    linkcolor = black,
    citecolor = black,
    linkbordercolor = {white},
    urlcolor = black
}
\usepackage[normalem]{ulem}

\usepackage{cleveref}
\crefname{equation}{Eq.}{Eqs.}
\Crefname{equation}{Equation}{Equations}
\crefname{figure}{Fig.}{Figs.}
\Crefname{figure}{Figure}{Figures}
\crefname{section}{Sec.}{Sects.}
\Crefname{section}{Section}{Sections}
\crefname{table}{Table}{Tables}
\crefname{appendix}{Appendix}{Apps.}
\Crefname{appendix}{Appendix}{Apps.}

\newcommand{\ie}{i.e.~}
\newcommand{\twd}{\tilde\omega_d}
\newcommand{\heff}{\hbar_\text{eff}}
\newcommand{\Nt}{N_t}
\newcommand{\Nr}{N_r}

\begin{document}

\title{Reminiscence of classical chaos in driven transmons}

\author{Joachim Cohen}
\email{cohenj38@gmail.com}
\affiliation{Institut Quantique and D\'epartement de Physique, Universit\'e de Sherbrooke, Sherbrooke, Qu\'ebec, J1K 2R1, Canada}
\author{Alexandru Petrescu}
\email{alexandru.petrescu@minesparis.psl.eu}
\affiliation{Institut Quantique and D\'epartement de Physique, Universit\'e de Sherbrooke, Sherbrooke, Qu\'ebec, J1K 2R1, Canada}
\affiliation{LPENS, D\'{e}partement de physique, Ecole normale sup\'{e}rieure, Centre Automatique et Syst\`{e}mes (CAS), MINES ParisTech, Universit\'{e} PSL, Sorbonne Universit\'{e}, CNRS, Inria, 75005 Paris, France}
\author{Ross Shillito}
\affiliation{Institut Quantique and D\'epartement de Physique, Universit\'e de Sherbrooke, Sherbrooke, Qu\'ebec, J1K 2R1, Canada}
\author{Alexandre Blais}
\affiliation{Institut Quantique and D\'epartement de Physique, Universit\'e de Sherbrooke, Sherbrooke, Qu\'ebec, J1K 2R1, Canada}
\affiliation{Canadian Institute for Advanced Research, Toronto, M5G1M1 Ontario, Canada}

\date{\today} 

\begin{abstract}
Transmon qubits are ubiquitously used in superconducting quantum information processor architectures. Strong drives are required to realize fast, high-fidelity, gates and measurements, including parametrically activated processes. Here, we show that even off-resonant drives, in regimes routinely used in experiments, can cause strong modifications to the structure of the transmon spectrum rendering a large part of it chaotic.
Accounting for the full nonlinear dynamics of the transmon in a Floquet-Markov formalism, we find that these chaotic states, often neglected through the hypothesis that the anharmonicity is weak, strongly impact the lifetime of the transmon's computational states.
In particular, we observe that chaos-assisted quantum phase slips greatly enhance band dispersions. In the presence of a measurement resonator, we find that approaching chaotic behavior correlates with strong transmon-resonator hybridization, and an average resonator response centered on the bare resonator frequency. These results lead to a photon number threshold characterizing the appearance of chaos-induced quantum demolition effects during strong-drive operations such as dispersive qubit readout. The phenomena described here are expected to 
be present in all circuits based on low-impedance Josephson-junctions.
\end{abstract}

\maketitle
 
\setcounter{secnumdepth}{5}

\section{Introduction}

Low-impedance Josephson junction circuits,
where the Josephson energy dominates over the charging energy, are fundamental building blocks of superconducting quantum processors. Although the most widely used superconducting qubit is the transmon~\cite{koch_charge-insensitive_2007}, capacitively shunted Josephson junctions appear in other species of qubits, such as the heavy fluxonium~\cite{earnest_realization_2018, lin_demonstration_2018}, the $0-\pi$ qubit~\cite{gyenis_experimental_2021} and the capacitively shunted flux qubit~\cite{You2007,yan_flux_2016}. 
Josephson junctions can also be used as simple nonlinear elements for parametrically activated multi-wave mixing~\cite{ Mirrahimi_2014, Leghtas_science_2015, lescanne_exponential_2020, sivak_kerr-free_2019}, or as linear inductive elements in a Josephson junction array to realize superinductances~\cite{masluk_microwave_2012, manucharyan_evidence_2012}.

To meet the requirements of quantum information processing with fast and high-fidelity operations, strong driving fields that are off-resonant from the qubit are often used in parametrically activated coupling~\cite{Mirrahimi_2014, Lescanne_2019}, multi-qubit gates~\cite{sheldon_procedure_2016,Paik2016} and dispersive readout~\cite{Blais_2004}. However, off-resonant drives of even moderate amplitude are often observed to cause spurious qubit transitions~\cite{walter_rapid_2017,minev_catch_2019,Grimm2020}. This is the case of the dispersive readout whose quantum-non demolition (QND) character is observed only at very small drive amplitudes, corresponding to a few photons ($\bar n \sim 2$) populating the measurement resonator~\cite{walter_rapid_2017}. Models based on perturbative expansion in the qubit-resonator coupling~\cite{boissonneault_dispersive_2009} or in qubit anharmonicity and drive amplitude~\cite{Malekakhlagh_2020,Petrescu_2020} have been explored to understand the origin of these unwanted transitions. At large photon numbers ($\bar n \gtrsim 100$), qubit-resonator resonances \cite{sank_measurement-induced_2016} and structural instabilities \cite{VerneyPRApplied2019,Lescanne_2019} have been shown to result in spurious transitions. Recently, a numerical study of the full time dynamics of the transmon has shown that qubit-resonator resonances can lead to leakage of the transmon population to states lying above the Josephson junction cosine potential, something which has been referred to as ionization~\cite{Shillito2022}. In that study, ionization was shown to coincide with the loss of QNDness experimentally observed at low photon numbers in Ref.~\cite{walter_rapid_2017}.

Here, we show that the often-neglected highly excited states of the transmon can play an important role for current experimental parameters and drive amplitudes, even in the absence of ionization~\cite{sank_measurement-induced_2016,WalterPRL2017,Lescanne_2019}. 
We find that a subset of the highly-excited states in the spectrum of a driven transmon, the so-called chaotic layer, significantly and unexpectedly dresses the charge dispersion of the low-energy transmon spectrum, with detrimental effects on the qubit dephasing time. This phenomenon can be understood as chaos-assisted \cite{tomsovic_chaos-assisted_1994} quantum phase slips \cite{matveev_et_al_2002}. The increased dependence on the offset charge even for the low-energy states suggests that in the presence of strong drives, models which rely on a perturbed harmonic oscillator such as the Kerr nonlinear oscillator \cite{Nigg_2012} cannot
give an accurate description of the system, in particular because the selection rules derived from 
these models are no longer applicable~\cite{sank_measurement-induced_2016}.
In addition, we show that the chaotic layer makes steady-state populations deviate significantly from the Boltzmann distribution \cite{breuer_quasistationary_2000, ketzmerick_statistical_2010, VerneyPRApplied2019}.

We draw upon the Floquet theory \cite{GRIFONI1998229, breuer_quasistationary_2000,  ketzmerick_statistical_2010} of nonlinear oscillators \cite{chirikov_universal_1979} to distinguish between the chaotic and regular states \cite{percival_regular_1973, berry_regular_1977} of a driven transmon. We introduce a rescaling of the transmon Hamiltonian from which an effective Planck constant emerges, $\hbar_\textrm{eff}=\sqrt{8 E_C/E_J}$ ($\hbar = 1$). In the transmon regime, where $\hbar_\textrm{eff}$ is small, the chaotic dynamics of the classical driven transmon becomes more resolved in the quantum spectrum as the number of chaotic states increases.
In particular, we show that the spectrum of a single driven transmon is correlated \cite{haake_quantum_2010}. In transmon systems, this type of analysis has been performed in the many-body regime \cite{Berke_2022}. Our study reveals that, within the range of current experimental parameters, the size of the classical chaotic domain is strongly sensitive to the drive frequency, something which can lead to instabilities in the quantum dynamics even at low drive power. Our results suggest ways to avoid experimental realizations of transmons from being plagued by these instabilities.

Moreover, by simulating the full transmon plus resonator circuit QED model, we show that chaos develops along two directions, that of increasing resonator Fock state number and that of increasing drive power. In addition to validating the study of the single driven transmon, we show that the transmon and the resonator strongly hybridize in the chaotic phase. As a result, in the context of the dispersive readout where chaotic effects are present, spurious qubit transitions become possible. In particular, we introduce a critical photon number around which chaos-induced  non-QND effects are expected. We also predict that spurious effects below this threshold should be exponentially reduced with $\sqrt{E_J/ 8 E_C}$.

The remainder of this article is organized as follows. In \cref{sec:driven_transmon}, we introduce a rescaled version of the Hamiltonian along with the key parameters of the dynamics, discuss chaos in the classical driven transmon, and briefly discuss the impact of classical chaos on the quantum system. \cref{sec:spectral_properties} tackles the spectral properties of the system and the dependence of the instability on drive frequency. In \cref{sec:bath_coupling}, we consider the coupling of the transmon to a bath, and the impact of the chaotic layer on the coherence properties of the low-energy sector is analyzed. \cref{sec:cQEDsim} focuses on the interplay of chaos and the validity of the dispersive approximation in the full circuit QED setup. \cref{sec:Ncrit} discusses spurious non-QND effects originating from the interaction with chaotic states.

\section{Periodically driven transmon}
\label{sec:driven_transmon}

The Hamiltonian of a transmon in a typical circuit QED setup takes the form~\cite{koch_charge-insensitive_2007,Blais2021}
\begin{align}
\hhh(t) =   4 E_C (\nnn-n_g)^2-E_J\cos(\ppphi) +\nnn \textbf{F},
\end{align}
where $\nnn$ and $\ppphi$ are respectively the charge and phase operators, $E_C$ and $E_J$ are the charging and Josephson energies, and $n_g$ an offset charge. The phase $\ppphi$ is compact and takes its values in the range $(-\pi,\pi]$.
The operator $\textbf{F}$ represents a classical driving field on the resonator or the coupling to a measurement resonator in a circuit QED setup~\cite{Blais2021}. As such, 
$\textbf{F} = F_\text{c}(t)+\textbf{F}_\text{q}$ can be expressed as the sum of a classical part and of a quantum part.
The quantum part
represents the displaced quadrature of the readout resonator, \ie $\textbf{F}_\text{q} = i g(\aaa^\dagger-\aaa)$ where $g$ is the light-matter coupling.
On the other hand, the classical part is assumed to take the form $F_\text{c}(t)= \varepsilon_d\cos(\omega_d t)$ and represents either a direct capacitive drive on the transmon or the classical amplitude of the resonator field. 
In this work, we are concerned with the limit of strong drives, where the classical part dominates over quantum fluctuations, \ie $\varepsilon_d > g$.
Expressing the drive amplitude $\varepsilon_d = 2g\sqrt{\bar{n}}$ in terms of an equivalent number of resonator photons $\bar{n}$, this strong drive limit corresponds to $\bar{n} > 1$. In typical cQED setup, one has $g/2\pi \sim 250 \textrm{MHz}$ and $\bar{n} \geq 2$, resulting in an effective drive $\varepsilon_d/2\pi \geq 700 \textrm{MHz}$.

In what follows, we neglect quantum fluctuations $\textbf{F}_\text{q}$ to study the periodically driven transmon Hamiltonian
\begin{align}\label{eq:model}
\hhh(t) =   4 E_C (\nnn-n_g)^2-E_J\cos(\ppphi) +\varepsilon_d\cos(\omega_d t)\nnn.
\end{align}
We return to a full circuit QED model accounting for the presence of the resonator in \cref{sec:cQEDsim}.

\subsection{Classical model}

\begin{figure}
    \centering
    \includegraphics[width=1\columnwidth]{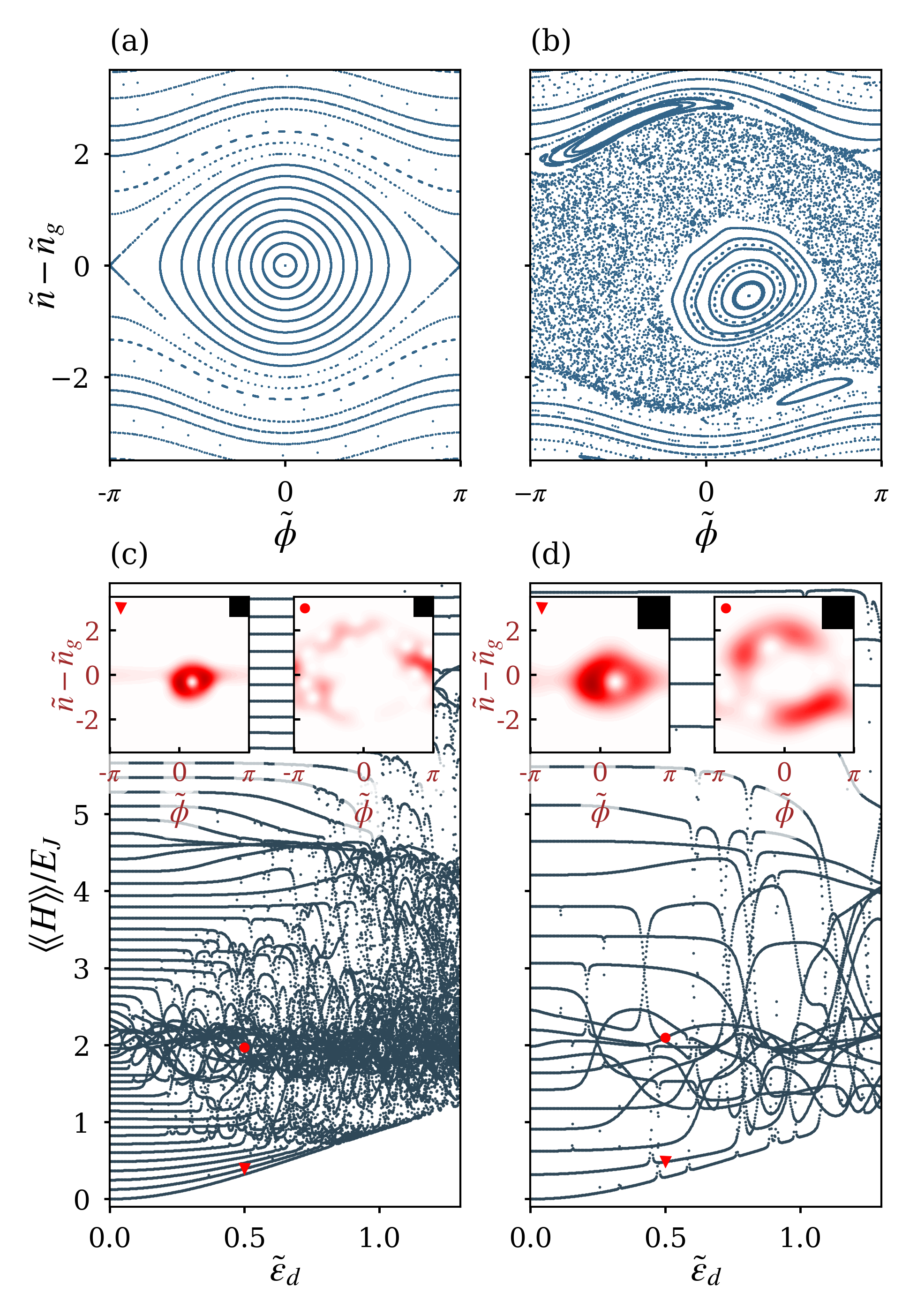}
    \caption{(a) Poincaré section of the undriven system and (b) of the system driven at $\tilde \varepsilon_d = 0.5$, $\tilde \omega_d = 1.34$ and with initial time $\tilde T/8$. In the driven case, a chaotic layer develops around the separatrix. Additional resonances within the unbounded states appear taking the form of tori located at the border of the separatrix. (c,d) Mean energy per cycle as a function of drive amplitude $\tilde \varepsilon_d$ for $\tilde \omega_d = 1.34$, $n_g = 0.25$, and for two different values of $\hbar_\text{eff}$: (c) $\hbar_\text{eff}^{-1} = 7.91$ ($E_J/E_C = 500$) and (d) $\hbar_\text{eff}^{-1} = 3$ ($E_J/E_C = 72$). In analogy to the classical case, strong hybridization of the states develops about the separatrix located at $\llangle H \rrangle /E_J = 2$. The insets marked by a red triangle show the Husimi functions of the Floquet modes of the first excited state at $\tilde\varepsilon_d = 0.5$. Its wavefunction remains globally regular and is reminiscent of the first Fock state. The insets marked by a red dot correspond to a state located close to the separatrix. The corresponding Husimi function is irregular and delocalized over the corresponding classical chaotic domain in (b). The black squares represent the phase-space area $\hbar_\text{eff}$ occupied by one state. 
    }
    \label{fig:mean_energy}
\end{figure}

We first consider the classical limit of the driven transmon Hamiltonian where we replace the conjugate operators $\{\ppphi,\nnn\}$ by the phase-space coordinates $\{\phi,n\}$. In doing so, it is useful to rescale energy and time using the relations $\tilde H = H/E_J$ and $\tilde t = \omega_p t$, where $\omega_p = \sqrt{8 E_J E_C}/\hbar$ is the  plasma frequency of the transmon.
Under this transformation, which preserves Hamilton's equations, the classical Hamiltonian takes the form
\begin{equation}\label{eq:resc_classical}
    \tilde H (\tilde t) = \frac{(\tilde n-\tilde n_g)^2}{2}-
    \cos{\tilde \phi}+\tilde \varepsilon_d\cos(\tilde \omega_d \tilde t)\tilde n.
\end{equation}
This corresponds to the Hamiltonian of a driven charged classical pendulum with dimensionless momentum $\tilde n= z n$ and position $\tilde \phi = \phi$~\cite{koch_charge-insensitive_2007}. Here, $z =\sqrt{8 E_C/E_J}$ is the characteristic impedance of the transmon. 
In this rescaled form, the three relevant parameters of the classical driven transmon are the rescaled drive amplitude 
$\tilde \varepsilon_d = \varepsilon_d/\omega_p$, the rescaled drive frequency $\tilde{\omega}_d = \omega_d/\omega_p$ and the rescaled offset charge $\tilde n_g =  z n_g$.

In the absence of a drive, two
different types of motion of the system can be distinguished.
For $\tilde H < 2$ (\ie $H < 2 E_J$), the system undergoes small and bounded phase oscillations. On the other hand, for $\tilde H > 2$, the system experiences unbounded full $\pm 2\pi$ rotations of the phase.
While manipulations of the transmon qubit are designed such as to only lead to small phase oscillations, the transmon can be promoted to states above $2E_J$ by strong drives~\cite{Shillito2022,Lescanne_2019,VerneyPRApplied2019,sank_measurement-induced_2016}. In the quantum case, the resulting full rotations correspond to quantum phase slips~\cite{matveev_et_al_2002,koch_charge-insensitive_2007}.
At the boundary of these two types of motion, defined by the trajectory of energy ${\tilde{H}} = 2$ also known as the separatrix, small perturbations can have a large impact, for example causing bounded oscillations to turn into unbounded rotations. This structural instability results in an irregular, chaotic motion of the pendulum at finite drive amplitude in the vicinity of the separatrix~\cite{chirikov_universal_1979}.

A useful representation of the system dynamics, both regular and chaotic, is provided by the stroboscopic Poincaré sections obtained by plotting the value of the phase space coordinates $\{\tilde \phi( \tilde t),\tilde n( \tilde t)\}$ at every period $\tilde T = \omega_p/\omega_d$ of the drive for some initial condition~\cite{zaslavskii1991}. 
\Cref{fig:mean_energy}(a) first shows this in the absence of a drive. There, the two expected types of motions are clearly visible:
the bounded oscillations leading to the closed orbits and the unbounded rotations to the nearly horizontal patterns. In the presence of the drive, see \cref{fig:mean_energy}(b),
the Poincaré sections break up into regular and chaotic regions. The regular regions consist of weakly perturbed Kolmogorov-Arnold-Moser tori, reminiscent of the motion of the unperturbed system, while the chaotic region develops around the separatrix~\cite{chirikov_universal_1979}.
The small tori located within the regular unbounded trajectories in \cref{fig:mean_energy}(b) are due to two resonances where the drive frequency $\pm \tilde \omega_d$ matches the energies of the trajectories which pass in the vicinity of $(\tilde \phi, \tilde n) = (0,\pm 2.5)$.

\subsection{Quantum model}

To compare the quantum dynamics to the classical one, we quantize the rescaled Hamiltonian of \cref{eq:resc_classical}. Importantly, because the  rescaling does not preserve the phase-space volume, it leads to a renormalization of the Planck constant upon quantization with $\hbar_\text{eff} = \hbar \omega_p/E_J = z$. As a result, the commutation relation of the rescaled operators is $[\tilde \ppphi,\tilde \nnn] = i\hbar_\text{eff}$.
Consequently, in addition to the three parameters that determine the classical dynamics enumerated above, $\tilde \varepsilon_d$, $\tilde \omega_d$, and $\tilde n_g$, the driven quantum dynamics of the transmon is characterized by a fourth parameter, $\hbar_\text{eff}$, characterizing quantum fluctuations.

The solutions to the time-dependent Schr\"odinger equation associated to the Hamiltonian in \cref{eq:model} are the time-dependent Floquet states $|\psi_k ( t)\rangle = \exp\big({-i  \varepsilon_k  t}\big) |\phi_k ( t)\rangle$, characterized by the rescaled quasienergies $\tilde \varepsilon_k$ and the time-periodic Floquet modes $|\phi_k ( t)\rangle$~\cite{GRIFONI1998229}. 
The quasienergies of the Floquet modes are defined up to integer multiples of the drive frequencies, and hence are not indicative of the amount of energy stored in the system. However, the mean energy per cycle for 
mode
$|\phi_k ( t)\rangle$ can be defined as~\cite{KetzmerickPRE2010}
\begin{equation}\label{eq:Mean_Energy}
    \llangle  H \rrangle = \frac{1}{ T}\int_0^{ T} d t \bra{\phi_k( t)}{ \hhh}( t)\ket{\phi_k( t)}, 
\end{equation}
where $T$ is the period of the drive.
This mean energy per cycle is plotted in \cref{fig:mean_energy}(c) as a function of the drive amplitude $\tilde \varepsilon_d$ and different Floquet states $k$ for a very weakly nonlinear transmon qubit with $\hbar_\text{eff}^{-1} = 7.91$ ($E_J/E_C = 500$). As the drive amplitude increases, the perturbation hybridized the states
around the energy $\llangle H \rrangle /E_J = 2$, corresponding to the energy of the separatrix in the classical system, and does so in an increasingly large bandwidth around that energy. 
Remarkably the main features subsist in the more experimentally relevant case of $\hbar_\text{eff}^{-1} = 3 $ ($E_J/E_C = 72$), although the hybridization is visually less pronounced because the level separations are larger, see \cref{fig:mean_energy}(d).

The diffusion of classical trajectories through the chaotic domain translates to delocalized Husimi functions of the Floquet modes for the driven quantum system~\cite{backer_regular--chaotic_2008,ketzmerick_statistical_2010}. This can be intuitively understood by the fact that the Floquet modes are eigenstates of the propagator over one period of the drive, which is the quantum analog of the stroboscopic Poincaré map defined above~\cite{graham_quantization_1988}.
In the insets of \cref{fig:mean_energy}(c) and \cref{fig:mean_energy}(d), we plot the Husimi functions at time $T/8$ of the Floquet modes indicated by the red markers on the spectra at $\tilde \varepsilon_d = 0.5$. Because the phase is defined on  $(-\pi,\pi]$, we use the definition of Ref.~\cite{Gonz_lez_1998} for a coherent state on a circle. The insets indicated by the red triangle correspond to the Husimi functions of the Floquet modes of the first excited state. Because it is outside of the region where the states are strongly mixed (\ie outside of the chaotic layer in the classical system), its wavefunction remains globally regular and is reminiscent of the first Fock state. On the other hand, the insets marked by a red dot correspond to a state located close to the separatrix. The corresponding Husimi functions are irregular and  delocalized over the region corresponding to the chaotic layer in the classical case, see \cref{fig:mean_energy}(b). The black squares in the upper right corner of the insets indicate the phase space area $\hbar_\text{eff}$ occupied by one state. A smaller $\hbar_\text{eff}$ results in a smaller amplitude of quantum fluctuations, and therefore in more resolved features in the Husimi functions.

\section{Quantum signatures of chaos in the driven spectrum}
\label{sec:spectral_properties}

\subsection{Level-spacing statistics}

\begin{figure}
    \centering
    \includegraphics[width=1\columnwidth]{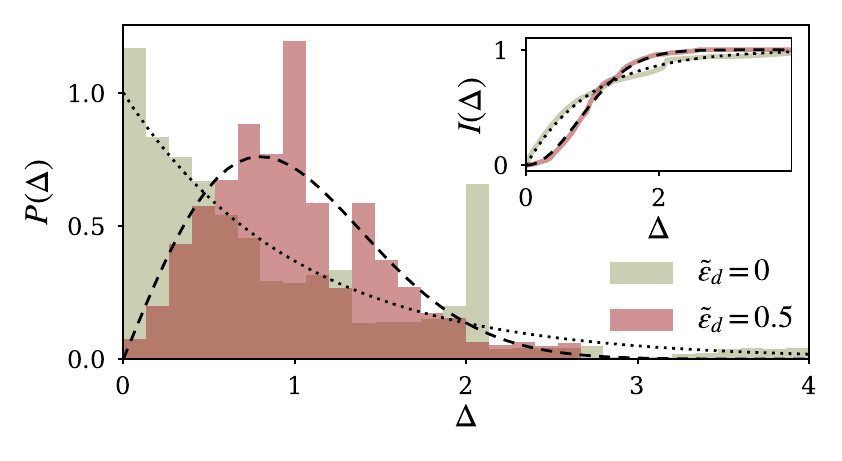}
    \caption{Level spacing statistics for $\hbar_\text{eff}^{-1} = 3$ and $\tilde \omega_d = 1.34$ for the undriven (green) and driven (red) spectrum. The set of states comprises those of mean energy satisfying $1.6 < \llangle H \rrangle /E_J < 2.5$. The statistics is generated using the Floquet spectra corresponding to 200 values of $n_g$ uniformly distributed over the interval $[0,0.5]$. The driven spectrum distribution follows the Wigner-Dyson distribution (dashed line), while the distribution of energies for the undriven system follows the Poisson distribution (dotted line). The inset shows the integrated distribution $I(\Delta) = \int_0^\Delta ds P(s)$.}
    \label{fig:level_spacing}
\end{figure}

The chaotic nature of the states
near the separatrix can be confirmed through the correlated nature of the Floquet spectrum, which manifests itself in the distribution of level spacings $P(\Delta)$. Indeed, the strong level hybridization observed near the separatrix results in level repulsion and, in turn, to a Floquet spectrum with strong correlations in the distribution of the spacing between levels. In that situation, $P(\Delta)$ is expected to follow the Wigner-Dyson distribution (dashed line in \cref{fig:level_spacing})~\cite{haake_quantum_2010}. In contrast, in the regular regime levels are uncorrelated and their distribution is Poissonian, as is characteristic of random uncorrelated events (dotted line in \cref{fig:level_spacing})~\cite{haake_quantum_2010}.

To display this distribution, we consider a set of $N$ states with sorted quasienergies $\varepsilon_1 \leq ... \leq \varepsilon_N$ lying within the first Brillouin zone, \ie $|\varepsilon_n|\leq \omega_d/2$ for all $n$. The spacing between adjacent quasienergy levels is defined by $\Delta_n = (\varepsilon_{n+1}-\varepsilon_{n})/\bar \Delta$ for $n = 1,...,N-1 $, where $\bar \Delta = \omega_d/N $ is the mean level spacing.
We also define $\Delta_N = (\varepsilon_{0}-\varepsilon_{N}+\omega_d)/\bar \Delta$ at the boundary of the Brillouin zone. In the limit of large $N$, the distribution $P(\Delta)$ is expected to follow the Wigner-Dyson statistics for correlated spectrum \cite{Mehta-1991}, whereas it is expected to follow a Poisson distribution for an uncorrelated spectrum. 

In analogy with the study of classical chaos, we define the chaotic domain using the mean energy per cycle $\llangle H \rrangle$. As can first be seen in \cref{fig:mean_energy}(c) for $\hbar_\text{eff}^{-1} = 7.91$, the mean energies of the chaotic states are concentrated around $2E_J$, and are separated from the regular states by a gap. The relevant energy bandwidth of the chaotic zone depends on the drive amplitude and, for $\tilde \varepsilon_d = 0.5$, we take the $N$ states whose energies satisfy $1.6 < \llangle H \rrangle /E_J < 2.5$. However, for the value of $\hbar_\text{eff}^{-1} = 3$ corresponding to a typical transmon qubit, the spectrum displays only a handful of chaotic states in this energy bandwidth. With about $N\sim 7$ levels, see \cref{fig:mean_energy}(d), this is far from enough to generate meaningful level-spacing statistics.
To circumvent this problem, we use the offset charge $n_g$ uniformly distributed in the interval  $[0,0.5]$ to generate statistics. This is possible because, even in the transmon regime, the chaotic states are not confined in the cosine potential well making them very sensitive to the offset charge. We return in \cref{sec:bath_coupling} to some of the observable implications of this $n_g$ dependence of the chaotic states. 

We plot in \cref{fig:level_spacing} the cumulative histograms of $P(\Delta)$ obtained from adding the distributions corresponding to 200 values of $n_g$ uniformly distributed in the interval $[0,0.5]$ for $\tilde \varepsilon_d = 0$ (green bars) and $\tilde \varepsilon_d = 0.5$ (red bars). In the absence of a drive, the system is regular and $P(\Delta)$ follows the Poisson distribution expected for an uncorrelated spectrum. This also suggests that the undriven spectrum is well randomized by the variation of $n_g$. In contrast, at $\tilde \varepsilon_d = 0.5$, $P(\Delta)$ approaches
the Wigner-Dyson distribution. The number of states $N$ in the selected energy bandwidth has increased in the chaotic case. This is due to the hybridization of states with mean energy close to the separatrix region $\llangle H \rrangle \sim 2 E_J$, as mentioned above. More generally, the extent of the chaotic layer is seen to increase with the drive amplitude, something which is particularly clear in \cref{fig:mean_energy}(c). The inset of ~\cref{fig:level_spacing} shows the integrated distribution $I(\Delta) = \int_0^\Delta ds P(s)$ for which the statistical variations are reduced because of the integration, thereby allowing for a clearer distinction between uncorrelated and correlated spectra~\cite{haake_quantum_2010}.

\subsection{Drive frequency dependence}
\label{sec:drive_frequency}

\begin{figure*}
    \centering
    \includegraphics[width=1\textwidth]{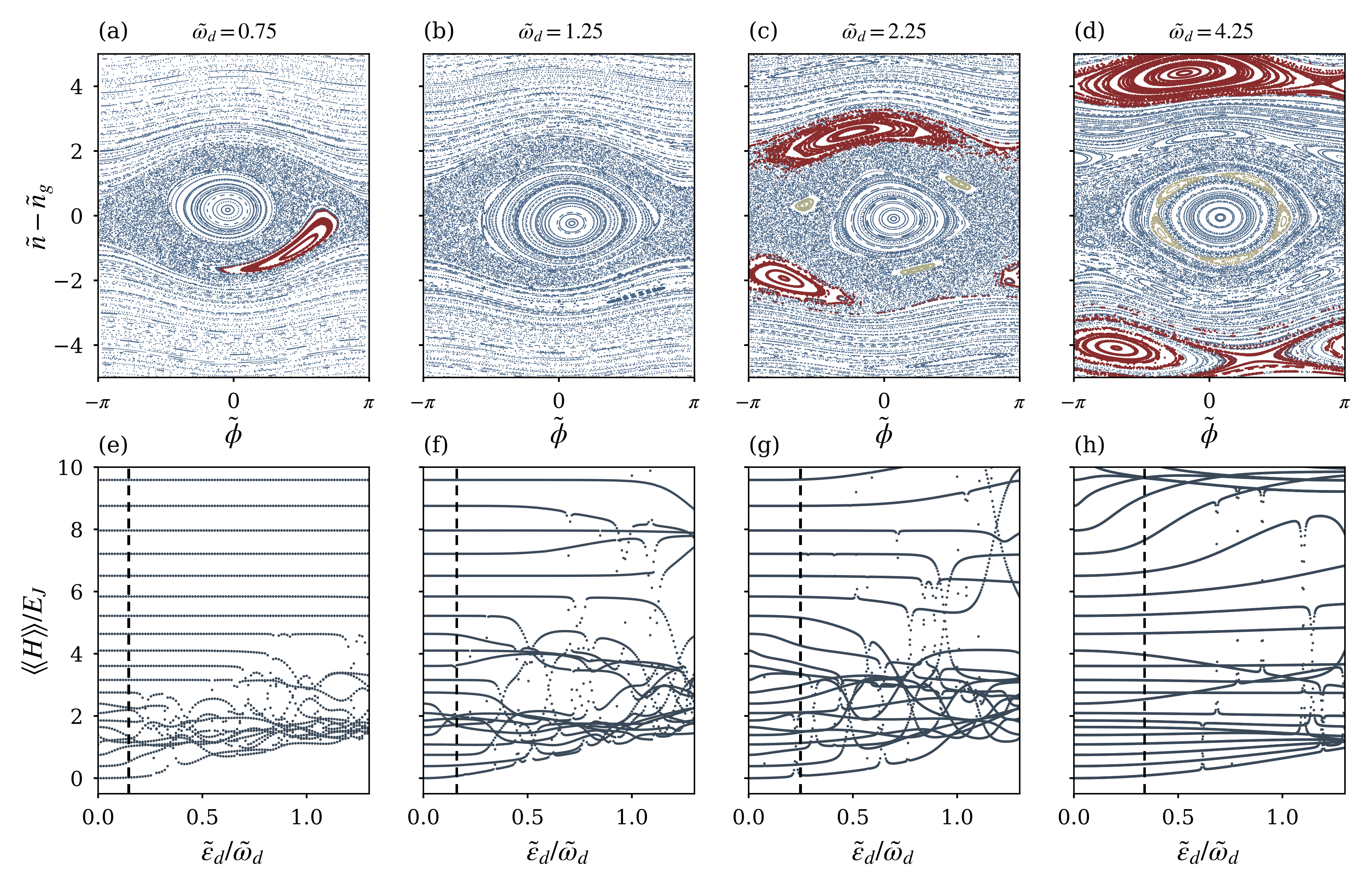}
    \caption[width=1\columnwidth]{
    (a)-(d): Poincaré sections at time $T/8$ for the drive frequencies (a) $\tilde \omega_d = 0.75$, (b) $\tilde \omega_d = 1.25$, (c) $\tilde \omega_d = 2.25$ and (d) $\tilde \omega_d = 4.25$. The drive amplitudes are chosen such that the frequency of the $0-1$ transition in the quantum system is ac-Stark shifted by $100 \text{ MHz}$ for all $\tilde \omega_d$. (e)-(f): Mean energy spectra  as a function of $\tilde \varepsilon_d/\tilde \omega_d$ at the same respective drive frequencies, with $ \hbar_\text{eff}^{-1} = 2.45$ and $n_g = 0.25$. In each case, the drive amplitude yielding an ac-Stark shift of $100 \text{ MHz}$ is indicated by the vertical black dashed line -- the above Poincaré sections have been computed for these amplitudes.
    For the parameters of  panels (a), (c) and (d), a 1:1 resonance occurs between the drive and the system. In (a), this resonance results in a second set of tori within the bounded states of the system (red region). The overlap of the small oscillation and this resonance results in a large chaotic layer. In the quantum system (e), this leads to strong level hybridization. In panels (c) and (d), the drive comes in resonance with the unbounded states, resulting in two additional out-of-phase sets of tori rather than one (red regions). From (c) to (d), these tori move away from the center as the drive frequency increases. In (c) and (d), 3:1 and 5:1 resonances emerge (green region). In (c), the proximity of the 1:1 (red) and 3:1 (green) resonances to the regular island also causes large instabilities, both in the classical and quantum systems. In (d), the resonances are far and the width of the chaotic layer is smaller, as expected from \cref{eq:Wc}. In (b), the resonance is absent.
    }
    \label{fig:freq_dependence_PS}
\end{figure*}

In addition to spreading over an increasingly large energy bandwidth with increasing drive amplitude, the size of the chaotic region also strongly depends on the drive frequency. This is illustrated in \cref{fig:freq_dependence_PS}(a)-(d) which shows the Poincaré sections for $\tilde \omega_d = 0.75,~1.25,~2.25$ and $4.25$. The mean energy spectra of the corresponding
quantum systems are shown in \cref{fig:freq_dependence_PS}(e)-(h) as a function of $\tilde\epsilon_d/\twd$ for $\hbar_\text{eff}^{-1} = 2.45$ and $n_g = 0.25$. To compare the effect of different drive frequencies, the ac-Stark shift computed from the quasienergies of the ground and first excited states is fixed to $100~\textrm{MHz}$ in panels (a)-(d) (see \cref{sec:ACStarkShift}). The corresponding drive amplitudes are represented by a vertical black dashed line in \cref{fig:freq_dependence_PS}(e)-(h).  
In panels (a)-(d), the width of the chaotic layer is observed to be maximal for intermediate drive frequencies $\twd \sim 1-2$. This observation is in qualitative agreement with the approximate expression for the width of the chaotic layer around the separatrix
\begin{equation}\label{eq:Wc}
W_c/E_J \approx \tilde \varepsilon_d \tilde \omega_d \sech\left(\frac{\pi \tilde \omega_d}{2}\right), 
\end{equation}
a result which is valid for $\tilde \omega_d > 1$~\cite{bubner_quantum_1991,zaslavskii1991}. Following this expression, 
the chaotic layer is expected to have a maximal width for $1<\tilde \omega_d<2$, and to exponentially decrease in width with increasing $\tilde \omega_d$ for $\tilde \omega_d > 2$.

The approximate expression for the width, however, does not account for resonances that result in the additional tori observed in \cref{fig:freq_dependence_PS}(a)-(d). In particular, in the range
$0.6 \lesssim \tilde\omega_d \lesssim 3$, which includes the regime of operation of current experiments, we find that resonances play an essential role in drive-induced instabilities. The origin of these resonances can be qualitatively understood by expressing the classical Hamiltonian of \cref{eq:resc_classical} in the equivalent form
\begin{equation}
    \tilde H(\tilde t) = \frac{(\tilde n-\tilde n_g)^2}{2} -\cos\left[\tilde \phi+ \frac{\tilde{\varepsilon}_d}{\tilde \omega_d} \sin(\tilde \omega_d \tilde t )\right].
\end{equation}
Using the Jacobi-Anger expansion to first order in $\tilde{\varepsilon}_d/\tilde \omega_d$, this can be approximated as
\begin{align}\label{eq:expansion}
    \tilde H(\tilde t) \approx & \frac{(\tilde n-\tilde n_g)^2}{2} - J_0\left(\frac{\tilde{\varepsilon}_d}{\tilde \omega_d}\right)\cos{\tilde \phi} \notag \\ 
    & + J_{1}\left(\frac{\tilde{\varepsilon}_d}{\tilde \omega_d}\right)
    \left[\cos(\tilde \phi -\tilde \omega_d \tilde t )-\cos(\tilde \phi +\tilde \omega_d \tilde t )\right],
\end{align}
where $J_k(z)$ is the $k$-th Bessel functions of the first kind. The first line of \cref{eq:expansion}, $\tilde H_{\tilde{\varepsilon}_d} = (\tilde n-\tilde n_g)^2/2 - J_0(\tilde{\varepsilon}_d/\tilde \omega_d)\cos{\tilde \phi}$, describes an undriven pendulum with a potential energy reduced by the factor $J_0(\tilde{\varepsilon}_d/\tilde \omega_d)$. The second line is  of smaller amplitudes and describes the first harmonic of the time-dependent perturbation.
Higher-order harmonics of the drive are
neglected because their amplitude is 
suppressed by the factor $J_{k}(\tilde{\varepsilon}_d/\tilde \omega_d)$ and because the
corresponding resonances occur at higher frequency, $\tilde \omega > 2\tilde \omega_d$.
The effects of the perturbation can be understood by inserting in the second line of \cref{eq:expansion} the solution $(\tilde \phi(t), \tilde n(t))$ of the system under the time-independent Hamiltonian $\tilde H_{\tilde{\varepsilon}_d}$ with initial conditions $(\tilde \phi_0, \tilde n_0 )$. 
Depending on the drive frequency, resonances can occur either within the bounded states or the unbounded states. 

The trajectories representing the bounded states of the pendulum can be generated in the Poincaré sections with the initial conditions $(\tilde \phi_0 < \pi , \tilde n_0 = 0)$. For small oscillations, the pendulum behaves as a slightly anharmonic oscillator, and its oscillation frequency decreases as the oscillation amplitude $\tilde \phi_0$ increases. For $|\tilde \phi_0|< 0.8\pi$, the nonlinearity does not play an important role and the pendulum frequency varies smoothly. This corresponds to a frequency range $0.65 \lesssim \tilde \omega < 1$ for the pendulum oscillation. In this case, under $\tilde H_{\tilde{\varepsilon}_d}$ the trajectory that passes through $(\tilde \phi_0, 0)$ takes the standard form $\tilde \phi (t) \approx \tilde \phi_0 \sin(\tilde \omega \tilde t)$ where we have neglected the higher harmonics of the motion. Inserting this expression for $\tilde \phi(t)$ in the second line of \cref{eq:expansion} leads to the slowly rotating terms $J_{1}\big(\tilde{\varepsilon}_d/\tilde \omega_d\big)J_{2k+1}(\phi_0)\cos[(\tilde \omega_d - (2k+1)\tilde \omega) \tilde t ]$. 

The case $k=0$ corresponds to a 1:1 resonance, \ie ${\tilde \omega_d = \tilde \omega \in [0.65,1]}$, which strongly impacts the low-energy states of the system. At $\tilde \omega_d = 0.75$
this resonance appears as a second set of tori located close to the central tori, 
see the red-coloured region in \cref{fig:freq_dependence_PS}(a). Large instability results from the overlap between these two tori, with a chaotic layer arising at their separatrices~\cite{chirikov_universal_1979}. The resulting increased width of the chaotic layer is not captured by the approximate expression of \cref{eq:Wc}.
The case $k=1$ leads to a 3:1 resonance for $\tilde \omega_d = 3\tilde \omega \in [2,3]$.  For $\tilde \omega_d = 2.25 $ this resonance results in 3 sets of tori surrounding the central regular island, see the three green-colored regions in \cref{fig:freq_dependence_PS}(c).
Although of smaller amplitudes than the 1:1 resonance, the overlap of this resonance with the main set of tori is likely to produce unstable motion at the boundary of these. Note also the appearance of a weak 5:1 resonance at $\tilde \omega_d = 4.25 $
resulting in 5 small tori within the regular island, see the green-colored region in \cref{fig:freq_dependence_PS}(d).

In contrast, no direct resonance occurs for the case $1.1 <\tilde \omega_d < 1.5$ which is common for the dispersive readout, see \cref{fig:freq_dependence_PS}(b) for $\tilde \omega_d = 1.25$. Nevertheless, as expected from \cref{eq:Wc}, the width of the chaotic layer is large at this drive frequency.

In the quantum case, the resulting large chaotic layer translates into a large hybridization of the states even at low drive amplitude, see the mean energy spectra \cref{fig:freq_dependence_PS}(e-g). This hybridization can be further observed from the modification of the rate matrices (see \cref{sec:ACStarkShift}).
This state hybridization can lead to loss of the QND character of the dispersive readout. As a concrete example, 
the value of $\hbar_\text{eff}^{-1}$ and $\tilde \omega_d = 0.75$ of \cref{fig:freq_dependence_PS}(f) was chosen to match the experimental parameters of \textcite{WalterPRL2017} where the dispersive readout fidelity was observed to degrade for measurement photon numbers $\bar n > 2.5$, something which was attributed to measurement-induced mixing of unknown origin. Using the light-matter coupling of $g/2\pi = 208 \text{ MHz}$ reported in Ref.~\cite{WalterPRL2017}, $\bar n  = 2.5$ can be converted to an effective drive $\tilde \varepsilon_d = 0.105$ on the transmon. As can be observed in \cref{fig:freq_dependence_PS}(e), the first excited state is ``absorbed" in the chaotic layer for $\varepsilon_d \gtrsim 0.1$. Since chaotic states often lead to strong hybridization between the transmon and the readout resonator (see \cref{sec:cQEDsim}), this hints at chaos-induced state mixing and non quantum-demolition effects in the readout beyond that drive amplitude. Although the above mean-field analysis can qualitatively predict unstable behavior of the transmon, the agreement cannot be expected to be quantitative at low resonator photon numbers since vacuum fluctuations of the resonator field and qubit-resonator parametric processes occurring in the presence of a readout tone can play an important role. We address some of these mechanisms in \cref{sec:cQEDsim}.

The drive can also come in resonance with unbounded states of the pendulum with energy ${\tilde H_{\tilde{\varepsilon}_d} > 2 J_0\big(\tilde{\varepsilon}_d/\tilde \omega_d\big) \sim 2 }$. The
trajectories at those energies take the approximate form $\tilde \phi(\tilde t) = \pm \tilde \omega \tilde t +F(\tilde t)$, where $F$ is a function of period $2\pi/\tilde \omega$~\cite{chirikov_universal_1979}. Using this expression in Hamilton's equation of motion for the phase, we find that the oscillation frequency satisfies $\tilde \omega = |\langle \tilde n - \tilde n_g \rangle | \gtrsim 1.5$, where $|\langle \tilde n - \tilde n_g \rangle |$ is the averaged momentum of the trajectory over one period. This lower bound can vary with the drive amplitude through the effect of the reduction factor $J_0\big(\tilde{\varepsilon}_d/\tilde \omega_d\big)$. To leading order, substituting this expression for $\tilde \phi(\tilde t)$ in the second line of \cref{eq:expansion} results in the term $A_{0} J_{1}\big(\tilde{\varepsilon}_d/\tilde \omega_d\big)\cos[\pm (\tilde \omega - \tilde \omega_d) \tilde t]$, where $A_0$ is the zeroth Fourier component of $\cos[F(\tilde t)]$. Hence, for $\tilde\omega_d = \tilde\omega \gtrsim 1.5$, this resonance appears in the form of two tori moving clockwise and anti-clockwise, corresponding to $\pm \tilde\omega_d$, as depicted by the red-coloured regions in \cref{fig:freq_dependence_PS}(c) for $\tilde \omega_d = 2.25$ and \cref{fig:freq_dependence_PS}(d) for $\tilde \omega_d = 4.25$. 
At $\tilde \omega_d = 2.25$, the proximity of the resonance with the regular island
results in a large chaotic domain which translates in strong state hybridization in the quantum case, see \cref{fig:freq_dependence_PS}(g).
Because the average momentum associated to this resonance is $|\langle \tilde n - \tilde n_g \rangle | = \tilde\omega_d$, the pair of tori move further away from the center of phase space with increasing drive frequency. For this reason, at $\tilde \omega_d = 4.25$,
this resonance is far from the separatrix of the undriven system and does not affect the chaotic layer. As a result, for $\tilde \omega_d = 4.25$ and at larger drive frequencies, the width of the chaotic layer is exponentially suppressed as expected from \cref{eq:Wc}.
In this situation, the absence of instability yields a mean energy spectrum with little state hybridization, see \cref{fig:freq_dependence_PS}(h). This fact is further illustrated by the regular structure of the charge matrix elements in \cref{sec:ACStarkShift}.

To summarize the above, the frequency ranges $0.65< \tilde \omega_d < 1$ and $1.5 \leq \tilde \omega_d \leq 3$  lead to strong instabilities, in particular for the former. The parameter regime $1.1< \tilde \omega_d < 1.5$ avoids resonances but has a large chaotic layer around the separatrix. For $ \tilde \omega_d \geq 3.5$, resonances do not affect the bounded states (or very weakly through resonances of order greater than 5), and the width of the chaotic layer is suppressed. As shown in the next sections, the presence of the chaotic layer impacts the coherence times of the transmon qubit, and it should therefore be minimized when operating the transmon with strong drive, e.g.~in dispersive qubit readout. The frequency ranges $1.1 \leq \tilde \omega_d \leq 1.5$  and $ \tilde \omega_d \geq 3.5$ seem to be more benign.

\section{Impact on coherence properties}
\label{sec:bath_coupling}

Having established that the driven capacitively shunted Josephson junction exhibits signatures of chaos even at the large $\hbar_\text{eff}$ corresponding to the transmon regime, we now turn to the impacts of this observation on coherence properties of the transmon in the presence of a bosonic bath. In this situation, the total Hamiltonian now takes the form 
\begin{align}\label{eq:bath}
\hhh(t) = &  4 E_C (\nnn-n_g)^2-E_J\cos(\ppphi)+\varepsilon_d\cos(\omega_d t) \nnn \\ 
& + i \nnn \sum_k g_k (\bbb_k^\dagger - \bbb_k) + \sum_k \omega_k \bbb_k^\dagger \bbb_k, \notag 
\end{align}
where $\bbb_k$ and $\bbb_k^\dagger$ are the annihilation and creation operators of the bath modes. Because of hybridization with states in the chaotic layer which have a strong charge dispersion, it is important to keep the gate charge $n_g$ in \cref{eq:bath} even when interested in the coherence properties of the low-lying eigenstates of the system. 

\subsection{Rate matrix and steady-state population}

\begin{figure}
    \centering
    \includegraphics[width=1\columnwidth]{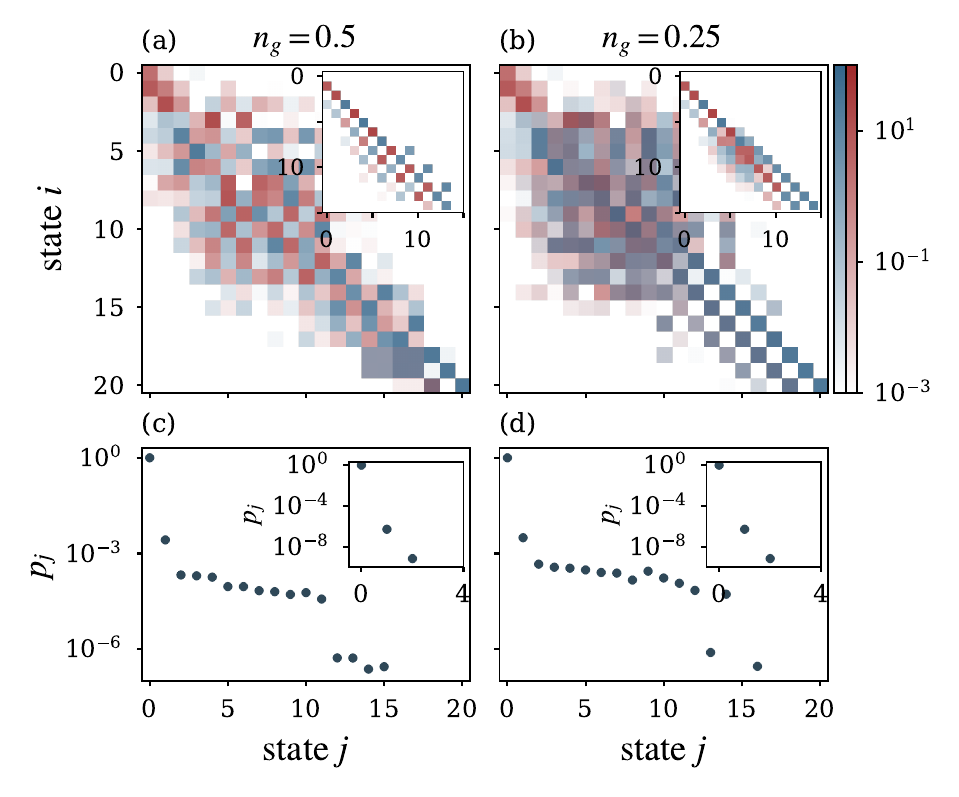}
    \caption[width=1\columnwidth]{Top panels : Rate matrices in units of the coupling strength to the bath, at $\tilde \varepsilon_d = 0.4 $, $\hbar_\text{eff}^{-1} = 3 $, $\tilde \omega_d = 1.34$ and $T = 10 \text{ mK}$ for (a) $n_g = 0.5 $ and (b) $n_g = 0.25 $. The states are sorted by their mean energy $\llangle H \rrangle$. The blue (red) squares correspond to the sum of rates involving an even (odd) number of drive photons $k$. The insets show the rate matrices of the undriven systems. At finite drive, the instability develops around the separatrix leading to an irregular block in the rate matrix also allowing for upward excitation rates. At the symmetric point $n_g = 0.5$, the red and blue squares do not overlap, while at $n_g = 0.25$, hybridization of parity sectors can be observed through the presence of purple squares even in the low-energy sector. The purple squares in (a) are due to numerical errors (see text). Bottom panels : Steady-state population as derived from the above rate matrices for (c) $n_g = 0.5 $ and (d) $n_g = 0.25 $. In the driven system, a plateau forms over the states that are part of the chaotic block of the rate matrices, instead of the regular exponential distribution in the undriven case, as shown in the insets.}
    \label{fig:rate_matrices} 
\end{figure}

Within the Floquet-Markov description of the system-bath coupling, the bath-induced transition rate from Floquet state $j$ to $i$ is given by~\cite{GRIFONI1998229}  
\begin{equation}\label{eq:Gamma}
\Gamma_{ij} = \sum_k |n_{ijk}|^2 [\Theta(\Delta_{ijk})+n_B(|\Delta_{ijk}|)]J(|\Delta_{ijk}|),
\end{equation}

where $\Theta(x)$ denotes the Heaviside function, $n_B(x)$ is the thermal occupation number of the bath, and $\Delta_{ijk} = \varepsilon_j - \varepsilon_i - k\omega_d$. The spectral function $J(x)$ is assumed to be that of an ohmic bath, \ie $J(x) \propto x \exp(-|x|/\omega_c)$, where $\omega_c$ is a high-frequency cut-off. In the expression for $\Gamma_{ij}$, $k$ can be interpreted as the number of drive photons participating positively or negatively to the transition.
We have also introduced the charge operator matrix elements 
\begin{equation}\label{eq:charge_matrix_elements}
n_{ijk} = \frac{\omega_d}{2\pi}\int_0^{2\pi/\omega_d} dt \bra{\phi_i(t)} \nnn \ket{\phi_j(t)}\exp(i\Delta_{ijk}t),
\end{equation}
where $\ket{\phi_j(t)}$ are the Floquet modes of the driven system.

The rate matrices of the driven transmon at $\tilde \varepsilon_d = 0.4$ and a temperature of $T = 10~\textrm{mK}$ are shown for $n_g = 0.5$ in \cref{fig:rate_matrices}(a) and for $n_g = 0.25$ in \cref{fig:rate_matrices}(b). As a comparison, the insets show the rate matrices in the undriven case. For a given transition $\Gamma_{ij}$, the blue (red) squares sum the contributions from even (odd) values of $k$, with purple squares indicating contributions from both even and odd values.
At zero drive, the upper triangular sector of the rate matrices (corresponding to upward transitions) contains negligible but non-zero elements due to the finite temperature (not visible in the insets). 
At finite drive,
the instability develops around the states located on the separatrix (typically around the 8th excited state), forming an irregular block in the rate matrix. In particular, states within the chaotic layer are all coupled to one another through the charge operator.
In addition, because of the drive photons, upward transitions are now apparent. The appearance of an irregular block in the rate matrix directly relates to the repulsive statistics of the quasienergies. In fact, chaotic systems can be accurately described by random matrices in the limit of a large number of chaotic states \cite{haake_quantum_2010}.  In \cref{sec:standard_deviation}, we leverage this property to estimate the average charge matrix element between two chaotic states when all the low-energy states are chaotic.

In the transmon regime, the low-energy sector is almost independent of the offset charge and, if one neglects its influence, the Hamiltonian becomes effectively symmetric under the parity transformation $\ppphi \rightarrow -\ppphi$ and $\nnn \rightarrow -\nnn$ (see \cref{sec:symmetry_appendix}).
Because it neglects the gate charge, this symmetry is implicit in the Kerr nonlinear oscillator model of the transmon.
Although this symmetry is exact only at $n_g = 0$ and $n_g = 0.5$, at zero drive
it results in a suppression of the matrix elements of the charge operator $n_{i,i+2}$ in the low-energy sectors for all values of $n_g$,
and forbids the transition $i\rightarrow i+2$, see insets of \cref{fig:rate_matrices}(a,b).
The fact that this is only an approximate symmetry at $n_g = 0.25$ is apparent for the states in the separatrix region and above. 

In the presence of a drive term $ \varepsilon_d \cos( \omega_d  t) \nnn$,
the inversion symmetry only holds together with the time translation $t\rightarrow t+\pi/\omega_d$~\cite{GRIFONI1998229}. This symmetry of the driven Hamiltonian defines even and odd parity sectors among the time-dependent Floquet states (see \cref{sec:symmetry_appendix}). 
Because the charge operator is anti-symmetric,
under this generalized parity symmetry, transitions through the charge operator between two states of the same (opposite) parity can only involve an odd (even) number of drive photons $k$.
As can be seen in  \cref{fig:rate_matrices}(a) for $n_g = 0.5$, blue and red squares do not mix (\ie there are no purple squares) indicating that the symmetry is respected for that gate charge. The purple squares appearing for states 18 and 19 only result from numerical precision errors. 
The situation is very different at $n_g = 0.25$ where purple squares appear
in the chaotic layer but also for transitions involving the low-energy states. This results in
transitions that are otherwise forbidden at the symmetric points $n_g = 0$ and $n_g = 0.5$. As discussed in further details in the next section, the breaking of this effective symmetry in the low-energy sector is a consequence of a strong increase of the band dispersion in the presence of drive. Interestingly, transitions forbidden by the apparent inversion symmetry of the transmon were experimentally observed under strong drives, but remained unexplained~\cite{sank_measurement-induced_2016}.

The rate matrix can also be used to compute the system's steady-state density matrix $\rrr_{ss}$.
Under the assumption of weak system-bath coupling, the steady state is diagonal in the Floquet basis
\begin{equation}
    \rrr_{ss}(t)= \sum_j p_j \ket{\phi_j(t)}\bra{\phi_j(t)},
\end{equation} 
where the populations $p_j$ satisfy the rate equations $ p_j = \sum_i \Gamma_{ji} p_i- \sum_i \Gamma_{ij} p_j $. The insets of \cref{fig:rate_matrices}(c) and \cref{fig:rate_matrices}(d) show these steady-state populations for the undriven systems for $\hbar_\text{eff}^{-1} = 3$ which is typical of the transmon regime.  As expected, the populations  follow the thermal distribution with populations quickly dropping below $10^{-10}$.
In contrast, in the driven systems (here with $\tilde \varepsilon_d = 0.4$), the steady-state populations form a plateau corresponding to the chaotic layer.
This behavior is typical of chaotic systems~\cite{breuer_quasistationary_2000, ketzmerick_statistical_2010}.
Increasing the drive amplitude, the chaotic layer and therefore the plateau grow until all the low-energy states are part of
the plateau, including the ground state. This results in a dramatic decrease of the purity of the transmon's steady state, an observation which is in agreement with numerical~\cite{VerneyPRApplied2019} and experimental results~\cite{Lescanne_2019}.

This discussion sheds light on the distinction between the time-dynamics of the transmon ionization numerically observed in \cite{Shillito2022}, and the ionization in the steady state \cite{VerneyPRApplied2019,Lescanne_2019}. In the former, one captures the dynamics of
the dispersive readout of a transmon qubit.
As the cavity rings up on a timescale $\kappa^{-1}$, where $\kappa$ is the photon loss rate, the effective field amplitude $\varepsilon_d$ on the transmon increases. Starting in the ground or first excited states, the system follows 
the corresponding first or second line in the mean energy spectrum of \cref{fig:mean_energy}(c). Ionization occurs when 
one of those lines crosses
a large resonance with a chaotic state, \ie when the mean energy suddenly increases in \cref{fig:mean_energy}(c), something which happens at $\tilde \varepsilon_d \approx 1.1 $ for the ground state and at $\tilde \varepsilon_d \approx 0.75 $ for the first excited state. Once a chaotic state is populated, it decays either through the transmon-bath coupling or through the transmon-resonator coupling due to its numerous possible transitions at various frequencies (see \cref{sec:cQEDsim}).  
On the other hand, 
ionization in the steady state 
is only a function of the
rate matrix, which in turn depends on the matrix elements of the charge operator in the Floquet basis.
Large matrix elements between the ground state and the chaotic states are likely to lead to ionization through bath-induced transitions.
Although a resonance leads to large matrix elements, the contrary is not necessarily true.

\subsection{Chaos-assisted tunneling : effects on T2}

\begin{figure}
    \centering
    \includegraphics[width=1\columnwidth]{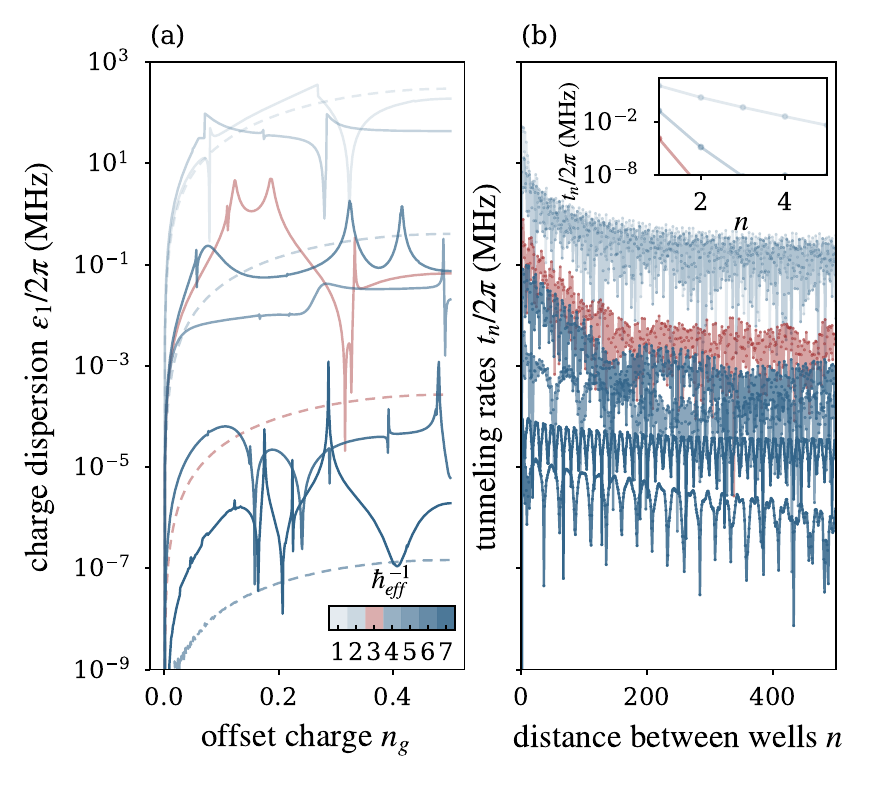}
    \caption{(a) Band dispersion of the first excited state for the undriven (dashed lines) and driven (solid lines) transmon with $\tilde \varepsilon_d = 0.4$, for $\hbar_\text{eff}^{-1} = \sqrt{E_J/8 E_C}$ in the range $[1,7]$. The experimentally relevant value of $\hbar_\text{eff}^{-1} =3$ is highlighted in red. The exponential suppression of the dispersion of the undriven case is strongly modified by the presence of the drive, showing both an overall increase and sharp features. The latter are due to small resonances with chaotic states that have a large charge dispersion. (b) Fourier coefficients of the energy $\varepsilon_1(n_g)$ of the first excited state at finite drive amplitude $\tilde \varepsilon_d = 0.4$ as a function of the Fourier index $n$. These Fourier components are understood as phase-slip rates of order $n$. The spikes or ``abrupt" resonances in (a) are responsible for the slowly decaying tail in (b) that translates to long-range tunneling between the Josephson potential wells. The inset shows the tunneling rates of the undriven systems with the expected exponential suppression with the distance between wells.}
    \label{fig:band_dispersion}
\end{figure}

A defining feature of the transmon is its charge dispersion which is exponentially suppressed with $8\hbar_\text{eff}^{-1} = \sqrt{8 E_J/E_C}$, making it almost insensitive to charge noise~\cite{koch_charge-insensitive_2007}.
Here, we show that this exponential suppression does not necessarily hold in the presence of a periodic drive. 

In the previous section, we have seen that the Hamiltonian of the driven transmon can be approximately divided into regular blocks of states: phase-like states at the bottom of the cosine potential well; charge-like states above that potential, on which the drive acts pertubatively; and one chaotic block with a strongly correlated spectrum. From perturbation theory performed on the regular block, one would expect a drive to lead to a slow hybridization amongst regular low-lying states. In this situation, the ground and first excited states would therefore weakly inherit an offset-charge sensitivity from weak dressing with higher energy states, leading to an overall small
increase of the band dispersion with the drive amplitude. In practice, we find that the presence of the chaotic layer results in the energy dispersion of these states to be
significantly modified when the system is driven. This is illustrated by the solid lines in \cref{fig:band_dispersion}(a) for the first excited state of the transmon with values of $\heff$ in the range 1 to 7 as labeled by the different colors. As a comparison, the dashed lines correspond to the undriven transmon and for which the exponential suppression of the charge dispersion is clearly observed.  
In contrast, in the driven transmon the energy bands are disrupted by peaks of multiple orders of magnitude in addition to being, on average, substantially larger than in the undriven case. The sharpness of these peaks is indicative of a weak resonance between the first excited state and strongly $n_g$-dependent states located in the chaotic layer. These results are obtained by identifying the first excited state in the undriven case, then tracking the corresponding Floquet mode  $\ket{\phi_1(t)}$ as a function of drive amplitude.

This phenomenon is closely related to chaos-assisted tunneling (CAT)~\cite{tomsovic_chaos-assisted_1994}. In CAT, tunneling between two sets of disjoint regular states is facilitated by their coupling to delocalized states in the chaotic layer. Moreover, because of the participation of states within the chaotic layer, the tunneling rates are expected to vary widely with the control parameters~\cite{tomsovic_chaos-assisted_1994}. For the transmon, chaos can assist tunneling between different wells of the cosine potential. Because the phase of the transmon is compact, tunneling between wells distant by $\delta\phi = 2\pi n $ translates to $n$ $2\pi-$swings of the transmon phase or, equivalently, to quantum phase-slips of order $n$. The transmon states acquire their $n_g$-dependence through these full phase rotations~\cite{koch_charge-insensitive_2007}. In the undriven transmon, the rate of these events decreases exponentially with $n$ for large $\heff^{-1}$.

To evaluate the phase slip rate in the presence of a drive, we compute the Fourier transform of the energy bands of the system~\cite{martinez_chaos-assisted_2021}.
Indeed, the Fourier components $t_n = \int_{-0.5}^{0.5} d n_g \varepsilon_1(n_g)e^{i2\pi n_g n}$ of the energy $\varepsilon_1(n_g)$ of the first excited state corresponds to the rate of phase slips of order $n$ when the system is in the first excited state~\cite{koch_charging_2009}. The components $t_n$ are plotted as a function of the index $n$ in \cref{fig:band_dispersion}(b). The inset shows the exponential suppression of $t_n$ expected for the undriven case. In the driven case (main panel), $t_n$ is no longer exponentially suppressed with $n$.
Instead, it shows a long tail, indicating that the sharp features
seen in \cref{fig:band_dispersion}(a) results from long-range hopping between the wells. Note that the tunnelling rates $t_n$ remain globally suppressed with $\hbar_\textrm{eff}^{-1}$, see \cref{fig:band_dispersion}(b). This is explained by an exponential suppression of the matrix elements $n_{ijk}$ between the regular states and the chaotic states with $\hbar_\textrm{eff}^{-1}$. We discuss this point in \cref{sec:Ncrit}.

In cold atoms trapped in a driven optical lattice, the signatures of CAT have been
observed by
measuring coherent oscillations between states localized in distinct wells~\cite{mouchet_chaos_2001,Arnal_sciadv_2020}.
Because the phase coordinate of the transmon is compact, tunneling oscillation between wells cannot be measured.
However, phase slips (due to CAT or not) lead to a phase accumulation which depends on the gate charge $n_g$ due to the Aharonov-Casher effect~\cite{manucharyan_evidence_2012}. Because the gate charge is a fluctuating function of time, this results in enhanced dephasing of the transmon. The increased phase slip rate due to CAT can thus in principle be witnessed through sharp variations of the pure dephasing rate $\gamma_\phi$ associated to the transmon's 0-1 transition, as a functions of $n_g$ and of the drive amplitude $\tilde \varepsilon_d$.

Within the Floquet-Markov theory, the dephasing rates takes the form~\cite{huang_engineering_2020}
\begin{align}\label{eq:dephasing}
    \gamma_{\phi} = A_e |2 g_{0,\phi}| \sqrt{|\log \omega_{IR} t_m|} + \sum_{k \neq 0} 2 S(k \omega_d ) |g_{k,\phi}|^2.
\end{align}
The first term represents $1/f$ charge noise, while the second term comes from the possible conversion of a photon loss to a dephasing event due to the hybridization of the logical states $\ket{0}$ and $\ket{1}$.
In the first term, $A_e$ is the amplitude of charge noise, 
$g_{k,\phi} = n_{11k}-n_{00k}$ where $n_{ijk}$ is a matrix element of the charge operator defined in \cref{eq:charge_matrix_elements}, $\omega_{IR}$ is the infrared cut-off, and $t_m$ is the characteristic measurement time. Typical values are $\sqrt{|\log \omega_{IR} t_m|}\sim 4$ and $A_e = 10^{-4} e$~\cite{IthierPRB2006}. In the second term, $S(\omega)$ is the spectral function of the relevant bath, here assumed to be associated to dielectric losses.

\begin{figure}
    \centering
    \includegraphics[width=1\columnwidth]{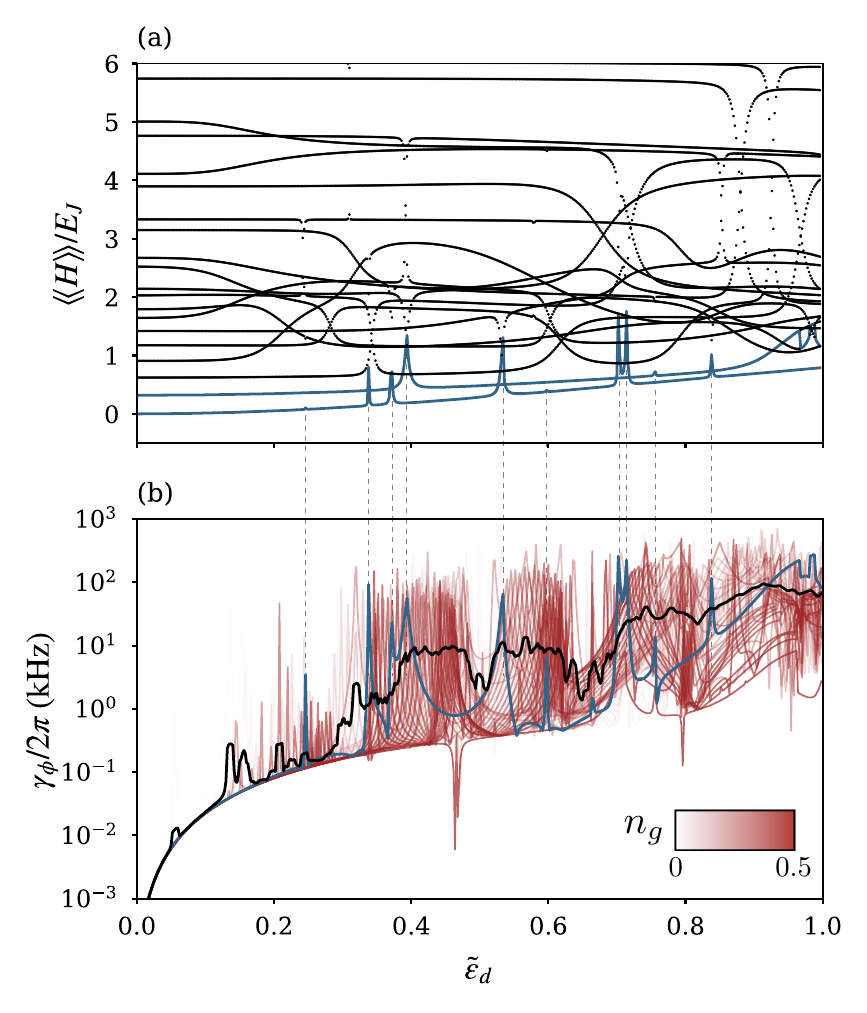}
    \caption{ (a) Mean energies of Floquet states of the driven transmon as a function of the drive amplitude, with $\hbar_\text{eff}^{-1} =3$, $\tilde \omega_d = 1.34$ and $n_g = 0.13$. While the ground and first excited states of the transmon are tracked as we vary the drive amplitude (blue lines), the rest of the states are not tracked (black dots).
    (b) Dephasing rate of the transmon as a function of drive amplitude due to $1/f$ charge noise and dielectric losses (see text) for 50 values of $n_g$ distributed uniformly between 0 and 0.5 (red lines) and the same parameters used in (a). The blue line corresponds to $n_g = 0.13$, as in (a). Small resonances between the computational states and the chaotic states that can be observed in (a) result in sharp peaks in the dephasing rate, due to $1/f$ charge noise. 
    The average dephasing rate over the offset charge (black line) is dominated by $1/f$ charge noise.}
    \label{fig:dephasing_rate}
\end{figure}

The dephasing rate obtained from \cref{eq:dephasing} and the numerically obtained Floquet modes $\ket{\phi_0(t)}$ and $\ket{\phi_1(t)}$ corresponding to the logical states in the presence of a drive is shown in \cref{fig:dephasing_rate}(b)
as a function of $\tilde \varepsilon_d$ for 50 values of $n_g$ uniformly spaced in the range $[0,0.5]$ (red lines).
All curves show a slow quadratic increase of $\gamma_\phi$ with the drive amplitude due to dielectric losses through the hybridization of $\ket{0}$ and $\ket{1}$ by the drive. This quadratic increase in the rate with the amplitude of the drive is a signature of a perturbative effect. 
More interestingly, the dephasing rate also displays sharp peaks whose position strongly depends on $n_g$ and $\tilde \varepsilon_d$, as is expected for CAT. To better understand the origin of these structures, the mean energy of the different Floquet mode is plotted in \cref{fig:dephasing_rate}(a) for $n_g = 0.13$ as a representative example. The value of $\gamma_\phi$ obtained for the same gate charge is highlighted in \cref{fig:dephasing_rate}(b) (blue line). By comparing the two plots (see the vertical dashed lines), it becomes clear that the sharp increases in $\gamma_\phi$ correspond to resonances between regular states (ground or first excited state) and chaotic states. As mentioned above, the resulting hybridization of the computational states with states that have a strong charge dispersion leads to a sharp increase in the dephasing rate. The black line is an average over all realizations of the gate charges. Depending on the time scale of the charge fluctuations~\cite{Christensen2019,Wilen2021} and of time needed to measure $\gamma_\phi$, this average may be more representative of potential experimental observations.

\section{Circuit QED: transmon coupled to a resonator}
\label{sec:cQEDsim}

\begin{figure}[t!]
    \centering
    \includegraphics{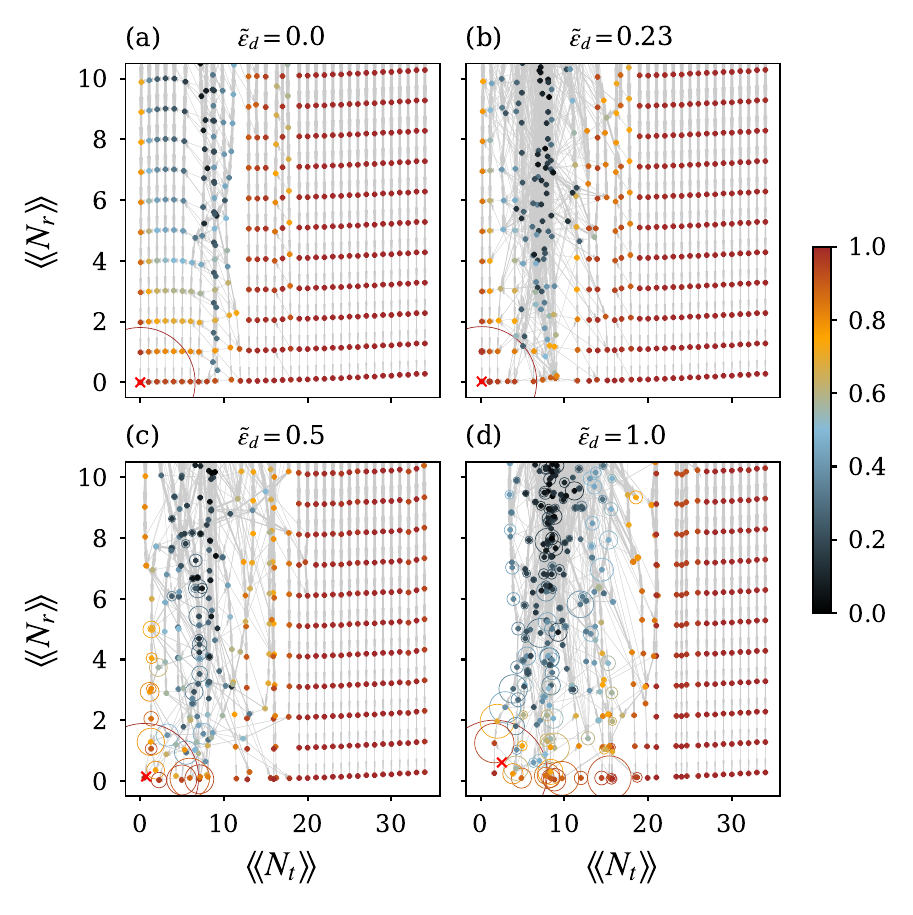}
    \caption{Grid of Floquet modes in the $(\langle \Nt \rangle_t,\langle \Nr \rangle_t)$ plane. The purity of the transmon reduced density matrix is encoded in the symbol color, for four different drive amplitudes: $\tilde{\varepsilon}=0$(a),  $0.25$(b), $0.5$(c), and $0.95$(d).
    In a zero-drive decoupled system, the points should form a square grid. In the coupled system, as $\langle \Nr \rangle$ and $\tilde{\varepsilon}_d$ increase, level hybridization leads to grid point displacement, and purity drops. With increasing drive, for $\langle \Nr \rangle_t \approx 0$, there is a purity drop in Floquet modes corresponding to the chaotic layer.  Steady state of Floquet-Markov master equation is represented by a red cross. The thickness of each arrow between a pair of Floquet modes is proportional to the corresponding rate $\Gamma_{ij}$ at zero temperature, apart from an offset to render the smallest rates visible. An arrow is plotted if $\Gamma_{ij} >10^{-3}\kappa$ for (a)-(b), or $10^{-2}\kappa$ for (c)-(d). Rates of the undriven system are dominated by single-photon relaxation (vertical lines), and Purcell decay (left-to-right arrows for $\langle \Nt \rangle_t \lesssim 6$) and rate flows tend to increase purity (a). Rates become increasingly non-local as the drive strength is increased (b)-(d) 
    Steady-state populations of each Floquet mode are encoded in the radii of circles around each grid point. For sufficiently low drive the steady state is dominated by the vacuum state (a)-(c), unlike steady states above the ionization threshold (d). See \cref{fig:purities_rates_grid_strong_drive} for the same data at higher drive power.
    }
    \label{fig:purities_rates_grid}
\end{figure}

We have so far treated the drive seen by the qubit as a purely classical field. To account for vacuum fluctuations and the richer structure of the energy levels in the presence of a cavity, we now consider a circuit QED setup where the transmon is capacitively coupled to a cavity. In this situation, the Hamiltonian takes the form~\cite{Blais2021}
\begin{align}
    \begin{split}
        \label{eq:cqed_model}
            \hhh(t) & =   4 E_C (\nnn-n_g)^2-E_J\cos(\ppphi) +\varepsilon_d\cos(\omega_d t)\nnn \\
            & + \omega_a \aaa^\dagger \aaa - i g \nnn (\aaa-\aaa^\dagger) \\
            &- (\aaa-\aaa^\dagger) \sum_{k} g_k (\bbb_k-\bbb_k^\dagger) +\sum_{k} \omega_k \bbb^\dag_k \bbb_k.
    \end{split}    
\end{align}
The first line of this expression is the Hamiltonian of the driven transmon qubit, as in \cref{eq:model}. The second line contains the free cavity Hamiltonian defined by a frequency $\omega_a$, together with the transmon-cavity capacitive coupling $g$. We take $g/2\pi=250$~MHz and $\omega_a / 2\pi= 8$~GHz, with the drive detuned from the cavity at $\omega_d / 2\pi = 7.5$~GHz,  while keeping the same parameters for the transmon qubit as in the previous sections. Finally, the last line represents the capactive coupling of the cavity to a bosonic bath. The drive term on the qubit can be either seen as directly acting on the qubit in the laboratory frame, or as resulting from a drive on the resonator. In that latter situation, the Hamiltonian of \cref{eq:cqed_model} is to be understood as expressed in a displaced frame where the drive on the cavity has been removed, \ie $\varepsilon_d = 2 g \sqrt{\bar{n}}$ where $\bar{n}$ is the cavity steady-state photon number in the laboratory frame.
In this section, we find the Floquet spectrum of \cref{eq:cqed_model} in the absence of coupling to the bath, and then calculate transition rates in linear response theory~\cite{VerneyPRApplied2019}. Details on the Floquet simulations are provided in \cref{sec:floquet_sim}.

\subsection{Structure of the Floquet spectrum}

To characterize the structure of the Floquet spectrum, for each Floquet mode we compute expectation values of a pair of operators that are good quantum numbers in the undriven and decoupled Hamiltonian: the transmon excitation number $\textbf{N}_t = \sum_{i,n} i\ket{in}\bra{in}$ and the resonator excitation number $\textbf{N}_r = \sum_{i,n} n\ket{in}\bra{in}$, where $\ket{in}$ is a bare state of the joint transmon-cavity system. In \cref{fig:purities_rates_grid} we show these quantities on a two-dimensional grid in the $(\llangle \Nt \rrangle,\llangle \Nr \rrangle)$ plane, the panels (a)-(d) representing different drive amplitudes $\tilde{\varepsilon}_d$. As in \cref{eq:Mean_Energy}, the double angle brackets represent the time-averaged expectation value of the operator in a given Foquet mode over a period of the drive.
In the absence of coupling and drive, this grid is expected to be rectangular. In the coupled case, for high enough resonator photon number $\llangle \Nr \rrangle$ or for large enough drive power, some Floquet modes deviate strongly from the regular rectangular grid. These strong deviations are associated with state dressing, and a significant hybridization of the two subsystems. More precisely, horizontal deviations from the rectangular grid can be interpreted as hybridization of transmon states, while vertical deviations originate from hybridization of Fock states. The hybridization of Fock states usually occurs through entanglement of the two systems, implying also a hybridization of the transmon states.

The degree of hybridization between the two subsystems is measured by the purity of the transmon reduced density matrix.  Encoding this purity into the color of the grid points in \cref{fig:purities_rates_grid}, deviations from the rectangular grid appear to correlate with drops in purity, \ie an increase of entanglement entropy between the transmon and the resonator. In particular, purity drops are drastic for states whose transmon excitation number corresponds to the chaotic layer identified in the previous sections for similar drive powers. 

We can gain further understanding of the purity drop by first diagonalizing the time-dependent Hamiltonian of the driven transmon [first line of \cref{eq:cqed_model}], and then expressing the transmon-resonator coupling $-i g \nnn (\aaa-\aaa^\dagger)$ in the joint basis of the transmon Floquet states $\{\ket{\tilde i}\}$ and cavity Fock states. In a frame rotating at the drive frequency for the resonator and at the quasienergies for the transmon, the Hamiltonian reduces to 
\begin{align}\label{eq:H_Floquetbasis_secV}
    \hhh(t) & =  \Delta \aaa^\dagger \aaa \notag \\
    & - i g \sum_{i,j,k} n_{ijk} \ket{\tilde i}\bra{\tilde j} (e^{{i}\Delta_{ij,k+1}t}\aaa -e^{{i}\Delta_{ij,k-1}t}\aaa^\dagger),
\end{align}
where {$\Delta = \omega_a - \omega_d$}. The charge matrix elements $n_{ijk}$ and the energy differences $\Delta_{ijk}$ are defined in \cref{sec:bath_coupling}, and the Floquet modes of the transmon are evaluated at $t=0$.
Large hybridization occurs between the states $\ket{\tilde i,n}$ and $\ket{\tilde j,n+1}$ if the coupling strength is of the order of the transition frequency, \ie $2g\sqrt{n+1}n_{ijk}\sim\Delta_{ij,k-1}-\Delta$. Note that, for small detuning $\Delta$, the transition frequencies $\Delta_{ij,k-1}-\Delta$ are small for $k=0,1$, and consequently hybridization of the states is mainly due to the matrix elements $n_{ij,k=0}$ and $n_{ij,k=1}$.
For chaotic states, we derive in \cref{sec:standard_deviation} an estimate of the matrix elements $n_{ijk}$ and the frequencies $\Delta_{ij,k-1}$. Using these estimates and the previous expression allows us to obtain a threshold value on the photon number $n$ for strong transmon-resonator hybridization (see details in \cref{sec:onset_hybridization}). For the parameters of \cref{fig:purities_rates_grid}, we find that the transmon states become strongly coupled even at the lowest Fock states. This strong hybridization can result in large decay rates for chaotic states, as explained in the paragraph below.

Transition rates, defined within linear response theory for the charge operator of the resonator $-i(\aaa-\aaa^\dagger)/\sqrt{2}$ in a similar way as in \cref{eq:Gamma}, are shown as gray arrows connecting the grid points in \cref{fig:purities_rates_grid}. At low drive power, rates are predominantly local, consisting of single-photon relaxation (vertical downward-pointing arrows connecting neighboring grid points), and
qubit Purcell decay (horizontal left-pointing arrows also connected neighboring grid points). The existence of one dominant rate allows to identify
states corresponding to definite transmon excitation number \cite{Shillito2022,Boissonneault_2010}. On the contrary, in the chaotic layer where the purity is low, at finite drive amplitude the rate matrix becomes non-local connecting non-adjacent points. Note that here, dissipation on the resonator is treated in linear response theory. In \cref{sec:dissipation}, we verify that typical loss rates of readout resonators ($\kappa/2\pi \sim 10~\textrm{MHz}$) do not alter the spectrum of the undriven system. In particular, hybridization is not weakened by the linewidth of the Fock states considered in \cref{fig:purities_rates_grid}.

Using the transitions rates, we can now find the steady state of the Floquet Markov master equation~\cite{VerneyPRApplied2019}. In order to avoid the saturation of the resonator Hilbert space (see \cref{sec:floquet_sim}), we impose a cutoff by setting all rates involving states with $\llangle \Nr \rrangle \geq 15$ to vanish. We find that this only causes a small quantitative change in the steady-state density matrix. 
The transmon and resonator excitation numbers in this steady state are represented on \cref{fig:purities_rates_grid} as a red cross for each drive power. Moreover, the occupation $p_{i}$ of each Floquet mode in the steady state
is encoded in the area of a circle centered at each grid point. For sufficiently low drive power, the steady state is dominated by the state with $\llangle N_{r,t}\rrangle \approx 0$, \ie the dressed-state closest to the vacuum state. Beyond a threshold drive amplitude, the steady state of the Floquet-Markov Lindblad master equation has significant weights on the chaotic states (see \cref{fig:purities_rates_grid_strong_drive} which shows similar results as \cref{sec:floquet_sim} but for larger drive amplitudes). This threshold corresponds to transmon ionization where states above the cosine potential of the transmon become populated~\cite{Lescanne_2019,VerneyPRApplied2019,Shillito2022}. Notably, strong hybridization leads to nonzero resonator photon number in the steady state, that is $\llangle \Nr \rrangle \gtrsim 2$ at $\tilde{\varepsilon}_d = 0.95$, see \cref{fig:threshold_ionization}.

\begin{figure}[t!]
    \centering\includegraphics{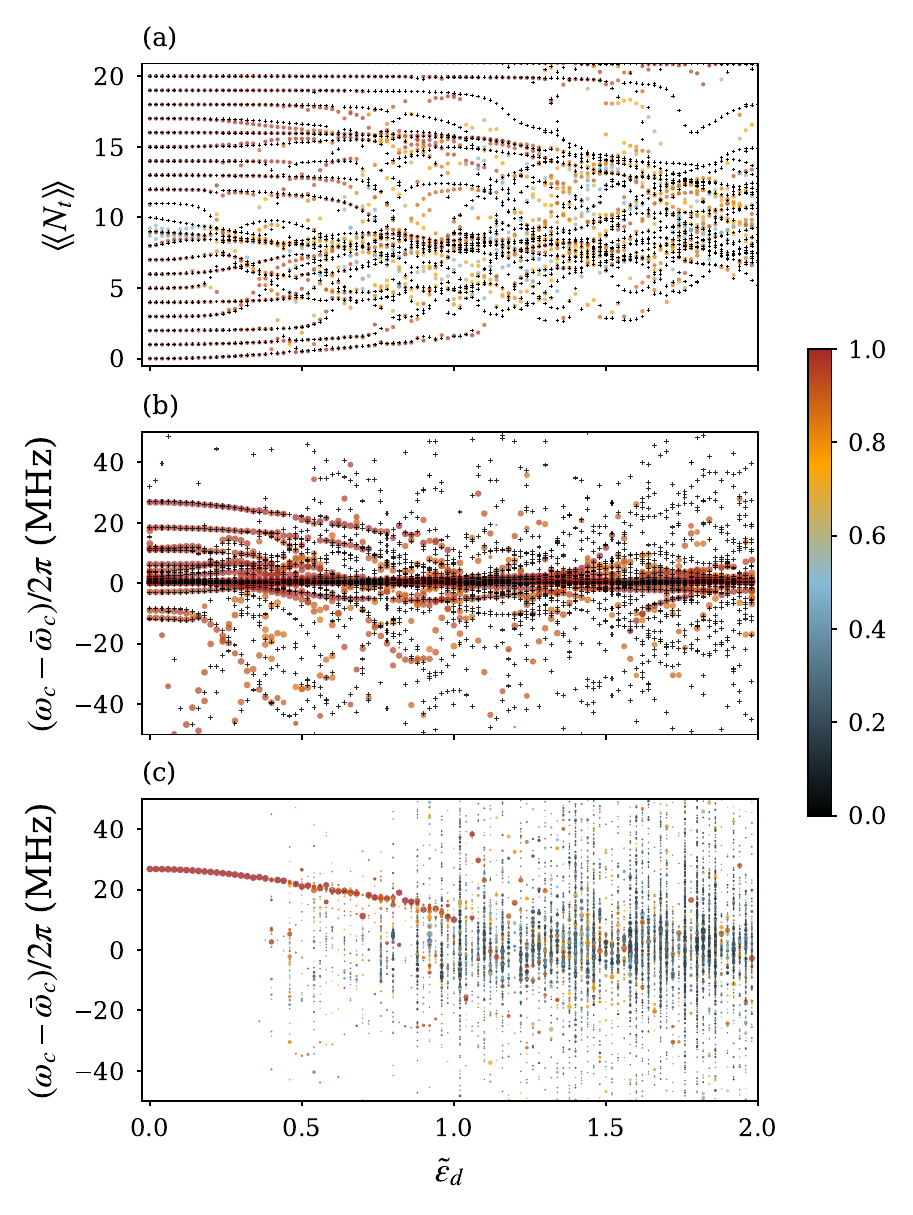}
    \caption{(a) Agreement of transmon population $\llangle \Nt \rrangle$ between the full circuit QED simulation (color encodes purity of transmon reduced density matrix; only states satisfying $\llangle \Nr \rrangle \leq 0.73$ are shown) of \cref{eq:cqed_model} and the driven transmon model of \cref{eq:model} (black crosses). 
    (b) Pulled resonator frequency as obtained from two-tone spectroscopy, with colors representing purities of Floquet modes corresponding to the resonator vacuum (only purities above 0.85 were retained, thus excluding cavity frequency pulls due to chaotic states). Pulls from perturbation theory \cref{eq:PT} are shown as black crosses. (c) Resonator frequency response in the steady state. Only states where the steady-state population times the square of the transition matrix element exceeded $0.001 \kappa$.
    }
    \label{fig:resonator_pull}
\end{figure}

Comparing to the mean-field study of a driven transmon of the previous sections, we find that the qualitative features of the spectrum of the off-resonantly driven circuit QED Hamiltonian match those of the driven transmon qubit. To illustrate this, we plot in \cref{fig:resonator_pull}(a) the transmon excitation number $\llangle \Nt\rrangle$ for the full circuit QED model of \cref{eq:cqed_model} (colored dots) versus drive amplitude and compare it to the corresponding observable in the Floquet spectrum of \cref{eq:model} with the same parameters, but  corresponding to a transmon driven by a classical field (black crosses). As in \cref{fig:purities_rates_grid}, for the former, color encodes the purity of the transmon reduced density matrix. While the agreement between the full circuit QED simulation and the mean field driven transmon is excellent for regular states, deviations appear within the chaotic layer where the purity drops. Differences within the chaotic layer between the two models are expected due to strong hybrization between the transmon and the resonator (not accounted in the mean field model), and due to fine sensitivity of chaotic spectra on external parameters \cite{haake_quantum_2010}. Nevertheless, in the two models, the threshold values of $\tilde{\varepsilon}_d$ for the onset of chaos agree.

\subsection{Frequency response of the resonator}

Performing the analog of a numerical two-tone spectroscopy experiment, in \cref{fig:resonator_pull}(b) we plot the pulled frequency response of the cavity versus drive amplitude $\tilde{\varepsilon}_d$. To obtain these results, we first identify the cavity vacuum states corresponding to each transmon occupation number. Then, for each of these vacuum states, labeled $i$, we identify the transition frequency corresponding to the largest matrix element $|\nnn_{ijk}|$. The corresponding transition rate is encoded in the symbol sizes. As in \cref{fig:resonator_pull}(a), the symbol color encodes the purity of the corresponding vacuum state. Only states with purity above $0.85$ are plotted, for which the cavity vacuum can be well identified as the states remain close to a tensor product state.

\Cref{fig:resonator_pull}(b) shows that, as expected, low-energy regular states of the transmon correspond to a traceable resonator pull that is monotonically decreasing with drive strength~\cite{Boissonneault_2010,Blais2021,Shillito2022}. In particular, there is a group of 4 levels (corresponding to pulls up to slightly below 30 MHz at zero drive) which exhibit ac-Stark shifts monotonically decreasing with drive strength $\tilde{\varepsilon}_d$ towards the bare frequency of the cavity, for sufficiently low drive. These levels correspond to resonator states pulled by the lowest regular transmon states
of \cref{fig:resonator_pull}(a), and the corresponding curves terminate for drive strengths where the corresponding transmon level strongly hybridizes with the chaotic layer. The pull is vanishing for the charge-like states of the transmon, as can be identified by an inspection of the transmon excitation numbers $\llangle \Nt \rrangle$ (not shown). 
On the other hand, the resonator frequency pull associated with chaotic states is not shown here because the two systems strongly hybridize when the transmon is in a chaotic state, rendering impossible the identification of the resonator vaccuum. Nonetheless, we address this point below when computing the steady-state density matrix.

Although it fails in the chaotic layer, perturbation theory in the transmon-cavity coupling can be used to compute the cavity pull in the regular regime, see the black symbols in \cref{fig:resonator_pull}(b). To second order, using \cref{eq:H_Floquetbasis_secV}, the cavity pull $\chi_i$ associated to the transmon Floquet state $i$ is given by~\cite{gramich_lamb-shift_2014} 
\begin{align}\label{eq:PT}
    \chi_i = \sum_{j,k} g^2 n_{ijk}^2\left(\frac{1}{\omega_a + \Delta_{ijk}}-\frac{1}{\omega_a-\Delta_{ijk}}\right). 
\end{align}
Here, $n_{ijk}$ and $\Delta_{ijk}$ can be numerically computed by diagonalizing the time-dependent Hamiltonian of the single driven transmon (see \cref{eq:H_Floquetbasis_secV} above), thereby fully accounting for the effect of the drive on the transmon.
In this way, we avoid the heavy computational cost of the Floquet operator on the full transmon-cavity system.
Perturbation theory (black symbols) agrees well with the full simulation only for regular states which can be identified in \cref{fig:resonator_pull}(b) by their large purity.
In particular, the perturbation theory for the cavity pull of the two computational states of the transmon agree with the full numerical calculations over the whole range of drive amplitudes for which the states remain regular.
This suggests using the Floquet basis as a starting point for perturbative treatments of multi-mode coupling used in two-qubit gate analysis \cite{sheldon_procedure_2016, malekakhlagh_first-principles_2020} or qubit readout \cite{Petrescu_2020}. This approach will be explored in future work. Note that for a small enough coupling strength $g$, the formula \cref{eq:PT} is expected to be valid also for chaotic transmon states. 

Additionally, we show in \cref{fig:resonator_pull}(c) the frequency response of the cavity in the steady state. For $\tilde{\varepsilon}_d \geq 0.95$ the average cavity response is centered on the bare cavity frequency, with a large variance. In Refs.~\cite{VerneyPRApplied2019,Lescanne_2019} the cavity response at the bare frequency was attributed to the Josephson potential essentially becoming transparent under strong drives, and the transmon being in a mixture of charge-like states. We rather observe here that this response is caused by the transmon entering a low-purity state with significant weight on a large number of chaotic Floquet modes, which corresponds to the transmon becoming strongly hybridized with the cavity and giving vanishing pull on average.

\section{Non-QND effects originating from the chaotic states}
\label{sec:Ncrit}

The purity drops and the connecting arrows in \cref{fig:purities_rates_grid} can be used to quantify
the quantum demolition character of the light-matter interaction in \cref{eq:cqed_model}.
For example, through its coupling with the resonator, the transmon can relax through the resonator bath. This effect, known as the Purcell decay, results in the left-pointing arrows connecting the low-energy states of the transmon in \cref{fig:purities_rates_grid}(a).
Moreover, approximating the transmon as a weakly nonlinear oscillator, recent works have shown that the small anharmonicity of the transmon can result in non-QNDness via correlated photon emission of the transmon-resonator systems~\cite{Malekakhlagh_2020,Petrescu_2020,hanai_intrinsic_2021}. However, the magnitude of these corrections, controlled by the relative transmon anharmonicity and the relative detuning of the transmon with the resonator, remains small compared to Purcell effect at sufficiently large detuning as considered in this paper. In addition, these correlated effects occur primarily at the qubit frequency, which could be efficiently suppressed by a Purcell filter. Here, we show that the presence of chaotic states in the transmon spectrum lead to non-QND effects that cannot be captured by a perturbative treatment.

In the previous sections, we have shown that part of the transmon spectrum is rendered chaotic when the system is driven and/or coupled to a resonator. Through the coupling with the resonator, see \cref{eq:H_Floquetbasis_secV}, regular states can decay to chaotic states through multi-photon transitions. This is further illustrated by the large nonlocal arrows in \cref{fig:purities_rates_grid} connecting distant dots in the $(\llangle N_t \rrangle, \llangle N_r \rrangle)$ plane. 
The inherently strong nonlinearity of these states cannot be accounted for by perturbative models taking the relative anharmonicity as a small parameter. Moreover, as shown in \cref{sec:bath_coupling}, chaotic states strongly depend on the gate charge $n_g$, while this parameter can be gauged away in oscillator models for the transmon where the phase becomes non-compact. We stress that, as compared to perturbative treatments where correlated effects occur at a few dominant frequencies, the transition frequencies involved in this decay are numerous and highly depend on the system parameters.

Spurious bath-assisted decay from the regular to the chaotic states can therefore result in a source of non-QNDness during strong-drive operations such as dispersive readout. In this section, in order to characterize such an effect, we introduce a `chaos-assisted' critical photon number $n_\textrm{ca}$, based on the onset of large hybridization between low-lying regular states and higher-energy chaotic states. Due to finite matrix elements between these states, transitions to the chaotic layer arise below $n_\textrm{ca}$. We propose to exponentially suppress this effect by increasing the ratio $E_J/E_C$.

\subsection{Chaos-assisted critical photon number}

Going back to our analysis of a driven transmon in the absence of quantum fluctuations from the resonator of \cref{sec:driven_transmon}, an `ionization threshold' between regular and chaotic levels occurs at a critical drive amplitude, and can be graphically identified from the plots of the mean energy of the transmon, see \cref{fig:mean_energy}(c,d). For each regular state, the critical drive corresponds to the drive amplitude $\tilde{\epsilon}_\textrm{d,ca}$ for which the level gets `absorbed' into the chaotic layer. As can be observed in \cref{fig:mean_energy}(c) and (d), as the drive amplitude increases, the chaotic layer grows into a cone-like shape whose tip corresponds to the separatrix energy, and thus the boundary between chaotic layer and regular states depends on drive amplitude. The closer a state is to the separatrix at zero drive, the smaller the critical drive amplitude necessary to have it completely hybridize with the chaotic layer. This critical drive amplitude then matches the so-called `ionization threshold' of the respective regular state~\cite{Shillito2022}.

Having associated the critical drive amplitude to the boundary between regular and chaotic states, we can observe this boundary in \cref{fig:mean_energy}(c) and (d) for two values of $\hbar_\textrm{eff}$. From these figures, we see that the dependence of the mean energy $\llangle H/E_J \rrangle$ \textit{at the boundary} as a function of $\tilde{\epsilon}_d$ is roughly independent of $\hbar_{\textrm{eff}}$, as this is ultimately tied to the energies of orbits at the regular-to-chaotic boundaries in the classical phase space in \cref{fig:mean_energy}(b). Remarkably, this implies that the mean energy $\llangle H/E_J \rrangle$ at the boundary solely depends on the parameters of the classical driven pendulum, \ie $\tilde \varepsilon_d = \varepsilon_d/\omega_p$ and $\tilde \omega_d = \omega_d/\omega_p$.

As emphasized in \cref{sec:driven_transmon}, what does depend on $\hbar_{\textrm{eff}}$ is the number of regular states below the chaotic layer. This number controls which mean energies  $\llangle H/E_J \rrangle$ are accessible by the quantum states, and ultimately the critical drive amplitudes at which regular states get absorbed into the chaotic layer.
As discussed in \cref{sec:drive_frequency}, the size of the regular central island in the Poincaré sections, e.g. in \cref{fig:freq_dependence_PS}(a)-(d), depends on the drive frequency and decreases with drive amplitude.
When quantizing the system, in the semiclassical limit, we expect the number of states contained in the low-energy regular subspace  to be directly proportional to the inverse of the effective Planck constant $\hbar_\textrm{eff}^{-1}$ and to the area of the regular island~\cite{percival_regular_1973,berry_regular_1977}. Hence, the critical drive amplitude for, say, the excited state, is defined as the amplitude at which the area of the regular island is equal to $2\hbar_\textrm{eff}$, \ie it contains exactly two states, the ground and the excited. The critical drive amplitude therefore increases with $\hbar_\textrm{eff}^{-1}$ for any given state: the larger the ratio $E_J/E_C$, the larger the critical drive amplitude needed to absorb the level into the chaotic layer.

Similarly, for the case of an undriven transmon coupled to a resonator, one can define a critical photon number $n_\textrm{ca} = ({\varepsilon}_\textrm{d,ca}/2g)^2$, where $\varepsilon_\textrm{d,ca}$ is the critical drive amplitude of the driven transmon without coupling to a resonator. We justify this definition by noting that the two systems, \ie the driven transmon and the undriven transmon coupled to a resonator, have similar spectra (see \cref{sec:static}). This hints towards the fact that one could replace the driving field by a quantum field without affecting the ionization threshold. For photon numbers $n>n_\textrm{ca}$, we expect the first excited state to be absorbed in the chaotic layer.

Note that this definition of $n_\textrm{ca}$ differs from the critical photon number $n_\textrm{crit}$ defined in  \cite{boissonneault_dispersive_2009, koch_charge-insensitive_2007}. In these works, $n_\textrm{crit}$ is used to ensure the validity of the dispersive appoximation through a small hybridization of the transmon (or two-level system in \cite{boissonneault_dispersive_2009}) with the resonator. Here, $n_\textrm{ca}$ is set as a threshold for chaos, and can be also seen as the onset of strong nonlinear effects.

\subsection{Exponential suppression of spurious decay to the chaotic layer with $\hbar_\textrm{eff}^{-1}$}

In this section, we argue that going to a deeper regime of the transmon, \ie larger ratio $E_J/E_C$, is a favorable regime in the presence of drive.
Indeed, another benefit of working in the large $E_J/E_C$ regime for the transmon is that the magnitude of the spurious interactions between the low-energy states and the chaotic states is exponentially suppressed with $\hbar_\textrm{eff}^{-1} = \sqrt{E_J/8 E_C}$.
This suppression mechanism is related to that of charge dispersion with $\sqrt{E_J/E_C}$~\cite{koch_charge-insensitive_2007}. Charge dispersion manifests itself through phase-slip events, which are in turn exponentially suppressed with $\sqrt{E_J/E_C}$ in an undriven transmon. In a driven transmon, chaos-assisted phase slips can occur through the interaction of the low-energy states with the chaotic states (see \cref{sec:bath_coupling}). Since phase slips involve the full nonlinearity of the cosine potential, they result in multiphoton transitions~\cite{houzet_critical_2020}. Nonetheless, as explained below, these chaos-assisted events are also exponentially suppressed with $\sqrt{E_J/8 E_C}$.

In a quantum system, the regular island and the chaotic layer are not dynamically separated as is the case in the classical system. Spurious transitions from the computational states to the chaotic states might occur well below the critical drive amplitude. The regular states have a finite overlap with the chaotic layer in phase space, resulting in finite coupling between regular and chaotic states. Note that CAT, studied in \cref{sec:bath_coupling}, is an example of such possible regular-to-chaos transition.
When a bath or a resonator is coupled to the driven transmon through the charge operator, this results in a finite decay from the regular states to the chaotic layer. Being concerned with high-fidelity operations such as qubit readout, one wished to minimize such
effects. 

First, to avoid these effects, one can maximize the size of the regular central island by driving at a suitable frequency, as discussed in \cref{sec:drive_frequency}. 
Second, one can maximize the number of states contained in the regular island. Previous works \cite{sheinman_decay_2006, backer_regular--chaotic_2008} have shown that for a state of a given excitation number, the tunneling rate to the chaotic layer is exponentially suppressed with $\hbar_\textrm{eff}^{-1}$. Intuitively, this follows from an exponential suppression, as a function of $\hbar_\textrm{eff}^{-1}$, of the phase-space overlap of a regular state with the chaotic layer decreases exponentially. In other words, the coupling of a regular state to the chaotic layer is mediated by the regular states in between, resulting in an exponential suppression with the number of intermediate states,  thus with $\hbar_\textrm{eff}^{-1}$. Note that at large values of $\hbar_\textrm{eff}^{-1}$, other processes such as resonance-assisted tunneling can alter this exponential law \cite{sheinman_decay_2006}.  

In practice, this translates to an exponential suppression of the matrix elements of the charge operator between regular and chaotic Floquet modes. As mentioned in \cref{sec:cQEDsim}, the matrix elements $n_{ijk}$ with $k =-1,0$ between Floquet modes determine the ability of the driven transmon to hybridize with the resonator and potentially decay into the resonator bath similarly to a Purcell effect. For a Floquet state $\ket{\tilde i}$ {of the driven transmon}, we define its {unitless} coupling to the subspace of {`high-energy'} Floquet states {above a threshold index $j>i$} as 
\begin{align}
    N_{ij, k\in \{-1,0\} } = \sqrt{\sum\limits_{{l \geq j; k=-1,0}} |n_{ilk}|^2}, \label{eq:nijk-10}
\end{align}
where the matrix elements $n_{ijk}$ are defined in \cref{eq:charge_matrix_elements}, and where the Floquet states are sorted through their mean energies defined in \cref{eq:Mean_Energy}. By its definition in \cref{eq:nijk-10}, the dimensionless coupling $N_{ij,k\in \{ -1,0\}}$ is a monotonically decreasing function. Relevant information about the coupling to the chaotic layer is encompassed in how fast this decrease is as a function of the threshold state index $j$.

\begin{figure}[t!]
    \centering
    \includegraphics{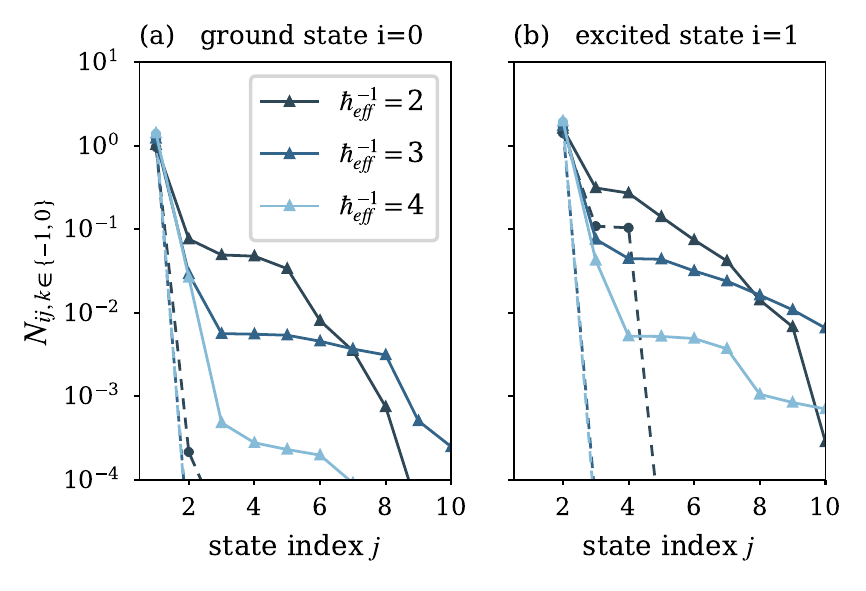}
    \caption{Unitless coupling $N_{ij, k\in \{-1,0\} }$ of the ground state (a) and first excited state (b) with the Floquet states of index $l\geq j$ over the Brillouin zones $k = -1,0$ (see \cref{eq:nijk-10}), for three values of $\hbar_\textrm{eff}^{-1} = 2,3,4$ corresponding to the ratios $E_J/E_C = 32, 72, 128$. The drive amplitude is set to $\tilde \varepsilon_d = 0.1$ (triangles and solid lines), and $\tilde \varepsilon_d = 0$ for comparison (filled circles and dashed lines). The values of the matrix elements are averaged over 50 values of the offset charge. The coupling to the chaotic layer is exponentially suppressed with $\hbar_\textrm{eff}^{-1}$. }
    \label{fig:exp_supp}
\end{figure}

In \cref{fig:exp_supp}(a) (respectively (b)), for three values of $\hbar_\textrm{eff}^{-1}$, and the same parameters as in \cref{fig:band_dispersion}, we plot the coupling $N_{ij, k\in\{-1,0\}}$ as a function of the index $j$ for the ground state $i=0$ (respectively the first excited state $i=1$), at zero drive (dotted line) and $\tilde \varepsilon_d = 0.1$ (solid line). As the matrix elements fluctuate as a function of the offset charge, they are  averaged over 50 values of $n_g$.
At zero drive, the matrix elements are {exponentially} suppressed with the state index $j$ for all three values of $\hbar_\textrm{eff}^{-1}$. 
At finite drive, after a rapid decay, all three curves stabilize on a plateau marked by comparatively slower decrease in both \cref{fig:exp_supp}(a) and (b). This plateau is due to the near-equality of the matrix elements $n_{ilk}$ when $l$ belongs to the chaotic layer. While the width of the plateau should grow with $\hbar_\textrm{eff}^{-1},$ as the chaotic states are more numerous, its value is exponentially reduced with $\hbar_\textrm{eff}^{-1}$. Due to its proximity to the chaotic states, the first excited state is more strongly coupled to the chaotic states than the ground state, which translates to larger values of the dimensionless coupling $N_{ij,k\in\{ -1,0\}}$ for the excited state than for the ground state, both within and away from the plateaus. 

Note that at this relatively small drive amplitude ($\tilde \varepsilon_d = 0.1$) and for a light-matter coupling strength $g/2\pi = 250~\textrm{MHz}$, the transmon-resonator interaction term involves transitions with the chaotic states at a coupling strength of approximately $20~\textrm{MHz}$ in the case $\hbar_\textrm{eff}^{-1} = 3$ ($E_J/E_C = 72$). The associated frequencies $\Delta_{ij,k-1}-\Delta$ strongly depend on the drive amplitude and the offset charge.
With the fluctuations of the offset charge and the ac-Stark shift leading to a sweep across the spectrum during readout \cite{Shillito2022, sank_measurement-induced_2016}, these transitions might result in non-QND effects during transmon readout.

\section{Conclusion}

In summary, we have shown that the presence of classical chaos has repercussions in the driven-dissipative quantum dynamics of transmon qubits. One consequence is that part of the spectrum of the driven transmon becomes strongly correlated, \ie it exhibits strong level hybridization, which favors chaos-assisted quantum phase slips that significantly enhance charge dispersion and qubit dephasing. This could also affect the coherence times of other Josephson-based qubits, such as the fluxonium, which involve arrays of hundreds of Josephson junctions~\cite{masluk_microwave_2012, manucharyan_evidence_2012, nguyen_high-coherence_2019}.
However, as pointed out in Ref.~\cite{burgelman_et_al_2022}, the inductively shunted transmon remains stable even at large drive powers.

With full circuit QED simulations, we have also shown that states corresponding to the chaotic layer have lower purity, which indicates that perturbation theories such as the dispersive approximation~\cite{Blais_2004} or black-box quantization~\cite{Nigg_2012} become inapplicable. In particular,
the Kerr nonlinear oscillator model of the transmon does not capture the large enhancement of charge dispersion and the strong correlations within the spectrum. This situation is analogous to recent studies of arrays of transmons~\cite{Berke_2022} which show that a dispersive theory cannot reproduce the spectrum in the chaotic phase. Moreover, we show that the response of the resonator at large drive amplitudes, commonly referred to as ``bright-stating’' the qubit, corresponds to entering a steady state massively populating chaotic states which exhibit high hybridization with the resonator, with the resonator pull averaging to zero~\cite{reed_high-fidelity_2010, Lescanne_2019}.

This important population of the chaotic states also mean that 
that, in order for numerical studies to capture chaos-induced effects on the low-energy states of a single driven transmon, it is necessary to use
a Hilbert space 
size
which contains the entire chaotic layer.
That is, in some instances it may be necessary to revisit the conventional wisdom that using a few transmon states is sufficient for accurate simulations of the driven transmon.
Moreover, based on the study of instabilities in the classical system, it may be possible to  identify favorable frequency placement for which the width of the chaotic layer is minimal, results which can be used to find optimal parameters for operations with strong drives such as dispersive qubit readout. The identification of the chaotic layer as a function of the classical parameters $\omega_d/\omega_p$ and $\epsilon_d/\omega_p$ leads to the definition of a chaos-induced critical photon number in the quantum system, which increases as a function of $E_J/E_C$. In particular, spurious transitions during strong-drive operations are expected to be minimized in the regime of large $E_J/E_C$.

\section*{Acknowledgments}
 We thank Michiel Burgelman, Mark Dykman, Mazyar Mirrahimi and Alain Sarlette for fruitful discussions. This work was supported by NSERC, the Canada First Research Excellence Fund and the U.S. Army Research Office Grant No.~W911NF-18-1-0411. This material is based upon work supported by the U.S. Department of Energy, Office of Science, National Quantum Information Science Research Centers, Quantum Systems Accelerator.

\appendix
\section{Suppression of chaotic behavior at high-frequency driving }
\label{sec:ACStarkShift}

\begin{figure}
    \centering
    \includegraphics[width=1\columnwidth]{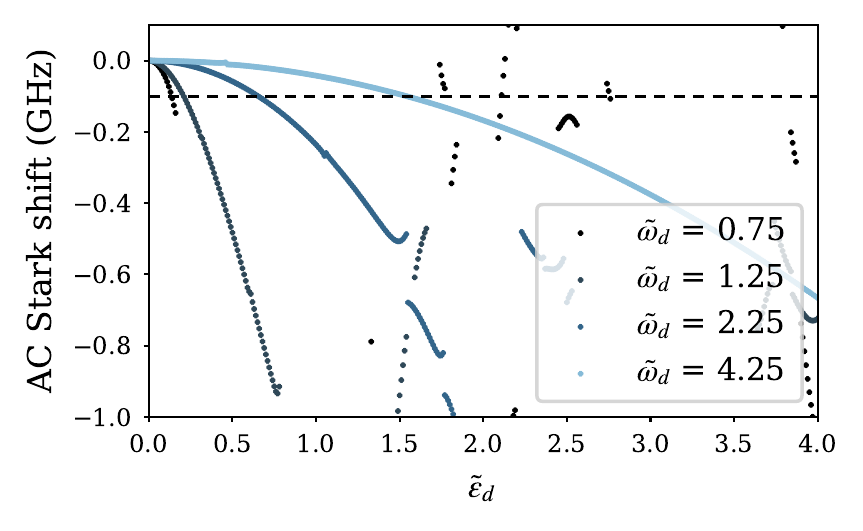}
    \caption[width=1\columnwidth]{ac-Stark shift of the 0 - 1 transition as a function of drive amplitude $\tilde \varepsilon_d$, for various frequencies, and at $\hbar_\textrm{eff}^{-1} = 2.45$. The dashed line indicates an ac-Stark shift of $-100 \textrm{MHz}$ and the corresponding amplitudes at which the Poincarés section in \cref{fig:freq_dependence_PS} are plotted. Tracking is lost beyond a certain drive amplitude which is well above that needed to work at a fixed shift of $-100 \textrm{MHz}$.}
    \label{fig:ACStarkShift}
\end{figure}

In \cref{sec:drive_frequency}, we study the dependence of the chaotic layer on the drive frequency. To have a fair comparison between the effects at different frequencies, we fixed the absolute value of the ac-Stark shift of the 0 - 1 transition to $100~\textrm{MHz}$, and correspondingly chose the drive amplitude. This is made possible by tracking the ground and first excited states as a function of drive amplitude to obtain the difference of the quasienergies, $\varepsilon_1 - \varepsilon_0$. The negative absolute value of the ac-Stark shift is plotted in \cref{fig:ACStarkShift} as a function of $\tilde \varepsilon_d$  for various frequencies.
Disruptions of the lines indicate that tracking is no longer possible.

\begin{figure*}
    \centering
    \includegraphics[width=1\textwidth]{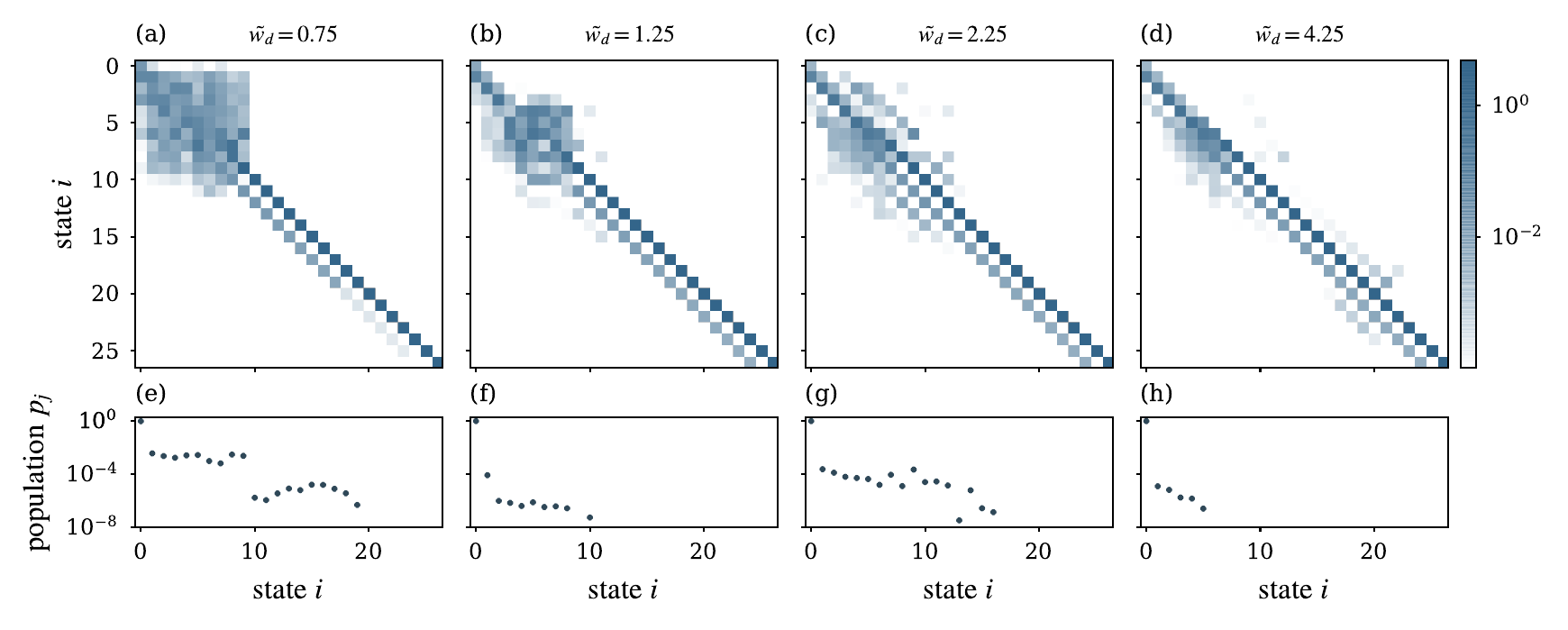}
    \caption[width=1\columnwidth]{ Rate matrices and steady-state populations at for the same parameters as in \cref{fig:freq_dependence_PS}, with the drive amplitude corresponding to an ac-Stark shift of $-100 \textrm{MHz}$. The rate matrices (b) and (d) are more regular, resulting in a smaller plateau in the steady-state distribution.}
    \label{fig:freq_ME2}
\end{figure*}

As a complementary study to \cref{sec:drive_frequency}, in \cref{fig:freq_ME2}(a)-(d), we show the rate matrices at an ac-Stark shift of $100~\textrm{MHz}$ for the same drive frequencies. The size of the chaotic block follows that of the chaotic layer in the Poincaré sections of \cref{fig:freq_dependence_PS}(a)-(d). In particular, the rate matrix corresponding to \cref{fig:freq_dependence_PS}(d) remains close to that of the undriven system [see inset of \cref{fig:rate_matrices}(b)], although the 0-1 transition frequency is shifted by $100~\textrm{MHz}$.

\section{Inversion symmetry sector of the driven system}
\label{sec:symmetry_appendix}

In this section, we provide further explanation on the inversion symmetry discussed in \cref{sec:bath_coupling}. First, let us consider the eigenvalue
equation $\hhh \ket{\psi} = E\ket{\psi}$ with $\hhh$ the Hamiltonian \cref{eq:model} of the undriven transmon ($\varepsilon_d = 0$). The eigenstates of $\hhh$ are
eigenstates of the parity operator $\mathbf{P}$, defined as $\mathbf{P}: \nnn \to - \nnn, \ppphi \to - \ppphi$, if and only if $n_g \equiv 0, 0.5 \text{ mod }1$. To prove this, we define the boost operator along charge $\mathbf{U} = \exp(-i n_g \ppphi)$. After transformation under this boost, the Hamiltonian $\hhh' = \mathbf{U} \hhh \mathbf{U}^\dag$ does not depend on $n_g$ and is therefore symmetric under $\mathbf{P}$. Thus parity acting onto an eigenstate must yield the same eigenstate if the spectrum is nondegenerate. That is, $\mathbf{P}\ket{\psi'}=e^{i\theta} \ket{\psi'}$, or $\langle -\phi|\psi'\rangle = e^{i\theta} \langle \phi|\psi'\rangle$, for all $\phi$, and some phase $\theta$ independent of $\phi$. Additionally, in the boosted frame, the eigenstates obey the `twisted-periodic' boundary condition $\langle \phi + 2\pi | \psi' \rangle = e^{-i 2\pi n_g}\langle \phi | \psi' \rangle$. Writing the two conditions above at $\phi=\pi$ gives $\langle -\pi | \psi' \rangle = e^{i\theta} \langle \pi | \psi'\rangle = e^{-i 2\pi n_g} \langle \pi | \psi'\rangle$, which implies that $e^{i\theta}=e^{i2\pi n_g}$. The wavefunctions are eigenstates of parity iff $\theta=0,\pi$, that is for $n_g = 0.0,0.5 \text{ mod }1$.

As explained in the main text, in the transmon regime, the low-energy sector is almost independent of the offset charge and the Hamiltonian becomes \textit{effectively} parity symmetric. When parity is a symmetry of $\hhh$, matrix elements of the charge operator $\langle i | \nnn|j \rangle \neq 0$ iff $i$ and $j$ belong to different parity sectors. Therefore if $\langle i | \nnn | i \rangle \neq 0 $ then parity symmetry is broken. By the Hellmann-Feynman theorem, $\langle i | \nnn | i \rangle \propto d E_i / d n_g$, so whenever a band has nonzero group velocity, parity symmetry is broken. This is why sweet spots (zero group velocity) occur at $n_g=0,0.5 \text{ mod }1$. Exponentially small group velocity, as in the transmon limit $E_J \gg E_C$, translates to exponentially weak breaking of parity. Thus, in the inset of \cref{fig:rate_matrices}(b) computed for $n_g=0.25$ and therefore nonzero group velocity, purple squares appear, but they are exponentially dim. By extension, in the low-energy manifold, transitions that are forbidden by parity selection rules at $n_g=0,0.5\text{ mod }1$ are exponentially suppressed when parity symmetry is broken, leading to an \textit{effective} parity symmetry. Note that it is this effective symmetry in the transmon limit $E_J\gg E_C$ that allows us to ignore charge dispersion effects and express the transmon $\hhh$ as a Kerr nonlinear oscillator.

At zero drive, this effective symmetry results in a suppression of the matrix elements of the charge operator $n_{i,i+2}$, and forbids the transition $i\rightarrow i+2$ in the low-energy sectors for both $n_g = 0.5$ and $n_g = 0.25$. This can be seen in the rate matrices of the undriven systems -- insets of \cref{fig:rate_matrices}(a,b)  -- where the squares corresponding to these transitions remain empty.
Importantly, this symmetry is not respected at $n_g = 0.25$ for states in the separatrix region where charge dispersion become non-negligible.

In the presence of a drive $\varepsilon_d \cos( \omega_d t)\nnn$, the inversion symmetry holds 
if one additionally applies the operation $t\rightarrow t+\pi/\omega_d$~\cite{GRIFONI1998229}. This symmetry of the driven Hamiltonian defines even and odd parity sectors among the time-dependent Floquet states. Note that the parity of a Floquet state depends on the Brillouin zone it belongs to in the undriven case. For example, in the undriven system, the eigenenergy of the first excited state decomposes as $E_1 = \varepsilon_1 + k\omega_d$, where the quasienergy satisfies $|\varepsilon_1|\leq \omega_d/2$, and $k=1$ for the parameters used in \cref{fig:rate_matrices}. Within the Floquet formalism, the Floquet state, corresponding to the eigenstate $\exp(-i E_1 t )\ket{1}$ of the time evolution operator, reads $\ket{\psi_1 (t)} = \exp(-i \varepsilon_1 t ) \ket{\phi_1 (t)}$, with the Floquet mode $\ket{\phi_1 (t)} = \exp(-i \omega_d t) \ket{1} $. Since the Floquet mode $\ket{\phi_1 (t)}$ is invariant under the transformation $t\rightarrow t+\pi/\omega_d$ and $\nnn \rightarrow -\nnn$, it belongs to the even sector. When increasing the drive, the state $\ket{\phi_1 (t)}$ will remain in the even sector, and expands only on even states $ e^{-i(2k+1)\omega_d t}\ket{2n+1}, e^{-i2k\omega_d t}\ket{2n}$, with $k\in\mathbb{Z}$ and $n\in\mathbb{N}$.

Under this symmetry, transitions through the charge operator between two states of the same (opposite) parity can only involve an odd (even) number of drive photons $k$, as the charge operator already changes the parity. This results in superposition of blue and red squares in at $n_g = 0.5$, see \cref{fig:rate_matrices}(a). On the contrary, at  $n_g = 0.25$, purple squares, indicative of red and blue contributions, light up not only in the chaotic block, but also for transitions involving the low-energy states.

Transitions which do not respect the apparent symmetry of the transmon were experimentally observed in \textcite{sank_measurement-induced_2016} but remained unexplained.
There, the situation is slightly more complicated as a harmonic mode is capacitively coupled to the transmon, making this effectively a 3-mode problem. Nonetheless, the situation is similar, and the same conclusions can be drawn. Indeed, the inversion symmetry for the full Hamiltonian holds upon adding the transformation $\nnn_r \to - \nnn_r$, where $\nnn_r=-i(\aaa-\aaa^\dagger)/\sqrt{2}$ is the charge operator of the resonator.
Because the resonator is low-Q, transitions between states are most likely to be induced by the bath coupled to the resonator through the resonator charge operator $\nnn_r$. Since $\nnn_r$ maps one parity sector to another similarly to $\nnn$, the single driven transmon analysis remains valid for the full driven circuit QED setup.

\section{Floquet simulations for circuit QED Hamiltonian}
\label{sec:floquet_sim}

\begin{figure}[t!]
    \centering\includegraphics{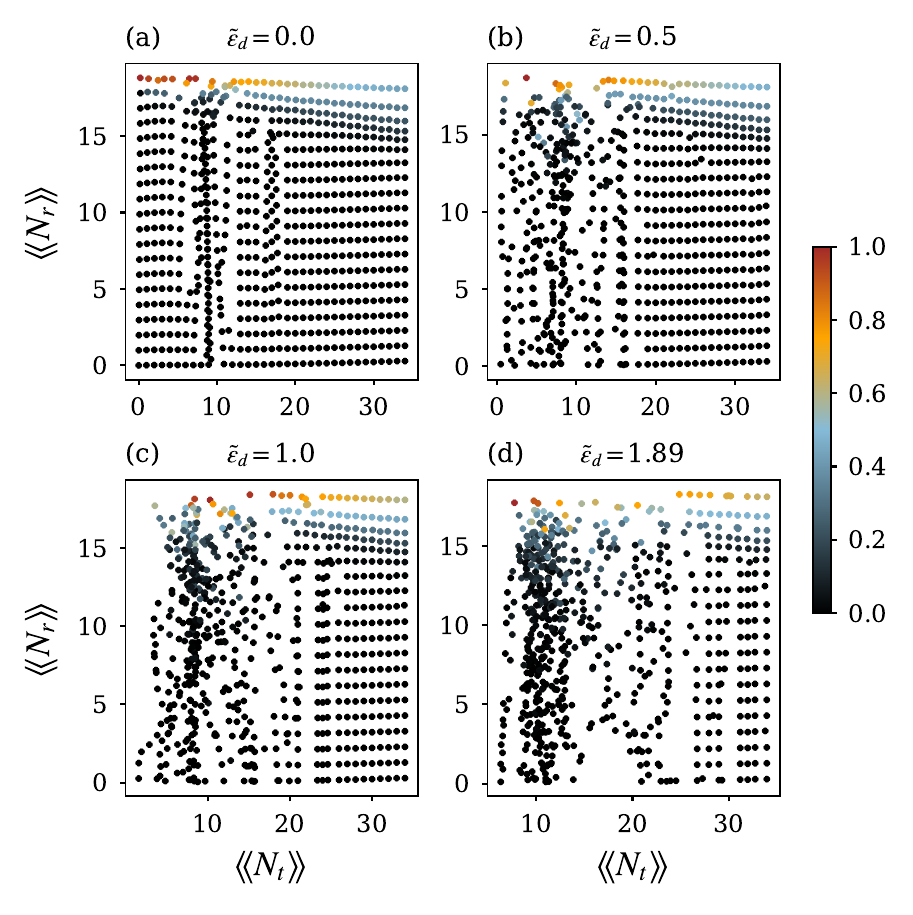}
    \caption{Commutator error, encoded in symbol color, for each Floquet mode, expressed as $|1-\llangle [ \aaa,\aaa^\dagger] \rrangle|$. }
    \label{fig:commerr_grid}
\end{figure}

In this Appendix, we provide more
details for the numerical simulations presented in \cref{sec:cQEDsim}. To capture the entirety of the chaotic layer even at strong drives $\tilde{\varepsilon}_d \geq 1$, we have used a local Hilbert space size of 35 for the transmon. For the resonator, we use 20 states, which is pertinent in the regime of off-resonant drives (here, the drive frequency is $500\text{ MHz}$ below the bare cavity frequency at $8\text{ GHz}$). We characterize
truncation errors
by plotting the error in the bosonic commutation relation in \cref{fig:commerr_grid}, and observe significant errors of the commutator only for states with $\llangle N_r \rrangle \geq 12$. We also observe
a `reflection' at the Hilbert space boundary \cite{Breuer1989Mar}, as indicated by monotonically decreasing values of $\llangle N_r \rrangle$ versus $\llangle N_t \rrangle$ (upper regions of the four panels of \cref{fig:commerr_grid}).  

\begin{figure}[t!]
    \centering
    \includegraphics[width=\linewidth]{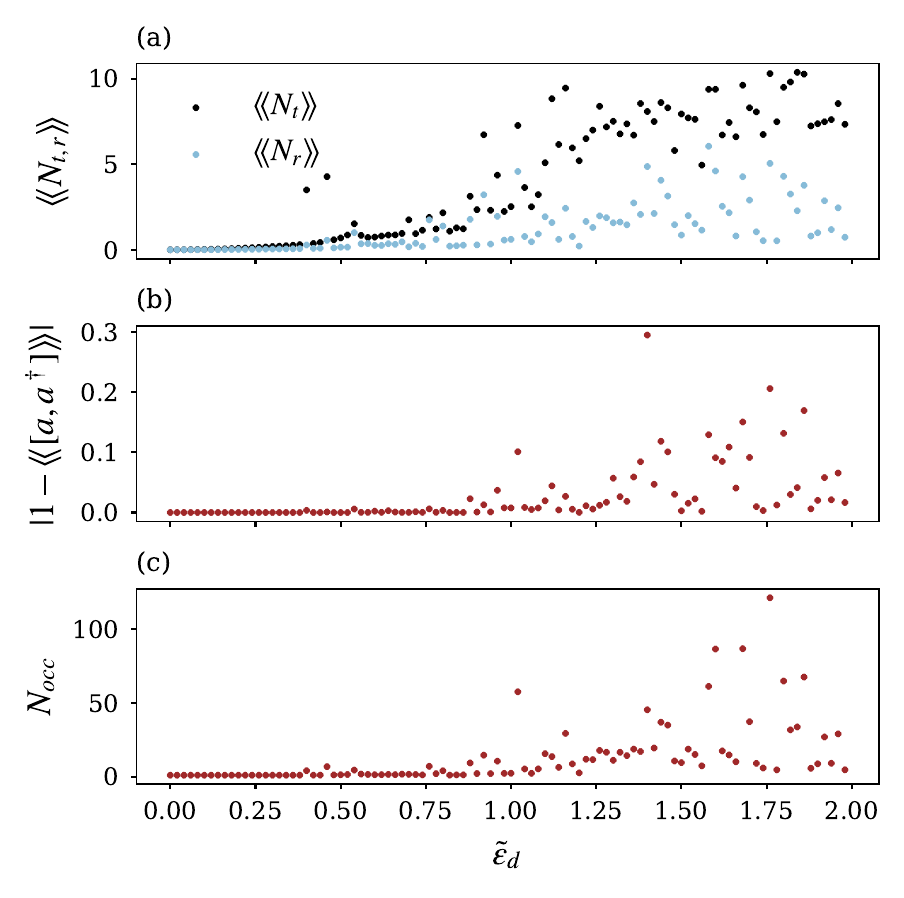}
    \caption{(a) Steady state population of the transmon and resonator as a function of drive strength. Both expectation values become non-monotonic beyond a threshold at $\tilde{\epsilon}_d \simeq 1$. (b) Commutator error (see also \cref{fig:commerr_grid}) in the steady state. (c) Number of occupied Floquet modes in the steady state \cite{burgelman_et_al_2022} as calculated from the entanglement entropy $N_{\textit{occ}} = \exp(-\sum_{i} p_i \log p_i)$.}
    \label{fig:threshold_ionization}
\end{figure}

Boundary effects in the Hilbert space become important at strong drives, where states with high $\llangle N_r \rrangle$ become populated, see \cref{fig:threshold_ionization}. In particular, averaging the commutator error in the steady state of the driven-dissipative evolution, we find that it correlates well with the number of occupied Floquet modes as determined from the entanglement entropy of the steady-state density matrix \cite{burgelman_et_al_2022}
[\cref{fig:threshold_ionization}(b,c)], which occurs at about the same threshold power $\tilde{\varepsilon}$ as ionization, defined as a significant increase in $\llangle N_t \rrangle$, see \cref{fig:threshold_ionization}(a).

When computing the steady state, a minimal bound for the required truncation of the resonator Hilbert space can be estimated based on energy considerations. As discussed in \cref{sec:cQEDsim}, one can first diagonalize the driven transmon Hamiltonian, and then write the transmon-resonator coupling in the new basis (Floquet basis for the transmon and Fock basis for the resonator). At strong drives, we need to account for the full chaotic layer of the driven transmon and its corresponding energy span. Although the mean energies of these states concentrate around $\llangle H \rrangle \sim 2E_J$, their Fourier decompositions extend over multiple Brillouin zones of width $\omega_d$. More precisely, the highest populated Brillouin zone is approximately given by the energy of the charge state that lies on the external boundary of the chaotic layer. For instance, on \cref{fig:mean_energy}(c) the border of the regular/chaotic domain is located around $\llangle H \rrangle_\textrm{max}= 3.5 E_J $ at   $\tilde \varepsilon_d  = 0.5$, and  around $\llangle H \rrangle_\textrm{max}= 5.5 E_J $ at   $\tilde \varepsilon_d  = 1$. 

Through the transmon-resonator coupling, the energy stored in the Floquet modes can be converted into resonator photons. In fact, the chaotic behaviour of the system with increasing Fock state number (strong hybridization) shows the ability for the resonator to absorb energy. 
The maximum energy $\llangle H \rrangle_\textrm{max}$ at  $\tilde \varepsilon_d  = 0.5$ (resp. $\tilde \varepsilon_d  = 1$), can be converted into $\llangle H \rrangle_\textrm{max}/\omega_a \approx 7 $ (resp. $11$) resonator photons. Interestingly, this is in agreement with the range of $\llangle N_r \rrangle$ of populated states in \cref{fig:purities_rates_grid} (highlighted with circles). Therefore, the dimension of the resonator Hilbert space has to be chosen larger than the estimated highest populated Fock state.

\begin{figure}[t!]
    \centering
    \includegraphics{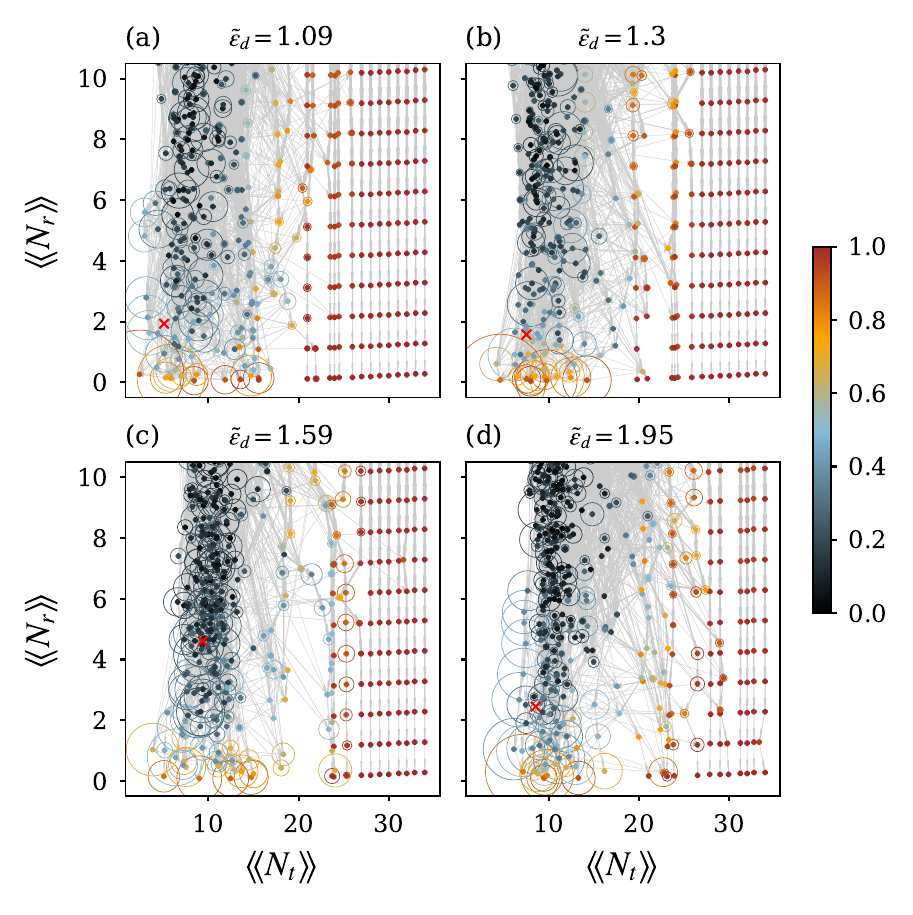}
    \caption{Same as \cref{fig:purities_rates_grid}, but for higher drive strengths. Steady-state expectation values of resonator photon number and transmon excitation number deviate significantly from zero. The steady state has significant weight over a large number of chaotic states. An arrow is plotted if $\Gamma_{ij} >10^{-3}\kappa$ for (a)-(b), or $10^{-2}\kappa$ for (c)-(d).}
    \label{fig:purities_rates_grid_strong_drive}
\end{figure}

\section{Chaos in the undriven cQED system}
\label{sec:static}

We consider an undriven cQED system composed of a transmon coupled to a resonator with Hamiltonian 
\begin{align}
    \begin{split}
        \label{eq:cqed_model_undriven}
            \hhh  =   4 E_C (\nnn-n_g)^2-E_J\cos(\ppphi) + \omega_a \aaa^\dagger \aaa - i g \nnn (\aaa-\aaa^\dagger).
    \end{split}    
\end{align}
This system is analog to a single driven transmon where the drive is replaced by a quantum field. We show that correlations develop in the spectrum with increasing Fock state number. By comparing the spectrum of this system with the Floquet spectrum of a driven transmon where an effective drive amplitude is varied, we find that the spectra agree at low photon number and for regular states, but show significant deviations for chaotic states.

\begin{figure}
    \centering
    \includegraphics[width=\linewidth]{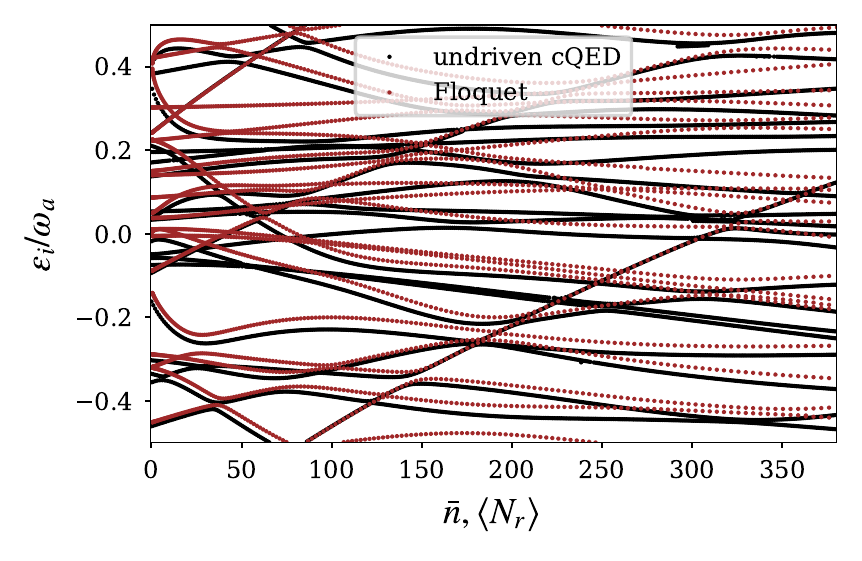}
    \caption{Black dots: eigenenergies of the undriven cQED Hamiltonian in \cref{eq:cqed_model_undriven}. For each eigenstate, the energy $\varepsilon_i$ and its mean photon number $\langle N_r \rangle_i$ are plotted in the $(\varepsilon_i, \langle N_r \rangle_i)$ plane. The energies $\varepsilon_i$ are the eigenenergies folded back into the first Brillouin zone defined by the resonator frequency. Only states satisfying $\langle N_t \rangle < 20$ are shown. The parameters are the same as in \cref{fig:purities_rates_grid}, with $\omega_a/2\pi = 8~\textrm{GHz}$ and $g/2\pi = 0.250~\textrm{GHz}$. 
    Red dots : quasienergies of the Floquet Hamiltonian \cref{eq:model} with $\omega_d/2\pi = 8~\textrm{GHz}$ as a function of $\bar{n} = (\epsilon_d/2g)^2$, for states satisfying $\langle N_t \rangle < 20$. The behaviour of the two spectra is qualitatively similar, but significant deviations appear at large photon number.}
    \label{fig:static_chaos}
\end{figure}

We bring the corresponding eigenenergies $E_i$ to the first Brillouin zone delimited by the resonator frequency $\omega_a$, by defining $\varepsilon_i = E_i[\omega_a]\in [-\omega_a/2,\omega_a/2]$. In \cref{fig:static_chaos} (black dots), for each eigenstate $\ket{i}$ satisfying $\langle N_t \rangle_i <20$, we show its energy  $\varepsilon_i$ and its mean photon number $\langle N_r \rangle_i$, in the $(\varepsilon, \langle N_r \rangle)$ plane. We note the appearance of seemingly continuous lines as a function of $\langle N_r \rangle$ (note that these are made of unlabeled dots), that can be associated to transmon states \cite{Shillito2022}. We refer to these lines as \textit{branches}.

For eigenstates with a given mean photon number $\bar{n}  = \langle N_r \rangle $, we can compare the spectrum of the undriven system with the quasispectrum of a single transmon driven (no resonator) at an amplitude $\varepsilon_d = 2g\sqrt{\bar n}$ (red dots in \cref{fig:static_chaos}). Note that large anti-crossings appear in both spectra, which is the signature of a correlated spectrum. Although the two spectra qualitatively agree at low photon number, significant deviations appear at large photon number. 

As recently observed numerically in Ref.~\cite{Shillito2022}, in the undriven system, the frequency pull on the resonator goes to zero as $\langle N_r \rangle$ increases. To see this, let us consider the effect of a probe on the system in a given state $\ket{i}$, with the probe frequency being close to that of the resonator, at $\omega_a+\delta$, where $\delta$ is small. A weak enough probe can only cause transitions to states with neighbouring mean photon number $\langle N_r \rangle_i \pm 1$. In addition, since the probe frequency is $\omega_a+\delta$, it can only connect $\ket{i}$ with states of energy $\varepsilon_i\pm\delta$, as all the energies are plotted modulo the resonator frequency. Hence, the probe causes transitions to neighbouring states in the $(\varepsilon_i, \langle N_r \rangle_i)$ plane, which correspond to neighbouring states on the same branch. The frequency pull is then $\varepsilon_j - \varepsilon_i$, where $\ket{j}$ is the state on the same branch with $\langle N_r \rangle_j \approx \langle N_r \rangle_i \pm 1$. 
Hence, the frequency pulls corresponds to the slopes of the apparent branches, which decrease with increasing photon number $\langle N_r \rangle$.

To explain this, let us consider an eigenstate state $\ket{i}$ on one of the energy branches,  at $\langle N_r \rangle_i$. We pick the state $\ket{j}$ that has maximum overlap with $\aaa^\dag \ket{i}/\norm{\aaa^\dag \ket{i}}$, so that $i \rightarrow j$ is the brightest transition upon driving the resonator when the system is in state $\ket{i}$. 
By using
\begin{align}
    \bra{j}\hhh \aaa^\dag \ket{i} = \bra{j}[\aaa^\dag \hhh + [\hhh,\aaa^\dag]]\ket{i}, 
\end{align}
we obtain the general relation
\begin{align}
    (E_j - E_i -\omega_a) = g\frac{\bra{j}\nnn \ket{i}}{\bra{j}\aaa^\dag \ket{i}}.
\end{align}
While $|\bra{j}\nnn \ket{i}|$ grows sublinearly with the square root of the photon number $\sqrt{\langle N_r \rangle_i}$ (see \cref{sec:standard_deviation}), $\bra{j}\aaa^\dag \ket{i}$ scales as $\sqrt{\langle N_r \rangle_i}$ by definition of $\ket{j}$. Hence, the frequency difference satisfies 
\begin{align}
   |\varepsilon_j - \varepsilon_i| = |E_j - E_i - \omega_a| \propto g/\sqrt{\langle N_r \rangle_i} \rightarrow 0.
\end{align}

\section{Standard deviation of the transmon dipole moment in the chaotic phase}
\label{sec:standard_deviation}

We consider a single transmon driven at high enough power to render all of the phase-like states chaotic. The Hilbert space is then simply composed of a finite number $N$ of chaotic states and an infinity of regular charge states. We define $U$ the unitary that diagonalizes the Floquet propagator $U_F = \mathcal{T}\exp[-i\int_0^T \hhh(t)dt]$, where $\mathcal{T}$ is the time-ordering operator. Because the system is chaotic, the matrix elements $U_{ij}$ of the unitary $U$ projected on the chaotic subspace can be modeled as random variables, distributed with variance $1/N$ and zero mean~\cite{haake_quantum_2010}.
The charge matrix element $n_{ij}$ in the Floquet basis (at a given time) are related to the charge matrix element $\tilde n_{ij}$ in the Floquet basis at zero drive via the unitary $U$, through $\nnn = U\tilde \nnn U^\dagger$.
We can therefore write
\begin{align}
\label{eq:matrix_element}
    n_{ij} = \sum_{r,s} U_{ir} \tilde n_{rs} U_{js}^*.
\end{align}
The $n_{ij}$'s can thus be modeled as random variables. Taking the average over these random variables of the modulus squared of the above equality, we obtain
\begin{equation}
     \begin{split}
        \mathbb{E}(|n_{ij}|^2) 
        & = \sum_{r,s,t,u}  \mathbb{E}(U_{ir}U_{js}^* U_{it}^* U_{ju})\tilde n_{rs} \tilde n_{tu}^* \\
        & = \sum_{r,s} \mathbb{E}(|U_{ir}|^2 |U_{js}|^2) |\tilde n_{rs}|^2 \\
        & \approx \sum_{r,s} \mathbb{E}(|U_{ir}|^2) \mathbb{E}(|U_{js}|^2) |\tilde n_{rs}|^2 \\
        & = \sum_{r,s}|\tilde n_{rs}|^2/N_\textrm{ch}^2.
    \end{split}
\end{equation}
The second line is obtained by observing that only squared matrix elements $|U_{ir}|^2$ survive when taking the average value. In going from the second to third line, we obtain an approximation by neglecting the correlations between $|U_{ir}|^2$ and $|U_{js}|^2$. Finite correlations exist, since the $U_{ir}$ obey orthonormality conditions. However, using the probability distributions derived in Ref.~\cite{haake_quantum_2010} for random matrices, we checked numerically that these second-order correlations are small if the number of chaotic states is large enough, \ie $N_\textrm{ch}>8$, and vanish in the large $N_\textrm{ch}$ limit. The last line is obtained using the equality $\mathbb{E}(|U_{ir}|^2) = 1/N_\textrm{ch}$. 

The term $\sum_{r,s}|\tilde n_{rs}|^2$ can be easily estimated. Indeed, one can write $\sum_{r,s} |\tilde n_{rs}|^2  = \norm{\bf{P}_{ch}\nnn \bf{P}_{ch}}^2$, where $\bf{P}_{ch}$ is the projector on the chaotic subspace. As the regular states are charge states, a basis of the chaotic subspace are the $2N_\textrm{ch}$ charge states $\ket{n_\textrm{charge}}$ with $n_\textrm{charge} < N_\textrm{ch}/2$. We can therefore calculate $\norm{\bf{P}_{ch}\nnn \bf{P}_{ch}}^2$ in the charge basis, giving  
\begin{align*}
    \norm{\bf{P}_{ch}\nnn \bf{P}_{ch}}^2 &=  \sum_{n_\textrm{charge}=-N_\textrm{ch}/2}^{N_\textrm{ch}/2} n_\textrm{charge}^2 \\&= \frac{(N_\textrm{ch}/2)(N_\textrm{ch}/2+1)(N_\textrm{ch}+1)}{3}.
\end{align*}
This leads to 
\begin{equation*}
    \sqrt{\mathbb{E}(|n_{ij}|^2)} \approx \sqrt{\frac{N_\textrm{ch}}{12}}.
\end{equation*}
In other words, the standard deviation of the dipole moment for a given transition increases with the size of the chaotic layer. Loosely speaking, the action of the unitary $U$ is to ``randomize'' the matrix elements of the charge operator.
As a consequence, the regular structure of the matrix elements of the undriven system, e.g the harmonic oscillator-like structure  $n_{ij} \propto \delta_{j,i+1}+\delta_{j,i-1}$ of the low-energy states, is not preserved.

For a numerical illustration of the above, we define the mean dipole moment for a Floquet state $i$ by 
\begin{equation}\label{eq:dipole_moment}
    \langle n_i \rangle = \sqrt{\frac{1}{M}\sum_{j=0}^{M} |n_{ij}|^2},
\end{equation}
where $M$ is an integer larger than the number of chaotic states $N_\textrm{ch}$. The matrix elements $n_{ij}$ are taken between Floquet modes at time $t = 0$.  We also define the mean dipole moment  
\begin{equation*}
    \langle n \rangle= \sqrt{\frac{1}{M}\sum_{i=0}^{M} \langle n_i \rangle^2}.
\end{equation*}
Using the same argument as above, we have $\langle n \rangle = \norm{\bf{P}_{M}\nnn \bf{P}_{M}}\approx \sqrt{{M}/{12}}$, where $\bf{P}_{M}$ is the projector on the manifold spanned by the charge states $\ket{m}$ such that $|m|\geq M/2$. 

\begin{figure}
    \centering
    \includegraphics[width=\linewidth]{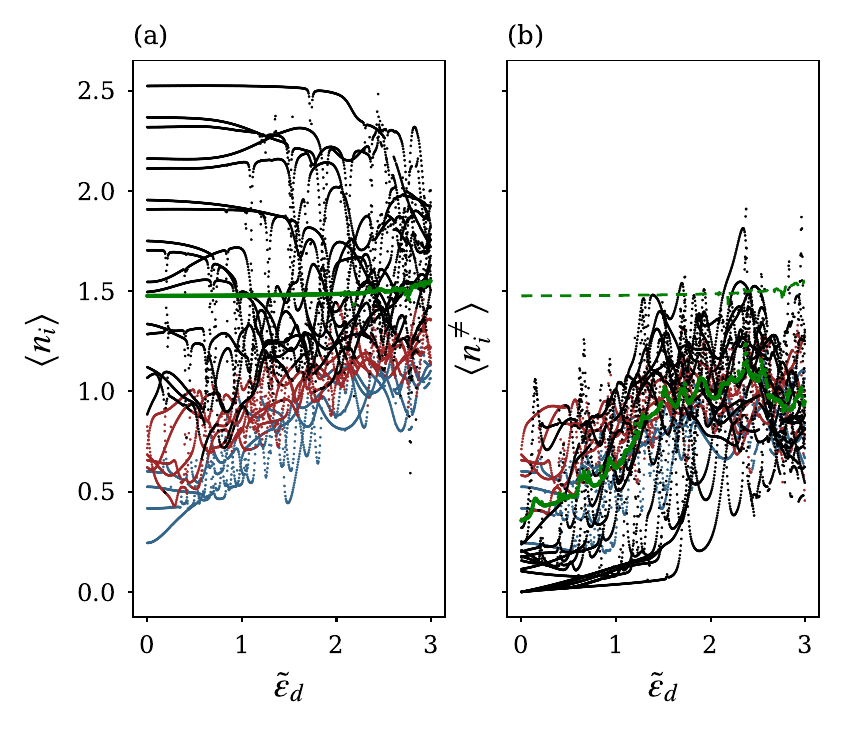}
    \caption{(a) Mean dipole moment ${\langle n_{i} \rangle}$ associated to each Floquet mode $i$, as a function of the drive amplitude, for the same parameters as in \cref{fig:mean_energy}(d), except $n_g = 0.12$.  The states are sorted through their mean energies. For each Floquet state $i$, the average in \cref{eq:dipole_moment} is performed over the first 25 transmon states $j$. For clarity, the low-energy states ($0 \leq i \leq 3$) are colored in blue, the states initially close to the separatrix ($4 \leq i \leq 8$) appear in red, and the above initially charge-like states are colored in black. The average dipole moment ${\langle n \rangle}$ is shown in green. Its constant value from $\tilde \varepsilon_d = 0$ to $2.5$ indicate that higher-energy states do not yet hybridize with the represented states here. For $\tilde \varepsilon_d > 2.5$, we see that ${\langle n \rangle}$ starts to increase. (b) Same as in (a) but using the definition \eqref{eq:dipole_moment_neq} for the dipole moment ${\langle n_{i}^\neq \rangle}$, where the average is performed over states with $j \neq i$. The dipole moment is low for charge-like states until they enter the chaotic layer. The average dipole moment ${\langle n^\neq \rangle}$ increases with the drive amplitude, due to the increasing number of charge-like states with initially high dipole moment ${\langle n_{i} \rangle}$ hybridizing with chaotic states. The average dipole moment ${\langle n \rangle}$ is plotted for comparison as a green dashed line. At $\tilde \varepsilon_d \approx 2.5$, the states are fully hybridized and ${\langle n^\neq \rangle}$ is close to ${\langle n \rangle}$. For $\tilde \varepsilon_d > 2.5$, hybridization with higher-energy states start to decrease ${\langle n^\neq \rangle}$, where the average is performed only on the first 25 states.  Besides, the dipole moment of low-energy states ${\langle n_{i}^\neq \rangle}$ (in blue) dramatically increase upon entering the chaotic layer.}
    \label{fig:dipole}
\end{figure}

In \cref{fig:dipole}(a), the mean dipole moment per Floquet state $\langle n_i \rangle$ is plotted for the first 25 Floquet states (sorted by mean energy) as a function of the drive amplitude. For clarity, we highlight in red states that are starting close to the separatrix, in blue the low-energy states, and in black the charge-like states. In green is plotted the mean dipole moment $\langle n \rangle $. Its constant value for $\tilde \varepsilon_d <2.5$ is a good indication that the first $M$ states do not hybridize with higher charge-like states for this range of drive amplitude. Note that $M = 25$ gives $\langle n \rangle = \sqrt{{M}/{12}} \approx 1.44 $, which agrees with the green curve. For $\tilde \varepsilon_d >2.5$, we note that $\langle n \rangle $ increases, indicating the increasing hybridization of higher-energy states with the 25 states represented here. This is also an indication that at $\tilde \varepsilon  = 2.5$, the first 25 states are hybridized. 
As the drive amplitude increases, the mean dipole moments per Floquet states $\langle n_i \rangle$ converge slowly towards $\langle n \rangle$.

However, in the definition \cref{eq:dipole_moment} of the mean dipole, the diagonal matrix elements $n_{ii}$ also contribute to the average, yielding large dipole moments for charge-like states. As $n_{ii}$ does not result in transitions to other states, we define the more meaningful dipole moment for the state $i$,
\begin{equation}\label{eq:dipole_moment_neq}
    \langle n_i^\neq \rangle = \sqrt{\frac{1}{M-1}\sum_{j=0,j\neq i}^{M} |n_{ij}|^2},
\end{equation}
and the mean dipole moment 
\begin{equation*}
    \langle n^\neq \rangle= \sqrt{\frac{1}{M}\sum_{i=0}^{M} \langle n_i^\neq \rangle^2}.
\end{equation*}
These quantities are plotted in \cref{fig:dipole}(b), with the same color code used in panel (a). If the values of $\langle n_i^\neq \rangle$ are qualitatively the same as in \cref{fig:dipole}(a) for the blue and red states, the behaviour is dramatically different for the charge-like states, for which $\langle n_i^\neq \rangle$ is small at low drive power. This is a consequence of the fact that the charge operator is mostly diagonal for charge-like states. Crucially, we note that as the drive strength increases, the mean dipole moments $\langle n_i^\neq \rangle$ converge toward $\langle n^\neq \rangle$, which increases (green curve). This is due to the increasing participation of charge-like states in the chaotic layer. Note that at $\tilde{\varepsilon}_d = 2.5$, all 25 states are hybridized, and $\langle n^\neq \rangle$ has come close to $\langle n \rangle$ (green dashed line in \cref{fig:dipole}(b)).

Finally, note that the mean dipole moment of the ground state, corresponding to the lowest blue line in \cref{fig:dipole}(b), undergoes a 4-fold increase upon entering the chaotic layer for $\tilde \varepsilon_d > 1$.

\section{Onset of strong hybridization between the transmon and the resonator in the presence of drive}
\label{sec:onset_hybridization}

In this section, we study in detail the onset of the hybridization of the transmon and resonator when the (driven) transmon is in a chaotic state. We start with the Hamiltonian of the driven transmon coupled to a resonator, \cref{eq:cqed_model} in the main text, reproduced here in the absence of the coupling to the bath
\begin{align*}
    \hhh(t) & =   4 E_C (\nnn-n_g)^2-E_J\cos(\ppphi) +\varepsilon_d\cos(\omega_d t)\nnn \\
    & + \omega_a \aaa^\dagger \aaa - i g \nnn (\aaa-\aaa^\dagger).
\end{align*}
Moving to the Floquet basis $\{\ket{\tilde i}\}$ for the driven transmon, and in the frame rotating at the drive frequency, this Hamiltonian reduces to 
\begin{align}\label{eq:H_Floquetbasis_appendix}
    \hhh(t) & =  \sum_i \varepsilon_i \ket{\tilde i}\bra{\tilde i} + \Delta \aaa^\dagger \aaa \notag \\
    & - i g \sum_{i,j,k} n_{ijk} \ket{\tilde i}\bra{\tilde j} (e^{i(k-1)\omega_d t}\aaa -e^{i(k+1)\omega_d t}\aaa^\dagger),
\end{align}
where $\Delta = \omega_a-\omega_d$, and $\varepsilon_i$ are the quasienergies of the driven transmon. Assuming that $|\Delta|\ll \omega_d$ and as $|\varepsilon_i| < \omega_d/2$, the interaction term (second line) can come close to resonance for $k=0,1$ (term in $\aaa$) and $k = -1,0$ (term in $\aaa^\dag$). 
In order to study the impact of chaotic transmon states on the resonator, for simplicity we assume that the drive strength is large enough for all the phase-like states to be ionized, as in \cref{sec:standard_deviation}.

\begin{figure}
    \centering
    \includegraphics[width=\linewidth]{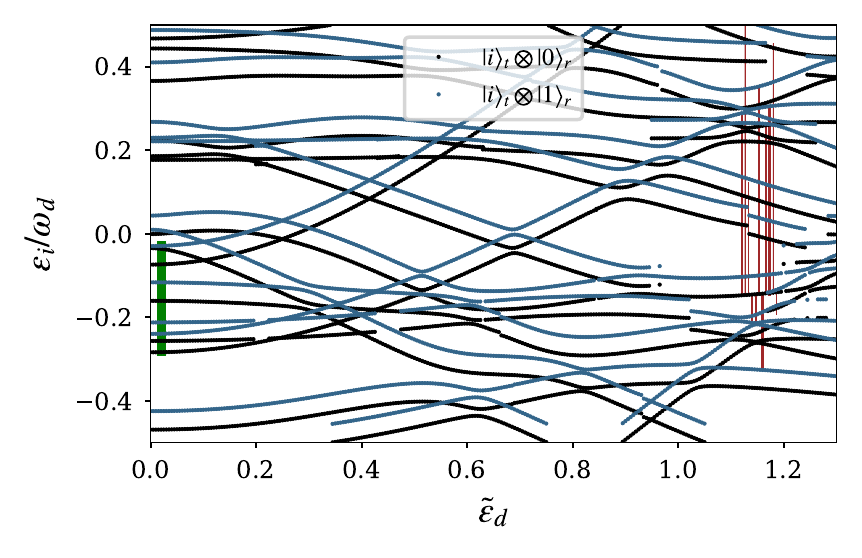}
    \caption{Spectrum of the uncoupled driven system corresponding to the Hamiltonian in  \cref{eq:H_Floquetbasis_appendix} with $g = 0$, $\Delta/2\pi = 200~\textrm{MHz}$, and the same parameters as in \cref{fig:dipole}. The energies $\varepsilon_i$ represent states of the form $\ket{\tilde i} \otimes \ket{n}$, where only the first 13 transmon states $\ket{i}$ (sorted through mean energies) are represented here, along with the $n = 0,1$ Fock states of the resonator. The states $\ket{\tilde i} \otimes \ket{0}$ (black dots) and $\ket{\tilde i} \otimes \ket{1}$ (blue dots) are translated by $\Delta$. The dots form lines which are interrupted when hybridization with higher-energy levels occurs (accompanied by sudden increase of mean energy). As explained in the text, at finite $g$, the two subspaces are coupled through the term $-i g n_{ij,-1}a^\dagger + h.c$.  At $\tilde \varepsilon_d = 0.05$, the green line represents the dominant coupling the state  $\ket{\tilde 1} \otimes \ket{0}$ (black dot) with the state $\ket{\tilde 0} \otimes \ket{1}$ (blue dots). The thickness of line is proportional to the coupling strength. The couplings to other states $\ket{\tilde j} \otimes \ket{1}$ were too small to be drawn (1000 times smaller). A dominant coupling results in an energy shift whose sign depends on the detuning.  At $\tilde \varepsilon_d = 1.15$, almost all states are strongly hybridized. As an example, we randomly pick one state $\ket{\tilde i} \otimes \ket{0}$ and we plot in red the lines representing the coupling to states of the one-photon subspace (with a slight horizontal offset for clarity). The couplings are numerous and of similar amplitude as shown by the thickness of the lines, with positive and negative detunings. All of these transitions contribute positively and negatively to shifting the energy of the state $\ket{\tilde i} \otimes \ket{0}$. }
    \label{fig:frequency_pull}
\end{figure}

We consider the tensor product state $\ket{\tilde i}\otimes\ket{n}$, composed of a chaotic transmon Floquet mode $\ket{\tilde i}$ and the n-th Fock state of the resonator. The term $-i g n_{ij,-1}a^\dagger $ couples this state to the state $\ket{\tilde j}\otimes\ket{n+1}$ with a coupling strength $g n_{ij,-1}\sqrt{n}$. Let us first give an intuitive picture of the physics by projecting the Hamiltonian in \cref{eq:H_Floquetbasis_appendix} on the vacuum and first few Fock states of the resonator. The corresponding situation is depicted in \cref{fig:frequency_pull}, where we plot the quasispectra of the uncoupled system corresponding to the Hamiltonian in \cref{eq:H_Floquetbasis_appendix} with $g = 0$, for the states $\ket{\tilde i} \otimes \ket{0}$ (black dots) and $\ket{\tilde i} \otimes \ket{1}$ (blue dots). Only the first 13 states $\ket{\tilde i}$ (sorted through mean energies) are represented, which are all part of the chaotic layer of the driven transmon at $\varepsilon_d \geq 1$. The two quasispectra are detuned by $\Delta$, due to the term $\Delta \aaa^\dagger \aaa$ in \cref{eq:H_Floquetbasis_appendix}. Here, we chose $\Delta/2\pi = 200~\textrm{MHz}$ for visual clarity. 

At $\epsilon_d = 0.05$, the possible transitions, due to the term $-i g n_{1j,-1}a^\dagger + h.c$, involving the state $\ket{\tilde 1} \otimes \ket{0}$ and the states $\ket{\tilde j} \otimes \ket{1}$, are shown as green vertical lines, where the thickness of the lines is proportional to the coupling strength. These lines couple one black dot and several blue dots. As expected from the harmonic oscillator-like structure of the matrix elements, one transition is dominant, corresponding to the transition with the state $\ket{\tilde 0} \otimes \ket{1}$. The coupling with the latter is the main contribution to the energy shift of the state $\ket{\tilde 1} \otimes \ket{0}$. Other green lines, representing coupling to other states, do not appear here as their coupling strength is more than a thousand times smaller than that of the dominant transition.  

At $\epsilon_d = 1.15$, almost all states represented are strongly hybridized. We pick one state of the form $\ket{\tilde i} \otimes \ket{0}$ (represented as a black dot), and similarly, the possible transitions and coupling strength are represented by red lines. To be able to visually distinguish between the numerous lines, we slightly offset these lines from each other on the x-axis. However, they represent the couplings of one black dot (corresponding to $\ket{\tilde i} \otimes \ket{0}$) to multiple blue dots at the same drive amplitude.  Contrary to the situation at low drive power, the transitions are numerous and of similar coupling strength, along with positive and negative detuning. The resulting energy shift on $\ket{\tilde i} \otimes \ket{0}$ is null on average (in the sense of random matrix model, see \cref{sec:standard_deviation}).

Strong hybridization between the resonator and the driven transmon can occur through the terms  $-i g n_{ij,k =-1,0}a^\dagger + h.c $, depending on the ratio between the coupling strength $g_\textrm{eff} = 2g n_{ij,k =-1,0}\sqrt{n+1}$ and the effective detuning $\delta_\textrm{eff}$ of the transition $i \leftrightarrow j$ \cite{Blais2021}. Below, we propose a possible estimate of the minimal Fock state number $n$ for which this happens \cite{Blais2021}.
Recall from \cref{sec:standard_deviation} that for chaotic states, we have $\sqrt{\mathbb{E}(|n_{ij,k = 0}|^2)} \approx \sqrt{{N_\textrm{ch}/12}}$, where $N_{ch}$ is the number of chaotic states (typically 10 at the ionization point of the ground state for the parameters in \cref{sec:cQEDsim}). This yields an effective coupling strength $g_\textrm{eff} = 2g\sqrt{n+1}\sqrt{{N_\textrm{ch}/12}}$. Note that this is an upper bound, as the contributions from the matrix elements $n_{ij,k =-1}$ are usually smaller. Besides, the mean level-spacing of the driven transmon is $\omega_d/N_\textrm{ch}$. Accounting for the detuning $\Delta$ of the resonator with respect to the drive, the effective detuning between the set of states $\ket{\tilde i_1}\otimes\ket{n}$ and the states $\ket{\tilde i_2}\otimes\ket{n+1}$, can be as low as $\delta_\textrm{eff} = \min(\omega_d/N_\textrm{ch},|\omega_d/N_\textrm{ch} - |\Delta||)$. Note that due to the repulsive statistics, the energy levels do not bunch, and the resulting variance of the effective detuning is rather small. On average, one has $\delta_\textrm{eff} = \omega_d/N_\textrm{ch}$.

The condition $g_\textrm{eff} \sim \delta_\textrm{eff}$ reads $2 g\sqrt{n+1}\sqrt{{N_\textrm{ch}/12}} \sim \omega_d/N_\textrm{ch}$, leading to the definition of a critical photon number, 
\begin{align*}
    n_\textrm{crit}^\textrm{ch} \sim \frac{3\omega_d^2}{g^2 N_\textrm{ch}^3},
\end{align*}
above which strong transmon-resonator hybridization occurs.
Note that this relation is valid for chaotic transmon states.
This criterion is sensitive to the fluctuations of the energy levels and of the charge matrix elements as a function of the system parameters, and therefore can only give a rough estimate.  The strong hybridization between the systems occurs in the steady state if this condition is satisfied for the lowest Fock states as well. 

\begin{figure}[b!]
    \centering\includegraphics{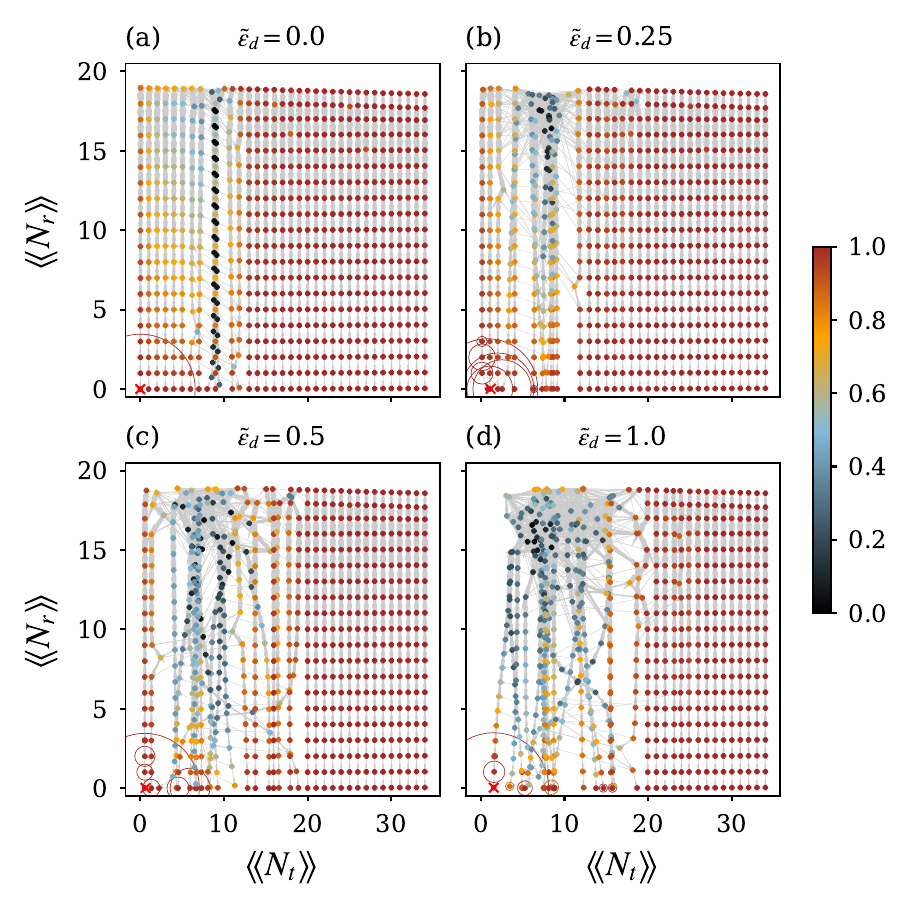}
    \caption{Same as \cref{fig:purities_rates_grid}, for $g/2\pi$ reduced by a factor $\sqrt{6}$.} 
    \label{fig:purities_rates_grid_lowg}
\end{figure}

As an example, for the parameters of \cref{sec:cQEDsim}, at a drive amplitude such that $N_\textrm{ch} \sim 12$ ($N_\textrm{ch}$ can be estimated by how many grid points of the first row come together in \cref{fig:purities_rates_grid}), we find $g_\textrm{eff}/2\pi \sim 0.5~\textrm{GHz}$ and $\delta_\textrm{eff}/2\pi \sim 0.6~\textrm{GHz}$, leading to $n_\textrm{crit}^\textrm{ch} \sim 0.5 $. A large drop in purity is expected to occur even for the lowest Fock states (see \cref{fig:purities_rates_grid}(d)).  As a second example, we consider the parameters of \cref{sec:cQEDsim} system with $g = 0.250/\sqrt{6}~\textrm{GHz}$ instead of $g = 0.250~\textrm{GHz}$ in \cref{fig:purities_rates_grid_lowg}. This reduction of the coupling $g$ makes the value of $n_{\textrm{crit}}^{\textrm{ch}}$ an order of magnitude higher. 

\section{Effect of the resonator dissipation on the spectrum}
\label{sec:dissipation}

\begin{figure}
    \centering
    \includegraphics[width=\linewidth]{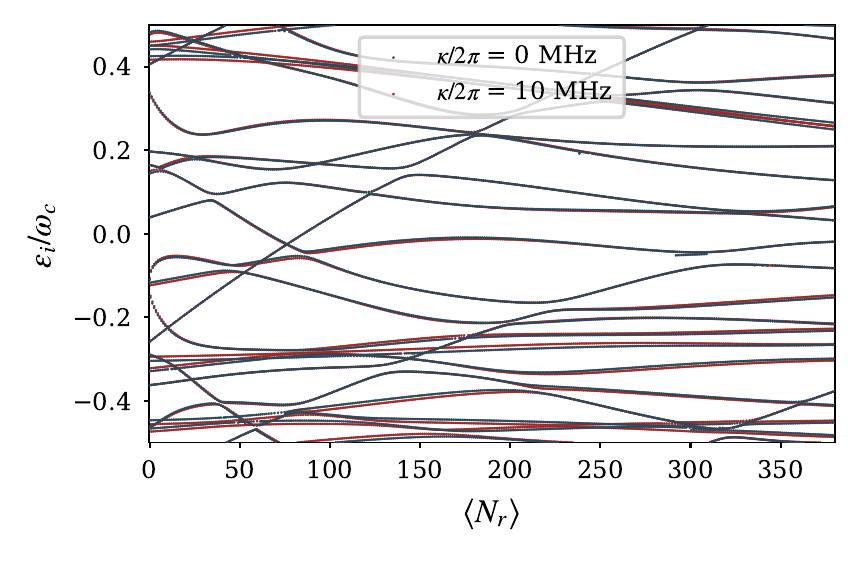}
    \caption{Eigenenergies of the undriven system described by the Hamiltonian in \cref{eq:cqed_model_undriven}, with the same parameters as in \cref{fig:purities_rates_grid}(a).  For each eigenstate, the energy $\varepsilon_i$ and its mean photon number $\langle N_r \rangle_i$ is plotted in the $(\varepsilon_i, \langle N_r \rangle_i)$ plane. The energies $\varepsilon_i$ are the eigenenergies modulo the resonator frequency. Only states satisfying $\langle N_t \rangle < 20$ are shown. Blue dots correspond to $\kappa/2\pi = 0$, and red dots to $\kappa/2\pi = 10 \textrm{MHz}$ (see \cref{sec:dissipation}). In both cases, the spectrum exhibits correlation at large $\langle N_r \rangle$. The spectrum is unchanged for this range of Fock states under weak dissipation. }
    \label{fig:static_dissipation}
\end{figure}

In \cref{sec:cQEDsim}, we have not accounted for the effect of the resonator dissipation on the spectrum. In the limit of very large $\kappa$, the two systems decouple and chaos disappears. However, for a given Fock state $\ket{n}$ of the resonator, the effect of dissipation on the system (not on resonance, \ie $\omega_d \neq \omega_a$) should be small as long as the decay rate of $n\kappa$ is small compared to the effective coupling strength $g\sqrt{n}$. This leads to the condition $\kappa \ll g\sqrt{n}$. For typical experimental parameters, $\kappa/2\pi = 10 \textrm{ MHz}$ and $g/2\pi = 250 \textrm{ MHz}$, dissipation will affect  Fock states with $ n \gtrsim (g/\kappa)^2 = 625$. As a numerical check, we diagonalize the non-hermitian undriven Hamiltonian
\begin{align}
    \begin{split}
        \label{eq:cqed_model_kappa}
            \hhh & =   4 E_C (\nnn-n_g)^2-E_J\cos(\ppphi) \\
            & + \big(\omega_a-i\frac{\kappa}{2}\big) \aaa^\dagger \aaa - i g \nnn (\aaa-\aaa^\dagger).
    \end{split}    
\end{align}
In \cref{fig:static_dissipation}, the real part of the energy $\varepsilon_i$ and the photon number $\langle N_r \rangle_i$ of each state are plotted in the $(\varepsilon_i, \langle N_r \rangle_i)$ plane for $\kappa = 0$ (blue dots) and $\kappa/2\pi = 10 \textrm{ MHz}$ (red dots). Qualitatively, the spectrum is the same in both cases. In particular, anti-crossings, which are signatures of resonances of the coupled system, do not seem to be attenuated.

\bibliographystyle{apsrev4-1}
\bibliography{references,mybib}

\begin{thebibliography}{63}%
\makeatletter
\providecommand \@ifxundefined [1]{%
 \@ifx{#1\undefined}
}%
\providecommand \@ifnum [1]{%
 \ifnum #1\expandafter \@firstoftwo
 \else \expandafter \@secondoftwo
 \fi
}%
\providecommand \@ifx [1]{%
 \ifx #1\expandafter \@firstoftwo
 \else \expandafter \@secondoftwo
 \fi
}%
\providecommand \natexlab [1]{#1}%
\providecommand \enquote  [1]{``#1''}%
\providecommand \bibnamefont  [1]{#1}%
\providecommand \bibfnamefont [1]{#1}%
\providecommand \citenamefont [1]{#1}%
\providecommand \href@noop [0]{\@secondoftwo}%
\providecommand \href [0]{\begingroup \@sanitize@url \@href}%
\providecommand \@href[1]{\@@startlink{#1}\@@href}%
\providecommand \@@href[1]{\endgroup#1\@@endlink}%
\providecommand \@sanitize@url [0]{\catcode `\\12\catcode `\$12\catcode
  `\&12\catcode `\#12\catcode `\^12\catcode `\_12\catcode `\%12\relax}%
\providecommand \@@startlink[1]{}%
\providecommand \@@endlink[0]{}%
\providecommand \url  [0]{\begingroup\@sanitize@url \@url }%
\providecommand \@url [1]{\endgroup\@href {#1}{\urlprefix }}%
\providecommand \urlprefix  [0]{URL }%
\providecommand \Eprint [0]{\href }%
\providecommand \doibase [0]{http://dx.doi.org/}%
\providecommand \selectlanguage [0]{\@gobble}%
\providecommand \bibinfo  [0]{\@secondoftwo}%
\providecommand \bibfield  [0]{\@secondoftwo}%
\providecommand \translation [1]{[#1]}%
\providecommand \BibitemOpen [0]{}%
\providecommand \bibitemStop [0]{}%
\providecommand \bibitemNoStop [0]{.\EOS\space}%
\providecommand \EOS [0]{\spacefactor3000\relax}%
\providecommand \BibitemShut  [1]{\csname bibitem#1\endcsname}%
\let\auto@bib@innerbib\@empty
\bibitem [{\citenamefont {Koch}\ \emph {et~al.}(2007)\citenamefont {Koch},
  \citenamefont {Yu}, \citenamefont {Gambetta}, \citenamefont {Houck},
  \citenamefont {Schuster}, \citenamefont {Majer}, \citenamefont {Blais},
  \citenamefont {Devoret}, \citenamefont {Girvin},\ and\ \citenamefont
  {Schoelkopf}}]{koch_charge-insensitive_2007}%
  \BibitemOpen
  \bibfield  {author} {\bibinfo {author} {\bibfnamefont {J.}~\bibnamefont
  {Koch}}, \bibinfo {author} {\bibfnamefont {T.~M.}\ \bibnamefont {Yu}},
  \bibinfo {author} {\bibfnamefont {J.}~\bibnamefont {Gambetta}}, \bibinfo
  {author} {\bibfnamefont {A.~A.}\ \bibnamefont {Houck}}, \bibinfo {author}
  {\bibfnamefont {D.~I.}\ \bibnamefont {Schuster}}, \bibinfo {author}
  {\bibfnamefont {J.}~\bibnamefont {Majer}}, \bibinfo {author} {\bibfnamefont
  {A.}~\bibnamefont {Blais}}, \bibinfo {author} {\bibfnamefont {M.~H.}\
  \bibnamefont {Devoret}}, \bibinfo {author} {\bibfnamefont {S.~M.}\
  \bibnamefont {Girvin}}, \ and\ \bibinfo {author} {\bibfnamefont {R.~J.}\
  \bibnamefont {Schoelkopf}},\ }\href {\doibase 10.1103/PhysRevA.76.042319}
  {\bibfield  {journal} {\bibinfo  {journal} {Physical Review A}\ }\textbf
  {\bibinfo {volume} {76}},\ \bibinfo {pages} {042319} (\bibinfo {year}
  {2007})}\BibitemShut {NoStop}%
\bibitem [{\citenamefont {Earnest}\ \emph {et~al.}(2018)\citenamefont
  {Earnest}, \citenamefont {Chakram}, \citenamefont {Lu}, \citenamefont
  {Irons}, \citenamefont {Naik}, \citenamefont {Leung}, \citenamefont {Ocola},
  \citenamefont {Czaplewski}, \citenamefont {Baker}, \citenamefont {Lawrence},
  \citenamefont {Koch},\ and\ \citenamefont
  {Schuster}}]{earnest_realization_2018}%
  \BibitemOpen
  \bibfield  {author} {\bibinfo {author} {\bibfnamefont {N.}~\bibnamefont
  {Earnest}}, \bibinfo {author} {\bibfnamefont {S.}~\bibnamefont {Chakram}},
  \bibinfo {author} {\bibfnamefont {Y.}~\bibnamefont {Lu}}, \bibinfo {author}
  {\bibfnamefont {N.}~\bibnamefont {Irons}}, \bibinfo {author} {\bibfnamefont
  {R.}~\bibnamefont {Naik}}, \bibinfo {author} {\bibfnamefont {N.}~\bibnamefont
  {Leung}}, \bibinfo {author} {\bibfnamefont {L.}~\bibnamefont {Ocola}},
  \bibinfo {author} {\bibfnamefont {D.}~\bibnamefont {Czaplewski}}, \bibinfo
  {author} {\bibfnamefont {B.}~\bibnamefont {Baker}}, \bibinfo {author}
  {\bibfnamefont {J.}~\bibnamefont {Lawrence}}, \bibinfo {author}
  {\bibfnamefont {J.}~\bibnamefont {Koch}}, \ and\ \bibinfo {author}
  {\bibfnamefont {D.}~\bibnamefont {Schuster}},\ }\href {\doibase
  10.1103/PhysRevLett.120.150504} {\bibfield  {journal} {\bibinfo  {journal}
  {Physical Review Letters}\ }\textbf {\bibinfo {volume} {120}},\ \bibinfo
  {pages} {150504} (\bibinfo {year} {2018})}\BibitemShut {NoStop}%
\bibitem [{\citenamefont {Lin}\ \emph {et~al.}(2018)\citenamefont {Lin},
  \citenamefont {Nguyen}, \citenamefont {Grabon}, \citenamefont {San~Miguel},
  \citenamefont {Pankratova},\ and\ \citenamefont
  {Manucharyan}}]{lin_demonstration_2018}%
  \BibitemOpen
  \bibfield  {author} {\bibinfo {author} {\bibfnamefont {Y.-H.}\ \bibnamefont
  {Lin}}, \bibinfo {author} {\bibfnamefont {L.~B.}\ \bibnamefont {Nguyen}},
  \bibinfo {author} {\bibfnamefont {N.}~\bibnamefont {Grabon}}, \bibinfo
  {author} {\bibfnamefont {J.}~\bibnamefont {San~Miguel}}, \bibinfo {author}
  {\bibfnamefont {N.}~\bibnamefont {Pankratova}}, \ and\ \bibinfo {author}
  {\bibfnamefont {V.~E.}\ \bibnamefont {Manucharyan}},\ }\href {\doibase
  10.1103/PhysRevLett.120.150503} {\bibfield  {journal} {\bibinfo  {journal}
  {Physical Review Letters}\ }\textbf {\bibinfo {volume} {120}},\ \bibinfo
  {pages} {150503} (\bibinfo {year} {2018})}\BibitemShut {NoStop}%
\bibitem [{\citenamefont {Gyenis}\ \emph {et~al.}(2021)\citenamefont {Gyenis},
  \citenamefont {Mundada}, \citenamefont {Di~Paolo}, \citenamefont {Hazard},
  \citenamefont {You}, \citenamefont {Schuster}, \citenamefont {Koch},
  \citenamefont {Blais},\ and\ \citenamefont
  {Houck}}]{gyenis_experimental_2021}%
  \BibitemOpen
  \bibfield  {author} {\bibinfo {author} {\bibfnamefont {A.}~\bibnamefont
  {Gyenis}}, \bibinfo {author} {\bibfnamefont {P.~S.}\ \bibnamefont {Mundada}},
  \bibinfo {author} {\bibfnamefont {A.}~\bibnamefont {Di~Paolo}}, \bibinfo
  {author} {\bibfnamefont {T.~M.}\ \bibnamefont {Hazard}}, \bibinfo {author}
  {\bibfnamefont {X.}~\bibnamefont {You}}, \bibinfo {author} {\bibfnamefont
  {D.~I.}\ \bibnamefont {Schuster}}, \bibinfo {author} {\bibfnamefont
  {J.}~\bibnamefont {Koch}}, \bibinfo {author} {\bibfnamefont {A.}~\bibnamefont
  {Blais}}, \ and\ \bibinfo {author} {\bibfnamefont {A.~A.}\ \bibnamefont
  {Houck}},\ }\href {\doibase 10.1103/PRXQuantum.2.010339} {\bibfield
  {journal} {\bibinfo  {journal} {PRX Quantum}\ }\textbf {\bibinfo {volume}
  {2}},\ \bibinfo {pages} {010339} (\bibinfo {year} {2021})}\BibitemShut
  {NoStop}%
\bibitem [{\citenamefont {You}\ \emph {et~al.}(2007)\citenamefont {You},
  \citenamefont {Hu}, \citenamefont {Ashhab},\ and\ \citenamefont
  {Nori}}]{You2007}%
  \BibitemOpen
  \bibfield  {author} {\bibinfo {author} {\bibfnamefont {J.~Q.}\ \bibnamefont
  {You}}, \bibinfo {author} {\bibfnamefont {X.}~\bibnamefont {Hu}}, \bibinfo
  {author} {\bibfnamefont {S.}~\bibnamefont {Ashhab}}, \ and\ \bibinfo {author}
  {\bibfnamefont {F.}~\bibnamefont {Nori}},\ }\href {\doibase
  10.1103/PhysRevB.75.140515} {\bibfield  {journal} {\bibinfo  {journal} {Phys.
  Rev. B}\ }\textbf {\bibinfo {volume} {75}},\ \bibinfo {pages} {140515}
  (\bibinfo {year} {2007})}\BibitemShut {NoStop}%
\bibitem [{\citenamefont {Yan}\ \emph {et~al.}(2016)\citenamefont {Yan},
  \citenamefont {Gustavsson}, \citenamefont {Kamal}, \citenamefont {Birenbaum},
  \citenamefont {Sears}, \citenamefont {Hover}, \citenamefont {Gudmundsen},
  \citenamefont {Rosenberg}, \citenamefont {Samach}, \citenamefont {Weber},
  \citenamefont {Yoder}, \citenamefont {Orlando}, \citenamefont {Clarke},
  \citenamefont {Kerman},\ and\ \citenamefont {Oliver}}]{yan_flux_2016}%
  \BibitemOpen
  \bibfield  {author} {\bibinfo {author} {\bibfnamefont {F.}~\bibnamefont
  {Yan}}, \bibinfo {author} {\bibfnamefont {S.}~\bibnamefont {Gustavsson}},
  \bibinfo {author} {\bibfnamefont {A.}~\bibnamefont {Kamal}}, \bibinfo
  {author} {\bibfnamefont {J.}~\bibnamefont {Birenbaum}}, \bibinfo {author}
  {\bibfnamefont {A.~P.}\ \bibnamefont {Sears}}, \bibinfo {author}
  {\bibfnamefont {D.}~\bibnamefont {Hover}}, \bibinfo {author} {\bibfnamefont
  {T.~J.}\ \bibnamefont {Gudmundsen}}, \bibinfo {author} {\bibfnamefont
  {D.}~\bibnamefont {Rosenberg}}, \bibinfo {author} {\bibfnamefont
  {G.}~\bibnamefont {Samach}}, \bibinfo {author} {\bibfnamefont
  {S.}~\bibnamefont {Weber}}, \bibinfo {author} {\bibfnamefont {J.~L.}\
  \bibnamefont {Yoder}}, \bibinfo {author} {\bibfnamefont {T.~P.}\ \bibnamefont
  {Orlando}}, \bibinfo {author} {\bibfnamefont {J.}~\bibnamefont {Clarke}},
  \bibinfo {author} {\bibfnamefont {A.~J.}\ \bibnamefont {Kerman}}, \ and\
  \bibinfo {author} {\bibfnamefont {W.~D.}\ \bibnamefont {Oliver}},\ }\href
  {\doibase 10.1038/ncomms12964} {\bibfield  {journal} {\bibinfo  {journal}
  {Nature Communications}\ }\textbf {\bibinfo {volume} {7}},\ \bibinfo {pages}
  {12964} (\bibinfo {year} {2016})}\BibitemShut {NoStop}%
\bibitem [{\citenamefont {Mirrahimi}\ \emph {et~al.}(2014)\citenamefont
  {Mirrahimi}, \citenamefont {Leghtas}, \citenamefont {Albert}, \citenamefont
  {Touzard}, \citenamefont {Schoelkopf}, \citenamefont {Jiang},\ and\
  \citenamefont {Devoret}}]{Mirrahimi_2014}%
  \BibitemOpen
  \bibfield  {author} {\bibinfo {author} {\bibfnamefont {M.}~\bibnamefont
  {Mirrahimi}}, \bibinfo {author} {\bibfnamefont {Z.}~\bibnamefont {Leghtas}},
  \bibinfo {author} {\bibfnamefont {V.~V.}\ \bibnamefont {Albert}}, \bibinfo
  {author} {\bibfnamefont {S.}~\bibnamefont {Touzard}}, \bibinfo {author}
  {\bibfnamefont {R.~J.}\ \bibnamefont {Schoelkopf}}, \bibinfo {author}
  {\bibfnamefont {L.}~\bibnamefont {Jiang}}, \ and\ \bibinfo {author}
  {\bibfnamefont {M.~H.}\ \bibnamefont {Devoret}},\ }\href {\doibase
  10.1088/1367-2630/16/4/045014} {\bibfield  {journal} {\bibinfo  {journal}
  {New Journal of Physics}\ }\textbf {\bibinfo {volume} {16}},\ \bibinfo
  {pages} {045014} (\bibinfo {year} {2014})}\BibitemShut {NoStop}%
\bibitem [{\citenamefont {Leghtas}\ \emph {et~al.}(2015)\citenamefont
  {Leghtas}, \citenamefont {Touzard}, \citenamefont {Pop}, \citenamefont {Kou},
  \citenamefont {Vlastakis}, \citenamefont {Petrenko}, \citenamefont {Sliwa},
  \citenamefont {Narla}, \citenamefont {Shankar}, \citenamefont {Hatridge},
  \citenamefont {Reagor}, \citenamefont {Frunzio}, \citenamefont {Schoelkopf},
  \citenamefont {Mirrahimi},\ and\ \citenamefont
  {Devoret}}]{Leghtas_science_2015}%
  \BibitemOpen
  \bibfield  {author} {\bibinfo {author} {\bibfnamefont {Z.}~\bibnamefont
  {Leghtas}}, \bibinfo {author} {\bibfnamefont {S.}~\bibnamefont {Touzard}},
  \bibinfo {author} {\bibfnamefont {I.~M.}\ \bibnamefont {Pop}}, \bibinfo
  {author} {\bibfnamefont {A.}~\bibnamefont {Kou}}, \bibinfo {author}
  {\bibfnamefont {B.}~\bibnamefont {Vlastakis}}, \bibinfo {author}
  {\bibfnamefont {A.}~\bibnamefont {Petrenko}}, \bibinfo {author}
  {\bibfnamefont {K.~M.}\ \bibnamefont {Sliwa}}, \bibinfo {author}
  {\bibfnamefont {A.}~\bibnamefont {Narla}}, \bibinfo {author} {\bibfnamefont
  {S.}~\bibnamefont {Shankar}}, \bibinfo {author} {\bibfnamefont {M.~J.}\
  \bibnamefont {Hatridge}}, \bibinfo {author} {\bibfnamefont {M.}~\bibnamefont
  {Reagor}}, \bibinfo {author} {\bibfnamefont {L.}~\bibnamefont {Frunzio}},
  \bibinfo {author} {\bibfnamefont {R.~J.}\ \bibnamefont {Schoelkopf}},
  \bibinfo {author} {\bibfnamefont {M.}~\bibnamefont {Mirrahimi}}, \ and\
  \bibinfo {author} {\bibfnamefont {M.~H.}\ \bibnamefont {Devoret}},\ }\href
  {\doibase 10.1126/science.aaa2085} {\bibfield  {journal} {\bibinfo  {journal}
  {Science}\ }\textbf {\bibinfo {volume} {347}},\ \bibinfo {pages} {853}
  (\bibinfo {year} {2015})},\ \Eprint
  {http://arxiv.org/abs/https://www.science.org/doi/pdf/10.1126/science.aaa2085}
  {https://www.science.org/doi/pdf/10.1126/science.aaa2085} \BibitemShut
  {NoStop}%
\bibitem [{\citenamefont {Lescanne}\ \emph {et~al.}(2020)\citenamefont
  {Lescanne}, \citenamefont {Villiers}, \citenamefont {Peronnin}, \citenamefont
  {Sarlette}, \citenamefont {Delbecq}, \citenamefont {Huard}, \citenamefont
  {Kontos}, \citenamefont {Mirrahimi},\ and\ \citenamefont
  {Leghtas}}]{lescanne_exponential_2020}%
  \BibitemOpen
  \bibfield  {author} {\bibinfo {author} {\bibfnamefont {R.}~\bibnamefont
  {Lescanne}}, \bibinfo {author} {\bibfnamefont {M.}~\bibnamefont {Villiers}},
  \bibinfo {author} {\bibfnamefont {T.}~\bibnamefont {Peronnin}}, \bibinfo
  {author} {\bibfnamefont {A.}~\bibnamefont {Sarlette}}, \bibinfo {author}
  {\bibfnamefont {M.}~\bibnamefont {Delbecq}}, \bibinfo {author} {\bibfnamefont
  {B.}~\bibnamefont {Huard}}, \bibinfo {author} {\bibfnamefont
  {T.}~\bibnamefont {Kontos}}, \bibinfo {author} {\bibfnamefont
  {M.}~\bibnamefont {Mirrahimi}}, \ and\ \bibinfo {author} {\bibfnamefont
  {Z.}~\bibnamefont {Leghtas}},\ }\href {\doibase 10.1038/s41567-020-0824-x}
  {\bibfield  {journal} {\bibinfo  {journal} {Nature Physics}\ }\textbf
  {\bibinfo {volume} {16}},\ \bibinfo {pages} {509} (\bibinfo {year}
  {2020})}\BibitemShut {NoStop}%
\bibitem [{\citenamefont {Sivak}\ \emph {et~al.}(2019)\citenamefont {Sivak},
  \citenamefont {Frattini}, \citenamefont {Joshi}, \citenamefont
  {Lingenfelter}, \citenamefont {Shankar},\ and\ \citenamefont
  {Devoret}}]{sivak_kerr-free_2019}%
  \BibitemOpen
  \bibfield  {author} {\bibinfo {author} {\bibfnamefont {V.}~\bibnamefont
  {Sivak}}, \bibinfo {author} {\bibfnamefont {N.}~\bibnamefont {Frattini}},
  \bibinfo {author} {\bibfnamefont {V.}~\bibnamefont {Joshi}}, \bibinfo
  {author} {\bibfnamefont {A.}~\bibnamefont {Lingenfelter}}, \bibinfo {author}
  {\bibfnamefont {S.}~\bibnamefont {Shankar}}, \ and\ \bibinfo {author}
  {\bibfnamefont {M.}~\bibnamefont {Devoret}},\ }\href {\doibase
  10.1103/PhysRevApplied.11.054060} {\bibfield  {journal} {\bibinfo  {journal}
  {Physical Review Applied}\ }\textbf {\bibinfo {volume} {11}},\ \bibinfo
  {pages} {054060} (\bibinfo {year} {2019})}\BibitemShut {NoStop}%
\bibitem [{\citenamefont {Masluk}\ \emph {et~al.}(2012)\citenamefont {Masluk},
  \citenamefont {Pop}, \citenamefont {Kamal}, \citenamefont {Minev},\ and\
  \citenamefont {Devoret}}]{masluk_microwave_2012}%
  \BibitemOpen
  \bibfield  {author} {\bibinfo {author} {\bibfnamefont {N.~A.}\ \bibnamefont
  {Masluk}}, \bibinfo {author} {\bibfnamefont {I.~M.}\ \bibnamefont {Pop}},
  \bibinfo {author} {\bibfnamefont {A.}~\bibnamefont {Kamal}}, \bibinfo
  {author} {\bibfnamefont {Z.~K.}\ \bibnamefont {Minev}}, \ and\ \bibinfo
  {author} {\bibfnamefont {M.~H.}\ \bibnamefont {Devoret}},\ }\href {\doibase
  10.1103/PhysRevLett.109.137002} {\bibfield  {journal} {\bibinfo  {journal}
  {Physical Review Letters}\ }\textbf {\bibinfo {volume} {109}},\ \bibinfo
  {pages} {137002} (\bibinfo {year} {2012})}\BibitemShut {NoStop}%
\bibitem [{\citenamefont {Manucharyan}\ \emph {et~al.}(2012)\citenamefont
  {Manucharyan}, \citenamefont {Masluk}, \citenamefont {Kamal}, \citenamefont
  {Koch}, \citenamefont {Glazman},\ and\ \citenamefont
  {Devoret}}]{manucharyan_evidence_2012}%
  \BibitemOpen
  \bibfield  {author} {\bibinfo {author} {\bibfnamefont {V.~E.}\ \bibnamefont
  {Manucharyan}}, \bibinfo {author} {\bibfnamefont {N.~A.}\ \bibnamefont
  {Masluk}}, \bibinfo {author} {\bibfnamefont {A.}~\bibnamefont {Kamal}},
  \bibinfo {author} {\bibfnamefont {J.}~\bibnamefont {Koch}}, \bibinfo {author}
  {\bibfnamefont {L.~I.}\ \bibnamefont {Glazman}}, \ and\ \bibinfo {author}
  {\bibfnamefont {M.~H.}\ \bibnamefont {Devoret}},\ }\href {\doibase
  10.1103/PhysRevB.85.024521} {\bibfield  {journal} {\bibinfo  {journal}
  {Physical Review B}\ }\textbf {\bibinfo {volume} {85}},\ \bibinfo {pages}
  {024521} (\bibinfo {year} {2012})}\BibitemShut {NoStop}%
\bibitem [{\citenamefont {Lescanne}\ \emph {et~al.}(2019)\citenamefont
  {Lescanne}, \citenamefont {Verney}, \citenamefont {Ficheux}, \citenamefont
  {Devoret}, \citenamefont {Huard}, \citenamefont {Mirrahimi},\ and\
  \citenamefont {Leghtas}}]{Lescanne_2019}%
  \BibitemOpen
  \bibfield  {author} {\bibinfo {author} {\bibfnamefont {R.}~\bibnamefont
  {Lescanne}}, \bibinfo {author} {\bibfnamefont {L.}~\bibnamefont {Verney}},
  \bibinfo {author} {\bibfnamefont {Q.}~\bibnamefont {Ficheux}}, \bibinfo
  {author} {\bibfnamefont {M.~H.}\ \bibnamefont {Devoret}}, \bibinfo {author}
  {\bibfnamefont {B.}~\bibnamefont {Huard}}, \bibinfo {author} {\bibfnamefont
  {M.}~\bibnamefont {Mirrahimi}}, \ and\ \bibinfo {author} {\bibfnamefont
  {Z.}~\bibnamefont {Leghtas}},\ }\href
  {http://dx.doi.org/10.1103/PhysRevApplied.11.014030} {\bibfield  {journal}
  {\bibinfo  {journal} {Physical Review Applied}\ }\textbf {\bibinfo {volume}
  {11}},\ \bibinfo {pages} {014030} (\bibinfo {year} {2019})}\BibitemShut
  {NoStop}%
\bibitem [{\citenamefont {Sheldon}\ \emph {et~al.}(2016)\citenamefont
  {Sheldon}, \citenamefont {Magesan}, \citenamefont {Chow},\ and\ \citenamefont
  {Gambetta}}]{sheldon_procedure_2016}%
  \BibitemOpen
  \bibfield  {author} {\bibinfo {author} {\bibfnamefont {S.}~\bibnamefont
  {Sheldon}}, \bibinfo {author} {\bibfnamefont {E.}~\bibnamefont {Magesan}},
  \bibinfo {author} {\bibfnamefont {J.~M.}\ \bibnamefont {Chow}}, \ and\
  \bibinfo {author} {\bibfnamefont {J.~M.}\ \bibnamefont {Gambetta}},\ }\href
  {\doibase 10.1103/PhysRevA.93.060302} {\bibfield  {journal} {\bibinfo
  {journal} {Physical Review A}\ }\textbf {\bibinfo {volume} {93}},\ \bibinfo
  {pages} {060302} (\bibinfo {year} {2016})},\ \bibinfo {note} {arXiv:
  1603.04821}\BibitemShut {NoStop}%
\bibitem [{\citenamefont {Paik}\ \emph {et~al.}(2016)\citenamefont {Paik},
  \citenamefont {Mezzacapo}, \citenamefont {Sandberg}, \citenamefont {McClure},
  \citenamefont {Abdo}, \citenamefont {C\'orcoles}, \citenamefont {Dial},
  \citenamefont {Bogorin}, \citenamefont {Plourde}, \citenamefont {Steffen},
  \citenamefont {Cross}, \citenamefont {Gambetta},\ and\ \citenamefont
  {Chow}}]{Paik2016}%
  \BibitemOpen
  \bibfield  {author} {\bibinfo {author} {\bibfnamefont {H.}~\bibnamefont
  {Paik}}, \bibinfo {author} {\bibfnamefont {A.}~\bibnamefont {Mezzacapo}},
  \bibinfo {author} {\bibfnamefont {M.}~\bibnamefont {Sandberg}}, \bibinfo
  {author} {\bibfnamefont {D.~T.}\ \bibnamefont {McClure}}, \bibinfo {author}
  {\bibfnamefont {B.}~\bibnamefont {Abdo}}, \bibinfo {author} {\bibfnamefont
  {A.~D.}\ \bibnamefont {C\'orcoles}}, \bibinfo {author} {\bibfnamefont
  {O.}~\bibnamefont {Dial}}, \bibinfo {author} {\bibfnamefont {D.~F.}\
  \bibnamefont {Bogorin}}, \bibinfo {author} {\bibfnamefont {B.~L.~T.}\
  \bibnamefont {Plourde}}, \bibinfo {author} {\bibfnamefont {M.}~\bibnamefont
  {Steffen}}, \bibinfo {author} {\bibfnamefont {A.~W.}\ \bibnamefont {Cross}},
  \bibinfo {author} {\bibfnamefont {J.~M.}\ \bibnamefont {Gambetta}}, \ and\
  \bibinfo {author} {\bibfnamefont {J.~M.}\ \bibnamefont {Chow}},\ }\href
  {\doibase 10.1103/PhysRevLett.117.250502} {\bibfield  {journal} {\bibinfo
  {journal} {Phys. Rev. Lett.}\ }\textbf {\bibinfo {volume} {117}},\ \bibinfo
  {pages} {250502} (\bibinfo {year} {2016})}\BibitemShut {NoStop}%
\bibitem [{\citenamefont {Blais}\ \emph {et~al.}(2004)\citenamefont {Blais},
  \citenamefont {Huang}, \citenamefont {Wallraff}, \citenamefont {Girvin},\
  and\ \citenamefont {Schoelkopf}}]{Blais_2004}%
  \BibitemOpen
  \bibfield  {author} {\bibinfo {author} {\bibfnamefont {A.}~\bibnamefont
  {Blais}}, \bibinfo {author} {\bibfnamefont {R.-S.}\ \bibnamefont {Huang}},
  \bibinfo {author} {\bibfnamefont {A.}~\bibnamefont {Wallraff}}, \bibinfo
  {author} {\bibfnamefont {S.~M.}\ \bibnamefont {Girvin}}, \ and\ \bibinfo
  {author} {\bibfnamefont {R.~J.}\ \bibnamefont {Schoelkopf}},\ }\href
  {\doibase 10.1103/physreva.69.062320} {\bibfield  {journal} {\bibinfo
  {journal} {Physical Review A}\ }\textbf {\bibinfo {volume} {69}} (\bibinfo
  {year} {2004}),\ 10.1103/physreva.69.062320}\BibitemShut {NoStop}%
\bibitem [{\citenamefont {Walter}\ \emph
  {et~al.}(2017{\natexlab{a}})\citenamefont {Walter}, \citenamefont {Kurpiers},
  \citenamefont {Gasparinetti}, \citenamefont {Magnard}, \citenamefont
  {Potočnik}, \citenamefont {Salathé}, \citenamefont {Pechal}, \citenamefont
  {Mondal}, \citenamefont {Oppliger}, \citenamefont {Eichler},\ and\
  \citenamefont {Wallraff}}]{walter_rapid_2017}%
  \BibitemOpen
  \bibfield  {author} {\bibinfo {author} {\bibfnamefont {T.}~\bibnamefont
  {Walter}}, \bibinfo {author} {\bibfnamefont {P.}~\bibnamefont {Kurpiers}},
  \bibinfo {author} {\bibfnamefont {S.}~\bibnamefont {Gasparinetti}}, \bibinfo
  {author} {\bibfnamefont {P.}~\bibnamefont {Magnard}}, \bibinfo {author}
  {\bibfnamefont {A.}~\bibnamefont {Potočnik}}, \bibinfo {author}
  {\bibfnamefont {Y.}~\bibnamefont {Salathé}}, \bibinfo {author}
  {\bibfnamefont {M.}~\bibnamefont {Pechal}}, \bibinfo {author} {\bibfnamefont
  {M.}~\bibnamefont {Mondal}}, \bibinfo {author} {\bibfnamefont
  {M.}~\bibnamefont {Oppliger}}, \bibinfo {author} {\bibfnamefont
  {C.}~\bibnamefont {Eichler}}, \ and\ \bibinfo {author} {\bibfnamefont
  {A.}~\bibnamefont {Wallraff}},\ }\href {\doibase
  10.1103/PhysRevApplied.7.054020} {\bibfield  {journal} {\bibinfo  {journal}
  {Physical Review Applied}\ }\textbf {\bibinfo {volume} {7}},\ \bibinfo
  {pages} {054020} (\bibinfo {year} {2017}{\natexlab{a}})}\BibitemShut
  {NoStop}%
\bibitem [{\citenamefont {Minev}\ \emph {et~al.}(2019)\citenamefont {Minev},
  \citenamefont {Mundhada}, \citenamefont {Shankar}, \citenamefont {Reinhold},
  \citenamefont {Gutiérrez-Jáuregui}, \citenamefont {Schoelkopf},
  \citenamefont {Mirrahimi}, \citenamefont {Carmichael},\ and\ \citenamefont
  {Devoret}}]{minev_catch_2019}%
  \BibitemOpen
  \bibfield  {author} {\bibinfo {author} {\bibfnamefont {Z.}~\bibnamefont
  {Minev}}, \bibinfo {author} {\bibfnamefont {S.}~\bibnamefont {Mundhada}},
  \bibinfo {author} {\bibfnamefont {S.}~\bibnamefont {Shankar}}, \bibinfo
  {author} {\bibfnamefont {P.}~\bibnamefont {Reinhold}}, \bibinfo {author}
  {\bibfnamefont {R.}~\bibnamefont {Gutiérrez-Jáuregui}}, \bibinfo {author}
  {\bibfnamefont {R.}~\bibnamefont {Schoelkopf}}, \bibinfo {author}
  {\bibfnamefont {M.}~\bibnamefont {Mirrahimi}}, \bibinfo {author}
  {\bibfnamefont {H.}~\bibnamefont {Carmichael}}, \ and\ \bibinfo {author}
  {\bibfnamefont {M.}~\bibnamefont {Devoret}},\ }\href {\doibase
  10.1038/s41586-019-1287-z} {\bibfield  {journal} {\bibinfo  {journal}
  {Nature}\ }\textbf {\bibinfo {volume} {570}},\ \bibinfo {pages} {200}
  (\bibinfo {year} {2019})}\BibitemShut {NoStop}%
\bibitem [{\citenamefont {Grimm}\ \emph {et~al.}(2020)\citenamefont {Grimm},
  \citenamefont {Frattini}, \citenamefont {Puri}, \citenamefont {Mundhada},
  \citenamefont {Touzard}, \citenamefont {Mirrahimi}, \citenamefont {Girvin},
  \citenamefont {Shankar},\ and\ \citenamefont {Devoret}}]{Grimm2020}%
  \BibitemOpen
  \bibfield  {author} {\bibinfo {author} {\bibfnamefont {A.}~\bibnamefont
  {Grimm}}, \bibinfo {author} {\bibfnamefont {N.~E.}\ \bibnamefont {Frattini}},
  \bibinfo {author} {\bibfnamefont {S.}~\bibnamefont {Puri}}, \bibinfo {author}
  {\bibfnamefont {S.~O.}\ \bibnamefont {Mundhada}}, \bibinfo {author}
  {\bibfnamefont {S.}~\bibnamefont {Touzard}}, \bibinfo {author} {\bibfnamefont
  {M.}~\bibnamefont {Mirrahimi}}, \bibinfo {author} {\bibfnamefont {S.~M.}\
  \bibnamefont {Girvin}}, \bibinfo {author} {\bibfnamefont {S.}~\bibnamefont
  {Shankar}}, \ and\ \bibinfo {author} {\bibfnamefont {M.~H.}\ \bibnamefont
  {Devoret}},\ }\href@noop {} {\bibfield  {journal} {\bibinfo  {journal}
  {Nature}\ }\textbf {\bibinfo {volume} {584}},\ \bibinfo {pages} {205}
  (\bibinfo {year} {2020})}\BibitemShut {NoStop}%
\bibitem [{\citenamefont {Boissonneault}\ \emph {et~al.}(2009)\citenamefont
  {Boissonneault}, \citenamefont {Gambetta},\ and\ \citenamefont
  {Blais}}]{boissonneault_dispersive_2009}%
  \BibitemOpen
  \bibfield  {author} {\bibinfo {author} {\bibfnamefont {M.}~\bibnamefont
  {Boissonneault}}, \bibinfo {author} {\bibfnamefont {J.~M.}\ \bibnamefont
  {Gambetta}}, \ and\ \bibinfo {author} {\bibfnamefont {A.}~\bibnamefont
  {Blais}},\ }\href {\doibase 10.1103/PhysRevA.79.013819} {\bibfield  {journal}
  {\bibinfo  {journal} {Physical Review A}\ }\textbf {\bibinfo {volume} {79}},\
  \bibinfo {pages} {013819} (\bibinfo {year} {2009})}\BibitemShut {NoStop}%
\bibitem [{\citenamefont {Malekakhlagh}\ \emph
  {et~al.}(2020{\natexlab{a}})\citenamefont {Malekakhlagh}, \citenamefont
  {Petrescu},\ and\ \citenamefont {T\"ureci}}]{Malekakhlagh_2020}%
  \BibitemOpen
  \bibfield  {author} {\bibinfo {author} {\bibfnamefont {M.}~\bibnamefont
  {Malekakhlagh}}, \bibinfo {author} {\bibfnamefont {A.}~\bibnamefont
  {Petrescu}}, \ and\ \bibinfo {author} {\bibfnamefont {H.~E.}\ \bibnamefont
  {T\"ureci}},\ }\href {\doibase 10.1103/PhysRevB.101.134509} {\bibfield
  {journal} {\bibinfo  {journal} {Phys. Rev. B}\ }\textbf {\bibinfo {volume}
  {101}},\ \bibinfo {pages} {134509} (\bibinfo {year}
  {2020}{\natexlab{a}})}\BibitemShut {NoStop}%
\bibitem [{\citenamefont {Petrescu}\ \emph {et~al.}(2020)\citenamefont
  {Petrescu}, \citenamefont {Malekakhlagh},\ and\ \citenamefont
  {T\"ureci}}]{Petrescu_2020}%
  \BibitemOpen
  \bibfield  {author} {\bibinfo {author} {\bibfnamefont {A.}~\bibnamefont
  {Petrescu}}, \bibinfo {author} {\bibfnamefont {M.}~\bibnamefont
  {Malekakhlagh}}, \ and\ \bibinfo {author} {\bibfnamefont {H.~E.}\
  \bibnamefont {T\"ureci}},\ }\href {\doibase 10.1103/PhysRevB.101.134510}
  {\bibfield  {journal} {\bibinfo  {journal} {Phys. Rev. B}\ }\textbf {\bibinfo
  {volume} {101}},\ \bibinfo {pages} {134510} (\bibinfo {year}
  {2020})}\BibitemShut {NoStop}%
\bibitem [{\citenamefont {Sank}\ \emph {et~al.}(2016)\citenamefont {Sank},
  \citenamefont {Chen}, \citenamefont {Khezri}, \citenamefont {Kelly},
  \citenamefont {Barends}, \citenamefont {Campbell}, \citenamefont {Chen},
  \citenamefont {Chiaro}, \citenamefont {Dunsworth}, \citenamefont {Fowler},
  \citenamefont {Jeffrey}, \citenamefont {Lucero}, \citenamefont {Megrant},
  \citenamefont {Mutus}, \citenamefont {Neeley}, \citenamefont {Neill},
  \citenamefont {O’Malley}, \citenamefont {Quintana}, \citenamefont
  {Roushan}, \citenamefont {Vainsencher}, \citenamefont {White}, \citenamefont
  {Wenner}, \citenamefont {Korotkov},\ and\ \citenamefont
  {Martinis}}]{sank_measurement-induced_2016}%
  \BibitemOpen
  \bibfield  {author} {\bibinfo {author} {\bibfnamefont {D.}~\bibnamefont
  {Sank}}, \bibinfo {author} {\bibfnamefont {Z.}~\bibnamefont {Chen}}, \bibinfo
  {author} {\bibfnamefont {M.}~\bibnamefont {Khezri}}, \bibinfo {author}
  {\bibfnamefont {J.}~\bibnamefont {Kelly}}, \bibinfo {author} {\bibfnamefont
  {R.}~\bibnamefont {Barends}}, \bibinfo {author} {\bibfnamefont
  {B.}~\bibnamefont {Campbell}}, \bibinfo {author} {\bibfnamefont
  {Y.}~\bibnamefont {Chen}}, \bibinfo {author} {\bibfnamefont {B.}~\bibnamefont
  {Chiaro}}, \bibinfo {author} {\bibfnamefont {A.}~\bibnamefont {Dunsworth}},
  \bibinfo {author} {\bibfnamefont {A.}~\bibnamefont {Fowler}}, \bibinfo
  {author} {\bibfnamefont {E.}~\bibnamefont {Jeffrey}}, \bibinfo {author}
  {\bibfnamefont {E.}~\bibnamefont {Lucero}}, \bibinfo {author} {\bibfnamefont
  {A.}~\bibnamefont {Megrant}}, \bibinfo {author} {\bibfnamefont
  {J.}~\bibnamefont {Mutus}}, \bibinfo {author} {\bibfnamefont
  {M.}~\bibnamefont {Neeley}}, \bibinfo {author} {\bibfnamefont
  {C.}~\bibnamefont {Neill}}, \bibinfo {author} {\bibfnamefont
  {P.}~\bibnamefont {O’Malley}}, \bibinfo {author} {\bibfnamefont
  {C.}~\bibnamefont {Quintana}}, \bibinfo {author} {\bibfnamefont
  {P.}~\bibnamefont {Roushan}}, \bibinfo {author} {\bibfnamefont
  {A.}~\bibnamefont {Vainsencher}}, \bibinfo {author} {\bibfnamefont
  {T.}~\bibnamefont {White}}, \bibinfo {author} {\bibfnamefont
  {J.}~\bibnamefont {Wenner}}, \bibinfo {author} {\bibfnamefont {A.~N.}\
  \bibnamefont {Korotkov}}, \ and\ \bibinfo {author} {\bibfnamefont {J.~M.}\
  \bibnamefont {Martinis}},\ }\href {\doibase 10.1103/PhysRevLett.117.190503}
  {\bibfield  {journal} {\bibinfo  {journal} {Physical Review Letters}\
  }\textbf {\bibinfo {volume} {117}},\ \bibinfo {pages} {190503} (\bibinfo
  {year} {2016})}\BibitemShut {NoStop}%
\bibitem [{\citenamefont {Verney}\ \emph {et~al.}(2019)\citenamefont {Verney},
  \citenamefont {Lescanne}, \citenamefont {Devoret}, \citenamefont {Leghtas},\
  and\ \citenamefont {Mirrahimi}}]{VerneyPRApplied2019}%
  \BibitemOpen
  \bibfield  {author} {\bibinfo {author} {\bibfnamefont {L.}~\bibnamefont
  {Verney}}, \bibinfo {author} {\bibfnamefont {R.}~\bibnamefont {Lescanne}},
  \bibinfo {author} {\bibfnamefont {M.~H.}\ \bibnamefont {Devoret}}, \bibinfo
  {author} {\bibfnamefont {Z.}~\bibnamefont {Leghtas}}, \ and\ \bibinfo
  {author} {\bibfnamefont {M.}~\bibnamefont {Mirrahimi}},\ }\href {\doibase
  10.1103/PhysRevApplied.11.024003} {\bibfield  {journal} {\bibinfo  {journal}
  {Phys. Rev. Applied}\ }\textbf {\bibinfo {volume} {11}},\ \bibinfo {pages}
  {024003} (\bibinfo {year} {2019})}\BibitemShut {NoStop}%
\bibitem [{\citenamefont {Shillito}\ \emph {et~al.}(2022)\citenamefont
  {Shillito}, \citenamefont {Petrescu}, \citenamefont {Cohen}, \citenamefont
  {Beall}, \citenamefont {Hauru}, \citenamefont {Ganahl}, \citenamefont
  {Lewis}, \citenamefont {Vidal},\ and\ \citenamefont {Blais}}]{Shillito2022}%
  \BibitemOpen
  \bibfield  {author} {\bibinfo {author} {\bibfnamefont {R.}~\bibnamefont
  {Shillito}}, \bibinfo {author} {\bibfnamefont {A.}~\bibnamefont {Petrescu}},
  \bibinfo {author} {\bibfnamefont {J.}~\bibnamefont {Cohen}}, \bibinfo
  {author} {\bibfnamefont {J.}~\bibnamefont {Beall}}, \bibinfo {author}
  {\bibfnamefont {M.}~\bibnamefont {Hauru}}, \bibinfo {author} {\bibfnamefont
  {M.}~\bibnamefont {Ganahl}}, \bibinfo {author} {\bibfnamefont {A.~G.~M.}\
  \bibnamefont {Lewis}}, \bibinfo {author} {\bibfnamefont {G.}~\bibnamefont
  {Vidal}}, \ and\ \bibinfo {author} {\bibfnamefont {A.}~\bibnamefont
  {Blais}},\ }\href {\doibase 10.48550/ARXIV.2203.11235} {\enquote {\bibinfo
  {title} {Dynamics of transmon ionization},}\ } (\bibinfo {year}
  {2022})\BibitemShut {NoStop}%
\bibitem [{\citenamefont {Walter}\ \emph
  {et~al.}(2017{\natexlab{b}})\citenamefont {Walter}, \citenamefont {Kurpiers},
  \citenamefont {Gasparinetti}, \citenamefont {Magnard}, \citenamefont
  {Poto\ifmmode~\check{c}\else \v{c}\fi{}nik}, \citenamefont {Salath\'e},
  \citenamefont {Pechal}, \citenamefont {Mondal}, \citenamefont {Oppliger},
  \citenamefont {Eichler},\ and\ \citenamefont {Wallraff}}]{WalterPRL2017}%
  \BibitemOpen
  \bibfield  {author} {\bibinfo {author} {\bibfnamefont {T.}~\bibnamefont
  {Walter}}, \bibinfo {author} {\bibfnamefont {P.}~\bibnamefont {Kurpiers}},
  \bibinfo {author} {\bibfnamefont {S.}~\bibnamefont {Gasparinetti}}, \bibinfo
  {author} {\bibfnamefont {P.}~\bibnamefont {Magnard}}, \bibinfo {author}
  {\bibfnamefont {A.}~\bibnamefont {Poto\ifmmode~\check{c}\else
  \v{c}\fi{}nik}}, \bibinfo {author} {\bibfnamefont {Y.}~\bibnamefont
  {Salath\'e}}, \bibinfo {author} {\bibfnamefont {M.}~\bibnamefont {Pechal}},
  \bibinfo {author} {\bibfnamefont {M.}~\bibnamefont {Mondal}}, \bibinfo
  {author} {\bibfnamefont {M.}~\bibnamefont {Oppliger}}, \bibinfo {author}
  {\bibfnamefont {C.}~\bibnamefont {Eichler}}, \ and\ \bibinfo {author}
  {\bibfnamefont {A.}~\bibnamefont {Wallraff}},\ }\href {\doibase
  10.1103/PhysRevApplied.7.054020} {\bibfield  {journal} {\bibinfo  {journal}
  {Phys. Rev. Applied}\ }\textbf {\bibinfo {volume} {7}},\ \bibinfo {pages}
  {054020} (\bibinfo {year} {2017}{\natexlab{b}})}\BibitemShut {NoStop}%
\bibitem [{\citenamefont {Tomsovic}\ and\ \citenamefont
  {Ullmo}(1994)}]{tomsovic_chaos-assisted_1994}%
  \BibitemOpen
  \bibfield  {author} {\bibinfo {author} {\bibfnamefont {S.}~\bibnamefont
  {Tomsovic}}\ and\ \bibinfo {author} {\bibfnamefont {D.}~\bibnamefont
  {Ullmo}},\ }\href {\doibase 10.1103/PhysRevE.50.145} {\bibfield  {journal}
  {\bibinfo  {journal} {Physical Review E}\ }\textbf {\bibinfo {volume} {50}},\
  \bibinfo {pages} {145} (\bibinfo {year} {1994})}\BibitemShut {NoStop}%
\bibitem [{\citenamefont {Matveev}\ \emph {et~al.}(2002)\citenamefont
  {Matveev}, \citenamefont {Larkin},\ and\ \citenamefont
  {Glazman}}]{matveev_et_al_2002}%
  \BibitemOpen
  \bibfield  {author} {\bibinfo {author} {\bibfnamefont {K.~A.}\ \bibnamefont
  {Matveev}}, \bibinfo {author} {\bibfnamefont {A.~I.}\ \bibnamefont {Larkin}},
  \ and\ \bibinfo {author} {\bibfnamefont {L.~I.}\ \bibnamefont {Glazman}},\
  }\href {\doibase 10.1103/PhysRevLett.89.096802} {\bibfield  {journal}
  {\bibinfo  {journal} {Phys. Rev. Lett.}\ }\textbf {\bibinfo {volume} {89}},\
  \bibinfo {pages} {096802} (\bibinfo {year} {2002})}\BibitemShut {NoStop}%
\bibitem [{\citenamefont {Nigg}\ \emph {et~al.}(2012)\citenamefont {Nigg},
  \citenamefont {Paik}, \citenamefont {Vlastakis}, \citenamefont {Kirchmair},
  \citenamefont {Shankar}, \citenamefont {Frunzio}, \citenamefont {Devoret},
  \citenamefont {Schoelkopf},\ and\ \citenamefont {Girvin}}]{Nigg_2012}%
  \BibitemOpen
  \bibfield  {author} {\bibinfo {author} {\bibfnamefont {S.~E.}\ \bibnamefont
  {Nigg}}, \bibinfo {author} {\bibfnamefont {H.}~\bibnamefont {Paik}}, \bibinfo
  {author} {\bibfnamefont {B.}~\bibnamefont {Vlastakis}}, \bibinfo {author}
  {\bibfnamefont {G.}~\bibnamefont {Kirchmair}}, \bibinfo {author}
  {\bibfnamefont {S.}~\bibnamefont {Shankar}}, \bibinfo {author} {\bibfnamefont
  {L.}~\bibnamefont {Frunzio}}, \bibinfo {author} {\bibfnamefont {M.~H.}\
  \bibnamefont {Devoret}}, \bibinfo {author} {\bibfnamefont {R.~J.}\
  \bibnamefont {Schoelkopf}}, \ and\ \bibinfo {author} {\bibfnamefont {S.~M.}\
  \bibnamefont {Girvin}},\ }\href {\doibase 10.1103/physrevlett.108.240502}
  {\bibfield  {journal} {\bibinfo  {journal} {Physical Review Letters}\
  }\textbf {\bibinfo {volume} {108}} (\bibinfo {year} {2012}),\
  10.1103/physrevlett.108.240502}\BibitemShut {NoStop}%
\bibitem [{\citenamefont {Breuer}\ \emph {et~al.}(2000)\citenamefont {Breuer},
  \citenamefont {Huber},\ and\ \citenamefont
  {Petruccione}}]{breuer_quasistationary_2000}%
  \BibitemOpen
  \bibfield  {author} {\bibinfo {author} {\bibfnamefont {H.-P.}\ \bibnamefont
  {Breuer}}, \bibinfo {author} {\bibfnamefont {W.}~\bibnamefont {Huber}}, \
  and\ \bibinfo {author} {\bibfnamefont {F.}~\bibnamefont {Petruccione}},\
  }\href {\doibase 10.1103/PhysRevE.61.4883} {\bibfield  {journal} {\bibinfo
  {journal} {Physical Review E}\ }\textbf {\bibinfo {volume} {61}},\ \bibinfo
  {pages} {4883} (\bibinfo {year} {2000})}\BibitemShut {NoStop}%
\bibitem [{\citenamefont {Ketzmerick}\ and\ \citenamefont
  {Wustmann}(2010{\natexlab{a}})}]{ketzmerick_statistical_2010}%
  \BibitemOpen
  \bibfield  {author} {\bibinfo {author} {\bibfnamefont {R.}~\bibnamefont
  {Ketzmerick}}\ and\ \bibinfo {author} {\bibfnamefont {W.}~\bibnamefont
  {Wustmann}},\ }\href {\doibase 10.1103/PhysRevE.82.021114} {\bibfield
  {journal} {\bibinfo  {journal} {Physical Review E}\ }\textbf {\bibinfo
  {volume} {82}},\ \bibinfo {pages} {021114} (\bibinfo {year}
  {2010}{\natexlab{a}})}\BibitemShut {NoStop}%
\bibitem [{\citenamefont {Grifoni}\ and\ \citenamefont
  {Hänggi}(1998)}]{GRIFONI1998229}%
  \BibitemOpen
  \bibfield  {author} {\bibinfo {author} {\bibfnamefont {M.}~\bibnamefont
  {Grifoni}}\ and\ \bibinfo {author} {\bibfnamefont {P.}~\bibnamefont
  {Hänggi}},\ }\href {\doibase https://doi.org/10.1016/S0370-1573(98)00022-2}
  {\bibfield  {journal} {\bibinfo  {journal} {Physics Reports}\ }\textbf
  {\bibinfo {volume} {304}},\ \bibinfo {pages} {229} (\bibinfo {year}
  {1998})}\BibitemShut {NoStop}%
\bibitem [{\citenamefont {Chirikov}(1979)}]{chirikov_universal_1979}%
  \BibitemOpen
  \bibfield  {author} {\bibinfo {author} {\bibfnamefont {B.~V.}\ \bibnamefont
  {Chirikov}},\ }\href {\doibase https://doi.org/10.1016/0370-1573(79)90023-1}
  {\bibfield  {journal} {\bibinfo  {journal} {Physics Reports}\ }\textbf
  {\bibinfo {volume} {52}},\ \bibinfo {pages} {263} (\bibinfo {year}
  {1979})}\BibitemShut {NoStop}%
\bibitem [{\citenamefont {Percival}(1973)}]{percival_regular_1973}%
  \BibitemOpen
  \bibfield  {author} {\bibinfo {author} {\bibfnamefont {I.~C.}\ \bibnamefont
  {Percival}},\ }\href {\doibase 10.1088/0022-3700/6/9/002} {\bibfield
  {journal} {\bibinfo  {journal} {Journal of Physics B: Atomic and Molecular
  Physics}\ }\textbf {\bibinfo {volume} {6}},\ \bibinfo {pages} {L229}
  (\bibinfo {year} {1973})}\BibitemShut {NoStop}%
\bibitem [{\citenamefont {Berry}(1977)}]{berry_regular_1977}%
  \BibitemOpen
  \bibfield  {author} {\bibinfo {author} {\bibfnamefont {M.~V.}\ \bibnamefont
  {Berry}},\ }\href@noop {} {\bibfield  {journal} {\bibinfo  {journal} {Journal
  of Physics A: Math. Gen.}\ }\textbf {\bibinfo {volume} {10}},\ \bibinfo
  {pages} {2083} (\bibinfo {year} {1977})}\BibitemShut {NoStop}%
\bibitem [{\citenamefont {Haake}(2010)}]{haake_quantum_2010}%
  \BibitemOpen
  \bibfield  {author} {\bibinfo {author} {\bibfnamefont {F.}~\bibnamefont
  {Haake}},\ }\href {https://link.springer.com/book/10.1007/978-3-642-05428-0}
  {\emph {\bibinfo {title} {{Quantum Signatures of Chaos}}}}\ (\bibinfo
  {publisher} {Springer},\ \bibinfo {address} {Berlin, Germany},\ \bibinfo
  {year} {2010})\BibitemShut {NoStop}%
\bibitem [{\citenamefont {Berke}\ \emph {et~al.}(2022)\citenamefont {Berke},
  \citenamefont {Varvelis}, \citenamefont {Trebst}, \citenamefont {Altland},\
  and\ \citenamefont {DiVincenzo}}]{Berke_2022}%
  \BibitemOpen
  \bibfield  {author} {\bibinfo {author} {\bibfnamefont {C.}~\bibnamefont
  {Berke}}, \bibinfo {author} {\bibfnamefont {E.}~\bibnamefont {Varvelis}},
  \bibinfo {author} {\bibfnamefont {S.}~\bibnamefont {Trebst}}, \bibinfo
  {author} {\bibfnamefont {A.}~\bibnamefont {Altland}}, \ and\ \bibinfo
  {author} {\bibfnamefont {D.~P.}\ \bibnamefont {DiVincenzo}},\ }\href
  {\doibase 10.1038/s41467-022-29940-y} {\bibfield  {journal} {\bibinfo
  {journal} {Nat. Commun.}\ }\textbf {\bibinfo {volume} {13}},\ \bibinfo
  {pages} {1} (\bibinfo {year} {2022})}\BibitemShut {NoStop}%
\bibitem [{\citenamefont {Blais}\ \emph {et~al.}(2021)\citenamefont {Blais},
  \citenamefont {Grimsmo}, \citenamefont {Girvin},\ and\ \citenamefont
  {Wallraff}}]{Blais2021}%
  \BibitemOpen
  \bibfield  {author} {\bibinfo {author} {\bibfnamefont {A.}~\bibnamefont
  {Blais}}, \bibinfo {author} {\bibfnamefont {A.~L.}\ \bibnamefont {Grimsmo}},
  \bibinfo {author} {\bibfnamefont {S.~M.}\ \bibnamefont {Girvin}}, \ and\
  \bibinfo {author} {\bibfnamefont {A.}~\bibnamefont {Wallraff}},\ }\href
  {\doibase 10.1103/RevModPhys.93.025005} {\bibfield  {journal} {\bibinfo
  {journal} {Rev. Mod. Phys.}\ }\textbf {\bibinfo {volume} {93}},\ \bibinfo
  {pages} {025005} (\bibinfo {year} {2021})}\BibitemShut {NoStop}%
\bibitem [{\citenamefont {Zaslavskiî}\ \emph {et~al.}(1991)\citenamefont
  {Zaslavskiî}, \citenamefont {Sagdeev}, \citenamefont {Usikov},\ and\
  \citenamefont {Chernikov}}]{zaslavskii1991}%
  \BibitemOpen
  \bibfield  {author} {\bibinfo {author} {\bibfnamefont {G.~M.}\ \bibnamefont
  {Zaslavskiî}}, \bibinfo {author} {\bibfnamefont {R.~Z.}\ \bibnamefont
  {Sagdeev}}, \bibinfo {author} {\bibfnamefont {D.~A.}\ \bibnamefont {Usikov}},
  \ and\ \bibinfo {author} {\bibfnamefont {A.~A.}\ \bibnamefont {Chernikov}},\
  }\href {\doibase 10.1017/CBO9780511599996} {\emph {\bibinfo {title} {Weak
  Chaos and Quasi-Regular Patterns}}},\ edited by\ \bibinfo {editor}
  {\bibfnamefont {A.~R.}\ \bibnamefont {Sagdeeva}},\ Cambridge Nonlinear
  Science Series\ (\bibinfo  {publisher} {Cambridge University Press},\
  \bibinfo {year} {1991})\BibitemShut {NoStop}%
\bibitem [{\citenamefont {Ketzmerick}\ and\ \citenamefont
  {Wustmann}(2010{\natexlab{b}})}]{KetzmerickPRE2010}%
  \BibitemOpen
  \bibfield  {author} {\bibinfo {author} {\bibfnamefont {R.}~\bibnamefont
  {Ketzmerick}}\ and\ \bibinfo {author} {\bibfnamefont {W.}~\bibnamefont
  {Wustmann}},\ }\href {\doibase 10.1103/PhysRevE.82.021114} {\bibfield
  {journal} {\bibinfo  {journal} {Phys. Rev. E}\ }\textbf {\bibinfo {volume}
  {82}},\ \bibinfo {pages} {021114} (\bibinfo {year}
  {2010}{\natexlab{b}})}\BibitemShut {NoStop}%
\bibitem [{\citenamefont {Bäcker}\ \emph {et~al.}(2008)\citenamefont
  {Bäcker}, \citenamefont {Ketzmerick}, \citenamefont {Löck},\ and\
  \citenamefont {Schilling}}]{backer_regular--chaotic_2008}%
  \BibitemOpen
  \bibfield  {author} {\bibinfo {author} {\bibfnamefont {A.}~\bibnamefont
  {Bäcker}}, \bibinfo {author} {\bibfnamefont {R.}~\bibnamefont {Ketzmerick}},
  \bibinfo {author} {\bibfnamefont {S.}~\bibnamefont {Löck}}, \ and\ \bibinfo
  {author} {\bibfnamefont {L.}~\bibnamefont {Schilling}},\ }\href {\doibase
  10.1103/PhysRevLett.100.104101} {\bibfield  {journal} {\bibinfo  {journal}
  {Physical Review Letters}\ }\textbf {\bibinfo {volume} {100}},\ \bibinfo
  {pages} {104101} (\bibinfo {year} {2008})}\BibitemShut {NoStop}%
\bibitem [{\citenamefont {Graham}(1988)}]{graham_quantization_1988}%
  \BibitemOpen
  \bibfield  {author} {\bibinfo {author} {\bibfnamefont {R.}~\bibnamefont
  {Graham}},\ }\href {\doibase 10.1209/0295-5075/7/8/001} {\bibfield  {journal}
  {\bibinfo  {journal} {Europhysics Letters (EPL)}\ }\textbf {\bibinfo {volume}
  {7}},\ \bibinfo {pages} {671} (\bibinfo {year} {1988})}\BibitemShut {NoStop}%
\bibitem [{\citenamefont {Gonz{\'{a}}lez}\ and\ \citenamefont {del
  Olmo}(1998)}]{Gonz_lez_1998}%
  \BibitemOpen
  \bibfield  {author} {\bibinfo {author} {\bibfnamefont {J.~A.}\ \bibnamefont
  {Gonz{\'{a}}lez}}\ and\ \bibinfo {author} {\bibfnamefont {M.~A.}\
  \bibnamefont {del Olmo}},\ }\href {\doibase 10.1088/0305-4470/31/44/012}
  {\bibfield  {journal} {\bibinfo  {journal} {Journal of Physics A:
  Mathematical and General}\ }\textbf {\bibinfo {volume} {31}},\ \bibinfo
  {pages} {8841} (\bibinfo {year} {1998})}\BibitemShut {NoStop}%
\bibitem [{\citenamefont {Mehta}(1991)}]{Mehta-1991}%
  \BibitemOpen
  \bibfield  {author} {\bibinfo {author} {\bibfnamefont {M.~L.}\ \bibnamefont
  {Mehta}},\ }\href {\doibase
  https://doi.org/10.1016/B978-0-12-488051-1.50006-8} {\emph {\bibinfo {title}
  {Random Matrices}}},\ \bibinfo {edition} {revised and enlarged 2nd}\ ed.,\
  edited by\ \bibinfo {editor} {\bibfnamefont {M.~L.}\ \bibnamefont {Mehta}}\
  (\bibinfo  {publisher} {Academic Press},\ \bibinfo {address} {San Diego},\
  \bibinfo {year} {1991})\ pp.\ \bibinfo {pages} {1--35}\BibitemShut {NoStop}%
\bibitem [{\citenamefont {Bubner}\ and\ \citenamefont
  {Graham}(1991)}]{bubner_quantum_1991}%
  \BibitemOpen
  \bibfield  {author} {\bibinfo {author} {\bibfnamefont {N.}~\bibnamefont
  {Bubner}}\ and\ \bibinfo {author} {\bibfnamefont {R.}~\bibnamefont
  {Graham}},\ }\href {\doibase 10.1103/PhysRevA.43.1783} {\bibfield  {journal}
  {\bibinfo  {journal} {Physical Review A}\ }\textbf {\bibinfo {volume} {43}},\
  \bibinfo {pages} {1783} (\bibinfo {year} {1991})}\BibitemShut {NoStop}%
\bibitem [{\citenamefont {Martinez}\ \emph {et~al.}(2021)\citenamefont
  {Martinez}, \citenamefont {Giraud}, \citenamefont {Ullmo}, \citenamefont
  {Billy}, \citenamefont {Guéry-Odelin}, \citenamefont {Georgeot},\ and\
  \citenamefont {Lemarié}}]{martinez_chaos-assisted_2021}%
  \BibitemOpen
  \bibfield  {author} {\bibinfo {author} {\bibfnamefont {M.}~\bibnamefont
  {Martinez}}, \bibinfo {author} {\bibfnamefont {O.}~\bibnamefont {Giraud}},
  \bibinfo {author} {\bibfnamefont {D.}~\bibnamefont {Ullmo}}, \bibinfo
  {author} {\bibfnamefont {J.}~\bibnamefont {Billy}}, \bibinfo {author}
  {\bibfnamefont {D.}~\bibnamefont {Guéry-Odelin}}, \bibinfo {author}
  {\bibfnamefont {B.}~\bibnamefont {Georgeot}}, \ and\ \bibinfo {author}
  {\bibfnamefont {G.}~\bibnamefont {Lemarié}},\ }\href {\doibase
  10.1103/PhysRevLett.126.174102} {\bibfield  {journal} {\bibinfo  {journal}
  {Physical Review Letters}\ }\textbf {\bibinfo {volume} {126}},\ \bibinfo
  {pages} {174102} (\bibinfo {year} {2021})}\BibitemShut {NoStop}%
\bibitem [{\citenamefont {Koch}\ \emph {et~al.}(2009)\citenamefont {Koch},
  \citenamefont {Manucharyan}, \citenamefont {Devoret},\ and\ \citenamefont
  {Glazman}}]{koch_charging_2009}%
  \BibitemOpen
  \bibfield  {author} {\bibinfo {author} {\bibfnamefont {J.}~\bibnamefont
  {Koch}}, \bibinfo {author} {\bibfnamefont {V.}~\bibnamefont {Manucharyan}},
  \bibinfo {author} {\bibfnamefont {M.~H.}\ \bibnamefont {Devoret}}, \ and\
  \bibinfo {author} {\bibfnamefont {L.~I.}\ \bibnamefont {Glazman}},\ }\href
  {\doibase 10.1103/PhysRevLett.103.217004} {\bibfield  {journal} {\bibinfo
  {journal} {Physical Review Letters}\ }\textbf {\bibinfo {volume} {103}},\
  \bibinfo {pages} {217004} (\bibinfo {year} {2009})}\BibitemShut {NoStop}%
\bibitem [{\citenamefont {Mouchet}\ \emph {et~al.}(2001)\citenamefont
  {Mouchet}, \citenamefont {Miniatura}, \citenamefont {Kaiser}, \citenamefont
  {Grémaud},\ and\ \citenamefont {Delande}}]{mouchet_chaos_2001}%
  \BibitemOpen
  \bibfield  {author} {\bibinfo {author} {\bibfnamefont {A.}~\bibnamefont
  {Mouchet}}, \bibinfo {author} {\bibfnamefont {C.}~\bibnamefont {Miniatura}},
  \bibinfo {author} {\bibfnamefont {R.}~\bibnamefont {Kaiser}}, \bibinfo
  {author} {\bibfnamefont {B.}~\bibnamefont {Grémaud}}, \ and\ \bibinfo
  {author} {\bibfnamefont {D.}~\bibnamefont {Delande}},\ }\href {\doibase
  10.1103/PhysRevE.64.016221} {\bibfield  {journal} {\bibinfo  {journal}
  {Physical Review E}\ }\textbf {\bibinfo {volume} {64}},\ \bibinfo {pages}
  {016221} (\bibinfo {year} {2001})},\ \bibinfo {note} {arXiv:
  nlin/0012013}\BibitemShut {NoStop}%
\bibitem [{\citenamefont {Arnal}\ \emph {et~al.}(2020)\citenamefont {Arnal},
  \citenamefont {Chatelain}, \citenamefont {Martinez}, \citenamefont {Dupont},
  \citenamefont {Giraud}, \citenamefont {Ullmo}, \citenamefont {Georgeot},
  \citenamefont {Lemarié}, \citenamefont {Billy},\ and\ \citenamefont
  {Guéry-Odelin}}]{Arnal_sciadv_2020}%
  \BibitemOpen
  \bibfield  {author} {\bibinfo {author} {\bibfnamefont {M.}~\bibnamefont
  {Arnal}}, \bibinfo {author} {\bibfnamefont {G.}~\bibnamefont {Chatelain}},
  \bibinfo {author} {\bibfnamefont {M.}~\bibnamefont {Martinez}}, \bibinfo
  {author} {\bibfnamefont {N.}~\bibnamefont {Dupont}}, \bibinfo {author}
  {\bibfnamefont {O.}~\bibnamefont {Giraud}}, \bibinfo {author} {\bibfnamefont
  {D.}~\bibnamefont {Ullmo}}, \bibinfo {author} {\bibfnamefont
  {B.}~\bibnamefont {Georgeot}}, \bibinfo {author} {\bibfnamefont
  {G.}~\bibnamefont {Lemarié}}, \bibinfo {author} {\bibfnamefont
  {J.}~\bibnamefont {Billy}}, \ and\ \bibinfo {author} {\bibfnamefont
  {D.}~\bibnamefont {Guéry-Odelin}},\ }\href {\doibase 10.1126/sciadv.abc4886}
  {\bibfield  {journal} {\bibinfo  {journal} {Science Advances}\ }\textbf
  {\bibinfo {volume} {6}},\ \bibinfo {pages} {eabc4886} (\bibinfo {year}
  {2020})},\ \Eprint
  {http://arxiv.org/abs/https://www.science.org/doi/pdf/10.1126/sciadv.abc4886}
  {https://www.science.org/doi/pdf/10.1126/sciadv.abc4886} \BibitemShut
  {NoStop}%
\bibitem [{\citenamefont {Huang}\ \emph {et~al.}(2020)\citenamefont {Huang},
  \citenamefont {Mundada}, \citenamefont {Gyenis}, \citenamefont {Schuster},
  \citenamefont {Houck},\ and\ \citenamefont {Koch}}]{huang_engineering_2020}%
  \BibitemOpen
  \bibfield  {author} {\bibinfo {author} {\bibfnamefont {Z.}~\bibnamefont
  {Huang}}, \bibinfo {author} {\bibfnamefont {P.~S.}\ \bibnamefont {Mundada}},
  \bibinfo {author} {\bibfnamefont {A.}~\bibnamefont {Gyenis}}, \bibinfo
  {author} {\bibfnamefont {D.~I.}\ \bibnamefont {Schuster}}, \bibinfo {author}
  {\bibfnamefont {A.~A.}\ \bibnamefont {Houck}}, \ and\ \bibinfo {author}
  {\bibfnamefont {J.}~\bibnamefont {Koch}},\ }\href
  {http://arxiv.org/abs/2004.12458} {\bibfield  {journal} {\bibinfo  {journal}
  {arXiv:2004.12458 [quant-ph]}\ } (\bibinfo {year} {2020})},\ \bibinfo {note}
  {arXiv: 2004.12458}\BibitemShut {NoStop}%
\bibitem [{\citenamefont {Ithier}\ \emph {et~al.}(2005)\citenamefont {Ithier},
  \citenamefont {Collin}, \citenamefont {Joyez}, \citenamefont {Meeson},
  \citenamefont {Vion}, \citenamefont {Esteve}, \citenamefont {Chiarello},
  \citenamefont {Shnirman}, \citenamefont {Makhlin}, \citenamefont {Schriefl},\
  and\ \citenamefont {Sch\"on}}]{IthierPRB2006}%
  \BibitemOpen
  \bibfield  {author} {\bibinfo {author} {\bibfnamefont {G.}~\bibnamefont
  {Ithier}}, \bibinfo {author} {\bibfnamefont {E.}~\bibnamefont {Collin}},
  \bibinfo {author} {\bibfnamefont {P.}~\bibnamefont {Joyez}}, \bibinfo
  {author} {\bibfnamefont {P.~J.}\ \bibnamefont {Meeson}}, \bibinfo {author}
  {\bibfnamefont {D.}~\bibnamefont {Vion}}, \bibinfo {author} {\bibfnamefont
  {D.}~\bibnamefont {Esteve}}, \bibinfo {author} {\bibfnamefont
  {F.}~\bibnamefont {Chiarello}}, \bibinfo {author} {\bibfnamefont
  {A.}~\bibnamefont {Shnirman}}, \bibinfo {author} {\bibfnamefont
  {Y.}~\bibnamefont {Makhlin}}, \bibinfo {author} {\bibfnamefont
  {J.}~\bibnamefont {Schriefl}}, \ and\ \bibinfo {author} {\bibfnamefont
  {G.}~\bibnamefont {Sch\"on}},\ }\href {\doibase 10.1103/PhysRevB.72.134519}
  {\bibfield  {journal} {\bibinfo  {journal} {Phys. Rev. B}\ }\textbf {\bibinfo
  {volume} {72}},\ \bibinfo {pages} {134519} (\bibinfo {year}
  {2005})}\BibitemShut {NoStop}%
\bibitem [{\citenamefont {Christensen}\ \emph {et~al.}(2019)\citenamefont
  {Christensen}, \citenamefont {Wilen}, \citenamefont {Opremcak}, \citenamefont
  {Nelson}, \citenamefont {Schlenker}, \citenamefont {Zimonick}, \citenamefont
  {Faoro}, \citenamefont {Ioffe}, \citenamefont {Rosen}, \citenamefont
  {DuBois}, \citenamefont {Plourde},\ and\ \citenamefont
  {McDermott}}]{Christensen2019}%
  \BibitemOpen
  \bibfield  {author} {\bibinfo {author} {\bibfnamefont {B.~G.}\ \bibnamefont
  {Christensen}}, \bibinfo {author} {\bibfnamefont {C.~D.}\ \bibnamefont
  {Wilen}}, \bibinfo {author} {\bibfnamefont {A.}~\bibnamefont {Opremcak}},
  \bibinfo {author} {\bibfnamefont {J.}~\bibnamefont {Nelson}}, \bibinfo
  {author} {\bibfnamefont {F.}~\bibnamefont {Schlenker}}, \bibinfo {author}
  {\bibfnamefont {C.~H.}\ \bibnamefont {Zimonick}}, \bibinfo {author}
  {\bibfnamefont {L.}~\bibnamefont {Faoro}}, \bibinfo {author} {\bibfnamefont
  {L.~B.}\ \bibnamefont {Ioffe}}, \bibinfo {author} {\bibfnamefont {Y.~J.}\
  \bibnamefont {Rosen}}, \bibinfo {author} {\bibfnamefont {J.~L.}\ \bibnamefont
  {DuBois}}, \bibinfo {author} {\bibfnamefont {B.~L.~T.}\ \bibnamefont
  {Plourde}}, \ and\ \bibinfo {author} {\bibfnamefont {R.}~\bibnamefont
  {McDermott}},\ }\href {\doibase 10.1103/PhysRevB.100.140503} {\bibfield
  {journal} {\bibinfo  {journal} {Phys. Rev. B}\ }\textbf {\bibinfo {volume}
  {100}},\ \bibinfo {pages} {140503} (\bibinfo {year} {2019})}\BibitemShut
  {NoStop}%
\bibitem [{\citenamefont {Wilen}\ \emph {et~al.}(2021)\citenamefont {Wilen},
  \citenamefont {Abdullah}, \citenamefont {Kurinsky}, \citenamefont {Stanford},
  \citenamefont {Cardani}, \citenamefont {D'Imperio}, \citenamefont {Tomei},
  \citenamefont {Faoro}, \citenamefont {Ioffe}, \citenamefont {Liu},
  \citenamefont {Opremcak}, \citenamefont {Christensen}, \citenamefont
  {DuBois},\ and\ \citenamefont {McDermott}}]{Wilen2021}%
  \BibitemOpen
  \bibfield  {author} {\bibinfo {author} {\bibfnamefont {C.~D.}\ \bibnamefont
  {Wilen}}, \bibinfo {author} {\bibfnamefont {S.}~\bibnamefont {Abdullah}},
  \bibinfo {author} {\bibfnamefont {N.~A.}\ \bibnamefont {Kurinsky}}, \bibinfo
  {author} {\bibfnamefont {C.}~\bibnamefont {Stanford}}, \bibinfo {author}
  {\bibfnamefont {L.}~\bibnamefont {Cardani}}, \bibinfo {author} {\bibfnamefont
  {G.}~\bibnamefont {D'Imperio}}, \bibinfo {author} {\bibfnamefont
  {C.}~\bibnamefont {Tomei}}, \bibinfo {author} {\bibfnamefont
  {L.}~\bibnamefont {Faoro}}, \bibinfo {author} {\bibfnamefont {L.~B.}\
  \bibnamefont {Ioffe}}, \bibinfo {author} {\bibfnamefont {C.~H.}\ \bibnamefont
  {Liu}}, \bibinfo {author} {\bibfnamefont {A.}~\bibnamefont {Opremcak}},
  \bibinfo {author} {\bibfnamefont {B.~G.}\ \bibnamefont {Christensen}},
  \bibinfo {author} {\bibfnamefont {J.~L.}\ \bibnamefont {DuBois}}, \ and\
  \bibinfo {author} {\bibfnamefont {R.}~\bibnamefont {McDermott}},\ }\href
  {\doibase 10.1038/s41586-021-03557-5} {\bibfield  {journal} {\bibinfo
  {journal} {Nature}\ }\textbf {\bibinfo {volume} {594}},\ \bibinfo {pages}
  {369} (\bibinfo {year} {2021})}\BibitemShut {NoStop}%
\bibitem [{\citenamefont {Boissonneault}\ \emph {et~al.}(2010)\citenamefont
  {Boissonneault}, \citenamefont {Gambetta},\ and\ \citenamefont
  {Blais}}]{Boissonneault_2010}%
  \BibitemOpen
  \bibfield  {author} {\bibinfo {author} {\bibfnamefont {M.}~\bibnamefont
  {Boissonneault}}, \bibinfo {author} {\bibfnamefont {J.~M.}\ \bibnamefont
  {Gambetta}}, \ and\ \bibinfo {author} {\bibfnamefont {A.}~\bibnamefont
  {Blais}},\ }\href {\doibase 10.1103/PhysRevLett.105.100504} {\bibfield
  {journal} {\bibinfo  {journal} {Phys. Rev. Lett.}\ }\textbf {\bibinfo
  {volume} {105}},\ \bibinfo {pages} {100504} (\bibinfo {year}
  {2010})}\BibitemShut {NoStop}%
\bibitem [{\citenamefont {Gramich}\ \emph {et~al.}(2014)\citenamefont
  {Gramich}, \citenamefont {Gasparinetti}, \citenamefont {Solinas},\ and\
  \citenamefont {Ankerhold}}]{gramich_lamb-shift_2014}%
  \BibitemOpen
  \bibfield  {author} {\bibinfo {author} {\bibfnamefont {V.}~\bibnamefont
  {Gramich}}, \bibinfo {author} {\bibfnamefont {S.}~\bibnamefont
  {Gasparinetti}}, \bibinfo {author} {\bibfnamefont {P.}~\bibnamefont
  {Solinas}}, \ and\ \bibinfo {author} {\bibfnamefont {J.}~\bibnamefont
  {Ankerhold}},\ }\href {\doibase 10.1103/PhysRevLett.113.027001} {\bibfield
  {journal} {\bibinfo  {journal} {Physical Review Letters}\ }\textbf {\bibinfo
  {volume} {113}},\ \bibinfo {pages} {027001} (\bibinfo {year}
  {2014})}\BibitemShut {NoStop}%
\bibitem [{\citenamefont {Malekakhlagh}\ \emph
  {et~al.}(2020{\natexlab{b}})\citenamefont {Malekakhlagh}, \citenamefont
  {Magesan},\ and\ \citenamefont {McKay}}]{malekakhlagh_first-principles_2020}%
  \BibitemOpen
  \bibfield  {author} {\bibinfo {author} {\bibfnamefont {M.}~\bibnamefont
  {Malekakhlagh}}, \bibinfo {author} {\bibfnamefont {E.}~\bibnamefont
  {Magesan}}, \ and\ \bibinfo {author} {\bibfnamefont {D.~C.}\ \bibnamefont
  {McKay}},\ }\href {\doibase 10.1103/PhysRevA.102.042605} {\bibfield
  {journal} {\bibinfo  {journal} {Phys. Rev. A}\ }\textbf {\bibinfo {volume}
  {102}},\ \bibinfo {pages} {042605} (\bibinfo {year}
  {2020}{\natexlab{b}})}\BibitemShut {NoStop}%
\bibitem [{\citenamefont {Hanai}\ \emph {et~al.}(2021)\citenamefont {Hanai},
  \citenamefont {McDonald},\ and\ \citenamefont
  {Clerk}}]{hanai_intrinsic_2021}%
  \BibitemOpen
  \bibfield  {author} {\bibinfo {author} {\bibfnamefont {R.}~\bibnamefont
  {Hanai}}, \bibinfo {author} {\bibfnamefont {A.}~\bibnamefont {McDonald}}, \
  and\ \bibinfo {author} {\bibfnamefont {A.}~\bibnamefont {Clerk}},\ }\href
  {\doibase 10.1103/PhysRevResearch.3.043228} {\bibfield  {journal} {\bibinfo
  {journal} {Physical Review Research}\ }\textbf {\bibinfo {volume} {3}},\
  \bibinfo {pages} {043228} (\bibinfo {year} {2021})}\BibitemShut {NoStop}%
\bibitem [{\citenamefont {Houzet}\ and\ \citenamefont
  {Glazman}(2020)}]{houzet_critical_2020}%
  \BibitemOpen
  \bibfield  {author} {\bibinfo {author} {\bibfnamefont {M.}~\bibnamefont
  {Houzet}}\ and\ \bibinfo {author} {\bibfnamefont {L.~I.}\ \bibnamefont
  {Glazman}},\ }\href {http://arxiv.org/abs/2012.02233} {\bibfield  {journal}
  {\bibinfo  {journal} {arXiv:2012.02233 [cond-mat]}\ } (\bibinfo {year}
  {2020})},\ \bibinfo {note} {arXiv: 2012.02233}\BibitemShut {NoStop}%
\bibitem [{\citenamefont {Sheinman}\ \emph {et~al.}(2006)\citenamefont
  {Sheinman}, \citenamefont {Fishman}, \citenamefont {Guarneri},\ and\
  \citenamefont {Rebuzzini}}]{sheinman_decay_2006}%
  \BibitemOpen
  \bibfield  {author} {\bibinfo {author} {\bibfnamefont {M.}~\bibnamefont
  {Sheinman}}, \bibinfo {author} {\bibfnamefont {S.}~\bibnamefont {Fishman}},
  \bibinfo {author} {\bibfnamefont {I.}~\bibnamefont {Guarneri}}, \ and\
  \bibinfo {author} {\bibfnamefont {L.}~\bibnamefont {Rebuzzini}},\ }\href
  {\doibase 10.1103/PhysRevA.73.052110} {\bibfield  {journal} {\bibinfo
  {journal} {Physical Review A}\ }\textbf {\bibinfo {volume} {73}},\ \bibinfo
  {pages} {052110} (\bibinfo {year} {2006})}\BibitemShut {NoStop}%
\bibitem [{\citenamefont {Nguyen}\ \emph {et~al.}(2019)\citenamefont {Nguyen},
  \citenamefont {Lin}, \citenamefont {Somoroff}, \citenamefont {Mencia},
  \citenamefont {Grabon},\ and\ \citenamefont
  {Manucharyan}}]{nguyen_high-coherence_2019}%
  \BibitemOpen
  \bibfield  {author} {\bibinfo {author} {\bibfnamefont {L.~B.}\ \bibnamefont
  {Nguyen}}, \bibinfo {author} {\bibfnamefont {Y.-H.}\ \bibnamefont {Lin}},
  \bibinfo {author} {\bibfnamefont {A.}~\bibnamefont {Somoroff}}, \bibinfo
  {author} {\bibfnamefont {R.}~\bibnamefont {Mencia}}, \bibinfo {author}
  {\bibfnamefont {N.}~\bibnamefont {Grabon}}, \ and\ \bibinfo {author}
  {\bibfnamefont {V.~E.}\ \bibnamefont {Manucharyan}},\ }\href {\doibase
  10.1103/PhysRevX.9.041041} {\bibfield  {journal} {\bibinfo  {journal}
  {Physical Review X}\ }\textbf {\bibinfo {volume} {9}},\ \bibinfo {pages}
  {041041} (\bibinfo {year} {2019})}\BibitemShut {NoStop}%
\bibitem [{\citenamefont {Burgelman}\ \emph {et~al.}(2022)\citenamefont
  {Burgelman}, \citenamefont {Rouchon}, \citenamefont {Sarlette},\ and\
  \citenamefont {Mirrahimi}}]{burgelman_et_al_2022}%
  \BibitemOpen
  \bibfield  {author} {\bibinfo {author} {\bibfnamefont {M.}~\bibnamefont
  {Burgelman}}, \bibinfo {author} {\bibfnamefont {P.}~\bibnamefont {Rouchon}},
  \bibinfo {author} {\bibfnamefont {A.}~\bibnamefont {Sarlette}}, \ and\
  \bibinfo {author} {\bibfnamefont {M.}~\bibnamefont {Mirrahimi}},\ }\href
  {\doibase 10.48550/ARXIV.2206.14631} {\enquote {\bibinfo {title}
  {Structurally stable subharmonic regime of a driven quantum josephson
  circuit},}\ } (\bibinfo {year} {2022})\BibitemShut {NoStop}%
\bibitem [{\citenamefont {Reed}\ \emph {et~al.}(2010)\citenamefont {Reed},
  \citenamefont {DiCarlo}, \citenamefont {Johnson}, \citenamefont {Sun},
  \citenamefont {Schuster}, \citenamefont {Frunzio},\ and\ \citenamefont
  {Schoelkopf}}]{reed_high-fidelity_2010}%
  \BibitemOpen
  \bibfield  {author} {\bibinfo {author} {\bibfnamefont {M.~D.}\ \bibnamefont
  {Reed}}, \bibinfo {author} {\bibfnamefont {L.}~\bibnamefont {DiCarlo}},
  \bibinfo {author} {\bibfnamefont {B.~R.}\ \bibnamefont {Johnson}}, \bibinfo
  {author} {\bibfnamefont {L.}~\bibnamefont {Sun}}, \bibinfo {author}
  {\bibfnamefont {D.~I.}\ \bibnamefont {Schuster}}, \bibinfo {author}
  {\bibfnamefont {L.}~\bibnamefont {Frunzio}}, \ and\ \bibinfo {author}
  {\bibfnamefont {R.~J.}\ \bibnamefont {Schoelkopf}},\ }\href {\doibase
  10.1103/PhysRevLett.105.173601} {\bibfield  {journal} {\bibinfo  {journal}
  {Physical Review Letters}\ }\textbf {\bibinfo {volume} {105}},\ \bibinfo
  {pages} {173601} (\bibinfo {year} {2010})},\ \bibinfo {note} {arXiv:
  1004.4323}\BibitemShut {NoStop}%
\bibitem [{\citenamefont {Breuer}\ and\ \citenamefont
  {Holthaus}(1989)}]{Breuer1989Mar}%
  \BibitemOpen
  \bibfield  {author} {\bibinfo {author} {\bibfnamefont {H.~P.}\ \bibnamefont
  {Breuer}}\ and\ \bibinfo {author} {\bibfnamefont {M.}~\bibnamefont
  {Holthaus}},\ }\href {\doibase 10.1007/BF01436579} {\bibfield  {journal}
  {\bibinfo  {journal} {Z. Phys. D: At. Mol. Clusters}\ }\textbf {\bibinfo
  {volume} {11}},\ \bibinfo {pages} {1} (\bibinfo {year} {1989})}\BibitemShut
  {NoStop}%
\end{thebibliography}%
\newpage

\end{document}